\documentclass[preprint,12pt, a4paper]{elsarticle}

\usepackage{tikz}
\usepackage{cleveref}
\usepackage{url}

\usepackage{quiver}
\usetikzlibrary{fit}
\newcommand{\aeh}{algebraic effects \& handlers }
\newcommand{\Aeh}{Algebraic effects \& handlers }

\newcommand{\step}[1]{\par\noindent \textbf{Step {#1}}}

\newtheorem{definition}{Definition}

\makeatletter
\@ifundefined{lhs2tex.lhs2tex.sty.read}%
  {\@namedef{lhs2tex.lhs2tex.sty.read}{}%
   \newcommand\SkipToFmtEnd{}%
   \newcommand\EndFmtInput{}%
   \long\def\SkipToFmtEnd#1\EndFmtInput{}%
  }\SkipToFmtEnd

\newcommand\ReadOnlyOnce[1]{\@ifundefined{#1}{\@namedef{#1}{}}\SkipToFmtEnd}
\usepackage{amstext}
\usepackage{amssymb}
\usepackage{stmaryrd}
\DeclareFontFamily{OT1}{cmtex}{}
\DeclareFontShape{OT1}{cmtex}{m}{n}
  {<5><6><7><8>cmtex8
   <9>cmtex9
   <10><10.95><12><14.4><17.28><20.74><24.88>cmtex10}{}
\DeclareFontShape{OT1}{cmtex}{m}{it}
  {<-> ssub * cmtt/m/it}{}

\DeclareFontShape{OT1}{cmtt}{bx}{n}
  {<5><6><7><8>cmtt8
   <9>cmbtt9
   <10><10.95><12><14.4><17.28><20.74><24.88>cmbtt10}{}
\DeclareFontShape{OT1}{cmtex}{bx}{n}
  {<-> ssub * cmtt/bx/n}{}
\newcommand{\Conid}[1]{\mathit{#1}}
\newcommand{\Varid}[1]{\mathit{#1}}
\newcommand{\anonymous}{\kern0.06em \vbox{\hrule\@width.5em}}
\newcommand{\plus}{\mathbin{+\!\!\!+}}
\newcommand{\bind}{\mathbin{>\!\!\!>\mkern-6.7mu=}}
\newcommand{\sequ}{\mathbin{>\!\!\!>}}

\renewcommand{\geq}{\geqslant}
\usepackage{polytable}

\@ifundefined{mathindent}%
  {\newdimen\mathindent\mathindent\leftmargini}%
  {}%

\def\resethooks{%
  \global\let\SaveRestoreHook\empty
  \global\let\ColumnHook\empty}
\newcommand*{\savecolumns}[1][default]%
  {\g@addto@macro\SaveRestoreHook{\savecolumns[#1]}}
\newcommand*{\restorecolumns}[1][default]%
  {\g@addto@macro\SaveRestoreHook{\restorecolumns[#1]}}
\newcommand*{\aligncolumn}[2]%
  {\g@addto@macro\ColumnHook{\column{#1}{#2}}}

\resethooks

\newcommand{\onelinecommentchars}{\quad-{}- }
\newcommand{\commentbeginchars}{\enskip\{-}
\newcommand{\commentendchars}{-\}\enskip}

\newcommand{\visiblecomments}{%
  \let\onelinecomment=\onelinecommentchars
  \let\commentbegin=\commentbeginchars
  \let\commentend=\commentendchars}

\newcommand{\invisiblecomments}{%
  \let\onelinecomment=\empty
  \let\commentbegin=\empty
  \let\commentend=\empty}

\visiblecomments

\newlength{\blanklineskip}
\newcommand{\hsindent}[1]{\quad}% default is fixed indentation
\let\hspre\empty
\let\hspost\empty

\EndFmtInput
\makeatother
\ReadOnlyOnce{polycode.fmt}%
\makeatletter

\newcommand{\hsnewpar}[1]%
  {{\parskip=0pt\parindent=0pt\par\vskip #1\noindent}}

\newcommand{\hscodestyle}{}

\newcommand{\sethscode}[1]%
  {\expandafter\let\expandafter\hscode\csname #1\endcsname
   \expandafter\let\expandafter\endhscode\csname end#1\endcsname}

  {\par\noindent
   \advance\leftskip\mathindent
   \hscodestyle
   \let\\=\@normalcr
   \let\hspre\(\let\hspost\)%
   \pboxed}%
  {\endpboxed\)%
   \par\noindent
   \ignorespacesafterend}

  {\hsnewpar\abovedisplayskip
   \advance\leftskip\mathindent
   \hscodestyle
   \let\hspre\(\let\hspost\)%
   \pboxed}%
  {\endpboxed%
   \hsnewpar\belowdisplayskip
   \ignorespacesafterend}

  {\hsnewpar\abovedisplayskip
   \advance\leftskip\mathindent
   \hscodestyle
   \let\\=\@normalcr
   \(\pboxed}%
  {\endpboxed\)%
   \hsnewpar\belowdisplayskip
   \ignorespacesafterend}

\newcommand{\plainhs}{\sethscode{plainhscode}}

\plainhs

  {\hsnewpar\abovedisplayskip
   \advance\leftskip\mathindent
   \hscodestyle
   \let\\=\@normalcr
   \(\parray}%
  {\endparray\)%
   \hsnewpar\belowdisplayskip
   \ignorespacesafterend}

  {\parray}{\endparray}

  {\(\parray}{\endparray\)}

\def\codeframewidth{\arrayrulewidth}
\RequirePackage{calc}

  {\parskip=\abovedisplayskip\par\noindent
   \hscodestyle
   \arrayrulewidth=\codeframewidth
   \tabular{@{}|p{\linewidth-2\arraycolsep-2\arrayrulewidth-2pt}|@{}}%
   \hline\framedhslinecorrect\\{-1.5ex}%
   \let\endoflinesave=\\
   \let\\=\@normalcr
   \(\pboxed}%
  {\endpboxed\)%
   \framedhslinecorrect\endoflinesave{.5ex}\hline
   \endtabular
   \parskip=\belowdisplayskip\par\noindent
   \ignorespacesafterend}

\newcommand{\framedhslinecorrect}[2]%
  {#1[#2]}

  {\(\def\column##1##2{}%
   \let\>\undefined\let\<\undefined\let\\\undefined
   \newcommand\>[1][]{}\newcommand\<[1][]{}\newcommand\\[1][]{}%
   \def\fromto##1##2##3{##3}%
   }{\) }%

  {\let\orighscode=\hscode
   \let\origendhscode=\endhscode
   \def\endhscode{\def\hscode{\endgroup\def\@currenvir{hscode}\\}\begingroup}
   \orighscode\def\hscode{\endgroup\def\@currenvir{hscode}}}%
  {\origendhscode
   \global\let\hscode=\orighscode
   \global\let\endhscode=\origendhscode}%

\makeatother
\EndFmtInput
\usepackage{enumitem}
\newlist{qwq}{itemize}{1}
\setlist[qwq]{label={}, nosep, leftmargin=-0.35cm}
\newcommand{\indentbegin}{\begin{qwq} \item}
  \newcommand{\indentend}{\end{qwq}}

\journal{Science of Computer Programming}

\begin{document}

\begin{frontmatter}



\title{A Framework for Higher-Order Effects \& Handlers}


\author{Birthe van den Berg and Tom Schrijvers}

\address{KU Leuven, Belgium}

\begin{abstract}
%%%%%%%%%%%%%%%%%%%%%%%%%%%%%%%%%%%%%%%%%%%
%%% Context: (Why now? Give background) %%%
%%%%%%%%%%%%%%%%%%%%%%%%%%%%%%%%%%%%%%%%%%%
%
Algebraic effects \& handlers are a modular approach
for modeling side-effects in functional programming.
Their syntax is defined in terms of a signature of effectful operations,
encoded as a functor, that are plugged into the free monad;
their denotational semantics is defined by fold-style handlers
that only interpret their part of the syntax and forward the rest.
%
%%%%%%%%%%%%%%%%%%%%%%%%%%%%%%%%%%%%%%%%%%%%%%%%%%%%%%%%%
%%% Need (Why you: Why is this useful to the reader?) %%%
%%%%%%%%%%%%%%%%%%%%%%%%%%%%%%%%%%%%%%%%%%%%%%%%%%%%%%%%%
%
%%%%%%%%%%%%%%%%%%%%%%%%%%%%%%%%%%%%%%%%%%%%%%
%%% Task (Why me/us: What are we solving?) %%%
%%%%%%%%%%%%%%%%%%%%%%%%%%%%%%%%%%%%%%%%%%%%%%
%
However, not all effects are algebraic: some need to access
an \emph{internal computation}.
For example, scoped effects distinguish between a computation in scope and
out of scope; parallel effects parallellize over a computation,
latent effects defer a computation.
Separate definitions have been proposed for these \emph{higher-order effects}
and their corresponding handlers, often leading to expedient and complex monad
definitions.
%
%%%%%%%%%%%%%%%%%%%%%%%%%%%%%%%%%%%%%%%%%%%%%%%%%%%%%%%%%%%%%%%%%%%
%%% Object (Why this document: How are we going to solve this?) %%%
%%%%%%%%%%%%%%%%%%%%%%%%%%%%%%%%%%%%%%%%%%%%%%%%%%%%%%%%%%%%%%%%%%%
%
In this work we propose a generic framework for higher-order effects,
generalizing algebraic effects \& handlers:
a generic free monad with higher-order effect signatures
and a corresponding interpreter.
%
%%%%%%%%%%%%%%%%%%%%%%%%%%%%%%%%%%%%%%%%%%
%%% Findings (What? We found that ...) %%%
%%%%%%%%%%%%%%%%%%%%%%%%%%%%%%%%%%%%%%%%%%
%
Specializing this higher-order syntax leads to various
definitions of previously defined (scoped, parallel, latent) and novel
(writer, bracketing) effects.
Furthermore, we formally show our framework theoretically correct,
also putting different effect instances on formal footing; a
significant contribution for parallel, latent, writer and bracketing effects.
%
%%%%%%%%%%%%%%%%%%%%%%%%%%%%%%%%%%%%%%%%%%%%%%%%%%%%%%%%%%%%%%%%%%
%%% Conclusions (So what? What do these results imply?) %%%%%%%%%%
%%%%%%%%%%%%%%%%%%%%%%%%%%%%%%%%%%%%%%%%%%%%%%%%%%%%%%%%%%%%%%%%%%
%
% We not only show this framework practically useful,
% we also show it theoretically correct.
%
%%%%%%%%%%%%%%%%%%%%%%%%%%%%%%%%%%%%%%%%%%%%%%%%%%%%%%%%%%%%%%%%%%
%%% Perspective: (What now? What does the future look like?) %%%%%
%%%%%%%%%%%%%%%%%%%%%%%%%%%%%%%%%%%%%%%%%%%%%%%%%%%%%%%%%%%%%%%%%%
%
\end{abstract}


\end{frontmatter}

%==================================================================
\section{Introduction}

Since the nineties, monads \cite{Moggi89,Moggi91,Wadler95} have been the standard for
modeling effects in a purely functional setting. \Aeh \cite{Plotkin09,Plotkin03},
however, have been gaining significant interest in recent years \cite{fused-effects,extensible-effects,eff-ocaml}.
The latter offer a modular approach to combining different kinds of effects.
In particular, it is their clean separation of \emph{algebraic operations} (syntax)
and \emph{handlers} (semantics) that makes them so attractive.
Composing different algebraic effects in a particular order
using the coproduct of their operations \cite{dtc} implies a different order of
handlers and consequently leads to a different interpretation \cite{zhixuan21}.
Handlers only know their part of the syntax, forwarding unknown effects to other
handlers.

Although this modular technique of algebraic effects \& handlers is desirable for
every effectful program, not all effects fit the approach.
The algebraicity property, which states that effectful operations
commute with sequencing, is not satisfied for all kinds of effects.
The proposal of Plotkin and Power \cite{Plotkin03} to model non-algebraic
effects as handlers, was shown \cite{scope14,esop22} to lose modularity. %, the coveted property of
% algebraic effects \& handlers.

In this work, we propose a generic syntax and denotational semantics for different
effects so that the desired modularity of separate syntax and semantics is retained.
In particular, we focus on higher-order effects, i.e., effects that have access to an internal
computation.
For example,
\emph{scoped effects \& handlers} \cite{esop22,lics18,scope14} (e.g.,
catching exceptions, local variables) distinguish between an operation in scope
and a continuation out of scope;
\emph{latent effects \& handlers} \cite{vandenBerg21} (e.g., lazy evaluation, staging)
defer an internal computation until a later point of execution;
\emph{parallel effects \& handlers} \cite{xie21} (e.g., for-loops)
parallellize an internal computation to improve performance or efficiency.

We achieve the desired modularity by extending the free monad definition of \aeh
to support higher-order functors, which incorporate the internal
computation in their effect signatures.
This, together with a generic fold-style interpreter, forms the framework for
modeling different effects.
In particular, the contributions of this article can be summarized as follows:

\begin{itemize}
  \item We define a generic free monad with higher-order effect signatures \cite{higher-order}
  and a corresponding fold-style interpreter to represent effects that can reason over an internal computation (\Cref{sec:generic}).
  \item We model different effects from the literature, such as \emph{algebraic effects \& handlers}, \emph{scoped effects \& handlers}, \emph{parallel effects \& handlers}, and \emph{latent effects \& handlers}, as instances of this framework
  (\Cref{sec:instances}).
  \item We model two novel classes of effects: \emph{writer effect \& handler} and \emph{bracketing effect \& handler}
  to denote the functionality of writing to a log and safely dealing with resources, respectively (\Cref{sec:instances}).
  \item We show that each class of (existing and novel) effects is modeled by a free monad with a corresponding recursion scheme (\Cref{sec:instances}).
  \item We provide practical programming examples for all effects,
  including combinations of different effects (\Cref{sec:instances}).
  \item We back our free monad definition by a categorical model that is based on
  a free-forgetful adjunction (\Cref{sec:formalization}).
  \item We deliver a full implementation in Haskell with more, detailed examples (Supplementary Material).
\end{itemize}

\noindent
We use Haskell % \footnote{We believe that other languages are equally suited for this approach.}
as a vehicle throughout this paper to illustrate our findings.

%==================================================================
\section{Background and Motivation}
\label{sec:background}

This section deals with the necessary preliminaries, starting with
functors, free monads, and their relation to algebraic effects \& handlers.

%-------------------------------------------------------------------------------
\subsection{Algebraic Effects \& Handlers}
\label{sec:aeh}

\Aeh \cite{Plotkin09,Plotkin03} use a free monad and an interpreter to
separate the syntax of effects from their semantics.
At the level of the syntax we distinguish \emph{algebraic operations}, determined by their signatures,
from \emph{computations}: a recursive structure over these operations via the free monad.
Furthermore, a recursion scheme (\emph{handler}) interprets these computations.

% - - - - - - - - - - - - - - - - - - - - - - - - - - - - - - - - - - - - - - - -
\paragraph{Algebraic Operations: Signatures}

Following the approach of Plotkin and Pretnar \cite{Plotkin2013}, we model effectful
operations by their \emph{signature} \ensuremath{\sigma}.
Effect signatures are modeled as functors. % and effectful operations as instances
% thereof.
\begin{hscode}\SaveRestoreHook
\column{B}{@{}>{\hspre}l<{\hspost}@{}}%
\column{3}{@{}>{\hspre}l<{\hspost}@{}}%
\column{5}{@{}>{\hspre}l<{\hspost}@{}}%
\column{E}{@{}>{\hspre}l<{\hspost}@{}}%
\>[3]{}\mathbf{class}\;\Conid{Functor}\;\sigma\;\mathbf{where}{}\<[E]%
\\
\>[3]{}\hsindent{2}{}\<[5]%
\>[5]{}\textsf{fmap}\mathbin{::}(\Varid{a}\to \Varid{b})\to \sigma\;\Varid{a}\to \sigma\;\Varid{b}{}\<[E]%
\ColumnHook
\end{hscode}\resethooks
% Algebraic structures, such as functors, come with laws they must obey.
% For functors, these are (1) preservation of identities, i.e., |fmap iden = iden|,
% and (2) preservation of composition, i.e., |fmap (f circ g) = fmap f circ fmap g|.
%
An effect signature groups several effectful operations.
For example, a signature for the state effect contains two operations: \ensuremath{\Conid{Get}} for
reading and returning the state, and \ensuremath{\Conid{Put}} for modifiying the state, overwriting
it with the given value.
\begin{hscode}\SaveRestoreHook
\column{B}{@{}>{\hspre}l<{\hspost}@{}}%
\column{3}{@{}>{\hspre}l<{\hspost}@{}}%
\column{E}{@{}>{\hspre}l<{\hspost}@{}}%
\>[3]{}\mathbf{data}\;State\;\Varid{s}\;\Varid{a}\mathrel{=}\Conid{Get}\;(\Varid{s}\to \Varid{a})\mid \Conid{Put}\;\Varid{s}\;\Varid{a}{}\<[E]%
\ColumnHook
\end{hscode}\resethooks
% - - - - - - - - - - - - - - - - - - - - - - - - - - - - - - - - - - - - - - - -
\paragraph{Computations: Free Monads}

Using a functor as signature, we can construct a recursive datatype to represent \emph{computations},
which may be either pure or effectful.
The computation may then call the operations included by the signature.
This recursive datatype is known as the \emph{free monad}:

% \noindent
% \begin{minipage}[t]{0.45\textwidth}
\begin{hscode}\SaveRestoreHook
\column{B}{@{}>{\hspre}l<{\hspost}@{}}%
\column{3}{@{}>{\hspre}l<{\hspost}@{}}%
\column{5}{@{}>{\hspre}l<{\hspost}@{}}%
\column{10}{@{}>{\hspre}l<{\hspost}@{}}%
\column{31}{@{}>{\hspre}l<{\hspost}@{}}%
\column{E}{@{}>{\hspre}l<{\hspost}@{}}%
\>[3]{}\mathbf{data}\;\Conid{Free}\;(\sigma\mathbin{::}\mathbin{*}\to \mathbin{*})\;\Varid{a}\;\mathbf{where}{}\<[E]%
\\
\>[3]{}\hsindent{2}{}\<[5]%
\>[5]{}\Conid{Var}{}\<[10]%
\>[10]{}\mathbin{::}\Varid{a}{}\<[31]%
\>[31]{}\to \Conid{Free}\;\sigma\;\Varid{a}{}\<[E]%
\\
\>[3]{}\hsindent{2}{}\<[5]%
\>[5]{}\Conid{Op}{}\<[10]%
\>[10]{}\mathbin{::}\sigma\;(\Conid{Free}\;\sigma\;\Varid{a}){}\<[31]%
\>[31]{}\to \Conid{Free}\;\sigma\;\Varid{a}{}\<[E]%
\ColumnHook
\end{hscode}\resethooks
% \end{minipage}%
% \begin{minipage}[t]{0.55\textwidth}
\begin{hscode}\SaveRestoreHook
\column{B}{@{}>{\hspre}l<{\hspost}@{}}%
\column{3}{@{}>{\hspre}l<{\hspost}@{}}%
\column{7}{@{}>{\hspre}l<{\hspost}@{}}%
\column{13}{@{}>{\hspre}l<{\hspost}@{}}%
\column{14}{@{}>{\hspre}l<{\hspost}@{}}%
\column{22}{@{}>{\hspre}l<{\hspost}@{}}%
\column{25}{@{}>{\hspre}l<{\hspost}@{}}%
\column{E}{@{}>{\hspre}l<{\hspost}@{}}%
\>[3]{}\mathbf{instance}\;{}\<[13]%
\>[13]{}\Conid{Functor}\;\sigma\Rightarrow \Conid{Monad}\;(\Conid{Free}\;\sigma){}\<[E]%
\\
\>[3]{}\hsindent{4}{}\<[7]%
\>[7]{}\textsf{return}{}\<[22]%
\>[22]{}\mathrel{=}\Conid{Var}{}\<[E]%
\\
\>[3]{}\hsindent{4}{}\<[7]%
\>[7]{}\Conid{Var}\;\Varid{x}{}\<[14]%
\>[14]{}\bind \Varid{k}{}\<[22]%
\>[22]{}\mathrel{=}{}\<[25]%
\>[25]{}\Varid{k}\;\Varid{x}{}\<[E]%
\\
\>[3]{}\hsindent{4}{}\<[7]%
\>[7]{}\Conid{Op}\;\Varid{t}{}\<[14]%
\>[14]{}\bind \Varid{k}{}\<[22]%
\>[22]{}\mathrel{=}\Conid{Op}\;(\textsf{fmap}\;(\bind \Varid{k})\;\Varid{t}){}\<[E]%
\ColumnHook
\end{hscode}\resethooks
% \end{minipage}

\noindent
Here, the \ensuremath{\Conid{Var}} constructor represents pure computations, whereas
\ensuremath{\Conid{Op}} recursively composes computations with a branching structure that is
determined by signature \ensuremath{\sigma}.
Isomorphically, we can present this \ensuremath{\Conid{Op}} constructor as taking two arguments:
an operation and a continuation \cite{freer}.%\footnote{This is referred to as the \emph{freer} monad.}
\begin{hscode}\SaveRestoreHook
\column{B}{@{}>{\hspre}l<{\hspost}@{}}%
\column{5}{@{}>{\hspre}l<{\hspost}@{}}%
\column{10}{@{}>{\hspre}l<{\hspost}@{}}%
\column{E}{@{}>{\hspre}l<{\hspost}@{}}%
\>[5]{}\Conid{Op}{}\<[10]%
\>[10]{}\mathbin{::}\forall\kern-2pt\;\Varid{b}\, .\,\underbrace{\sigma\;\Varid{b}}_{\Varid{operation}}\to \underbrace{(\Varid{b}\to \Conid{Free}\;\sigma\;\Varid{a})}_{\Varid{continuation}}\to \Conid{Free}\;\sigma\;\Varid{a}{}\<[E]%
\ColumnHook
\end{hscode}\resethooks
% The free monad is free in the sense that it satisfies the monad laws and no other.
% When plugging in a signature, representing an (effectful) operation, it gets a different
% meaning.
For example, the monad \ensuremath{\Conid{Free}\;(State\;\Conid{String})\;\Conid{Int}} represents computations
with a state of type \ensuremath{\Conid{String}} and a result of type \ensuremath{\Conid{Int}}.
Often, we define constructors to ease programming with effectful computations.
For example, for State \ensuremath{\Varid{s}}, we define \ensuremath{\textsf{get}} and \ensuremath{\textsf{put}} as computations that read and
set the state, respectively.

\noindent
\begin{minipage}{0.5\textwidth}
\begin{hscode}\SaveRestoreHook
\column{B}{@{}>{\hspre}l<{\hspost}@{}}%
\column{3}{@{}>{\hspre}l<{\hspost}@{}}%
\column{10}{@{}>{\hspre}l<{\hspost}@{}}%
\column{E}{@{}>{\hspre}l<{\hspost}@{}}%
\>[3]{}\textsf{get}{}\<[10]%
\>[10]{}\mathbin{::}\Conid{Free}\;(State\;\Varid{s})\;\Varid{s}{}\<[E]%
\\
\>[3]{}\textsf{get}{}\<[10]%
\>[10]{}\mathrel{=}\Conid{Op}\;(\Conid{Get}\;\textsf{return}){}\<[E]%
\ColumnHook
\end{hscode}\resethooks
\end{minipage}%
\begin{minipage}{0.5\textwidth}
\begin{hscode}\SaveRestoreHook
\column{B}{@{}>{\hspre}l<{\hspost}@{}}%
\column{3}{@{}>{\hspre}l<{\hspost}@{}}%
\column{10}{@{}>{\hspre}l<{\hspost}@{}}%
\column{E}{@{}>{\hspre}l<{\hspost}@{}}%
\>[3]{}\textsf{put}{}\<[10]%
\>[10]{}\mathbin{::}\Varid{s}\to \Conid{Free}\;(State\;\Varid{s})\;(){}\<[E]%
\\
\>[3]{}\textsf{put}\;\Varid{s}{}\<[10]%
\>[10]{}\mathrel{=}\Conid{Op}\;(\Conid{Put}\;\Varid{s}\;(\textsf{return}\;())){}\<[E]%
\ColumnHook
\end{hscode}\resethooks
\end{minipage}

\noindent
We use the monadic bind for sequencing effectful computations.

% - - - - - - - - - - - - - - - - - - - - - - - - - - - - - - - - - - - - - - - -
\paragraph{Interpretation: Folds}

% As free monads are defined in an inherently recursive way,
A recursion scheme is used to interpret free monads:
we write a fold-style interpreter that gives semantics to the computations
in the free monad by means of a generator and an algebra.
In particular, interpreting a free monad \ensuremath{\Conid{Free}\;\sigma\;\Varid{a}}
into semantic domain \ensuremath{\Varid{b}}
requires a generator \ensuremath{\Varid{gen}\mathbin{::}\Varid{a}\to \Varid{b}} to transform pure computations into
semantic domain \ensuremath{\Varid{b}} and
an algebra \ensuremath{\Varid{alg}\mathbin{::}\sigma\;\Varid{b}\to \Varid{b}} to interpret the signature's effectful
operations.
% In general, we define this fold as follows:
\begin{hscode}\SaveRestoreHook
\column{B}{@{}>{\hspre}l<{\hspost}@{}}%
\column{3}{@{}>{\hspre}l<{\hspost}@{}}%
\column{28}{@{}>{\hspre}l<{\hspost}@{}}%
\column{E}{@{}>{\hspre}l<{\hspost}@{}}%
\>[3]{}\textsf{fold}_{\textsf{Alg}}\mathbin{::}\Conid{Functor}\;\sigma\Rightarrow (\Varid{a}\to \Varid{b})\to (\sigma\;\Varid{b}\to \Varid{b})\to \Conid{Free}\;\sigma\;\Varid{a}\to \Varid{b}{}\<[E]%
\\
\>[3]{}\textsf{fold}_{\textsf{Alg}}\;\Varid{gen}\;\Varid{alg}\;(\Conid{Var}\;\Varid{x}){}\<[28]%
\>[28]{}\mathrel{=}\Varid{gen}\;\Varid{x}{}\<[E]%
\\
\>[3]{}\textsf{fold}_{\textsf{Alg}}\;\Varid{gen}\;\Varid{alg}\;(\Conid{Op}\;\Varid{op}){}\<[28]%
\>[28]{}\mathrel{=}\Varid{alg}\;(\textsf{fmap}\;(\textsf{fold}_{\textsf{Alg}}\;\Varid{gen}\;\Varid{alg})\;\Varid{op}){}\<[E]%
\ColumnHook
\end{hscode}\resethooks
For example, the following handler interprets \ensuremath{\Conid{Free}\;(\Conid{State}\;\Varid{s})\;\Varid{a}}
into semantic domain \ensuremath{\Varid{s}\to (\Varid{s},\Varid{a})} using this structural recursion scheme.

\begin{hscode}\SaveRestoreHook
\column{B}{@{}>{\hspre}l<{\hspost}@{}}%
\column{3}{@{}>{\hspre}l<{\hspost}@{}}%
\column{5}{@{}>{\hspre}l<{\hspost}@{}}%
\column{21}{@{}>{\hspre}l<{\hspost}@{}}%
\column{28}{@{}>{\hspre}c<{\hspost}@{}}%
\column{28E}{@{}l@{}}%
\column{32}{@{}>{\hspre}l<{\hspost}@{}}%
\column{E}{@{}>{\hspre}l<{\hspost}@{}}%
\>[3]{}h_{\textsf{State}}\mathbin{::}\Conid{Free}\;(State\;\Varid{s})\;\Varid{a}\to (\Varid{s}\to (\Varid{a},\Varid{s})){}\<[E]%
\\
\>[3]{}h_{\textsf{State}}\mathrel{=}\textsf{fold}_{\textsf{Alg}}\;(,)\;\Varid{alg}\;\mathbf{where}{}\<[E]%
\\
\>[3]{}\hsindent{2}{}\<[5]%
\>[5]{}\Varid{alg}\;(\Conid{Get}\;\Varid{k}){}\<[21]%
\>[21]{}\mathrel{=}\lambda \Varid{s}{}\<[28]%
\>[28]{}\to {}\<[28E]%
\>[32]{}\Varid{k}\;\Varid{s}\;\Varid{s}{}\<[E]%
\\
\>[3]{}\hsindent{2}{}\<[5]%
\>[5]{}\Varid{alg}\;(\Conid{Put}\;\Varid{s'}\;\Varid{k}){}\<[21]%
\>[21]{}\mathrel{=}\lambda \anonymous {}\<[28]%
\>[28]{}\to {}\<[28E]%
\>[32]{}\Varid{k}\;\Varid{s'}{}\<[E]%
\ColumnHook
\end{hscode}\resethooks
One can apply the above state handler to a computation that increments the state and returns the original state,
with an initial value of \ensuremath{\mathrm{0}}.\begin{hscode}\SaveRestoreHook
\column{B}{@{}>{\hspre}l<{\hspost}@{}}%
\column{3}{@{}>{\hspre}l<{\hspost}@{}}%
\column{E}{@{}>{\hspre}l<{\hspost}@{}}%
\>[3]{}\texttt{>>>}\;h_{\textsf{State}}\;(\textsf{get}\bind \lambda \Varid{s}\to \textsf{put}\;(\Varid{s}\mathbin{+}\mathrm{1})\sequ \textsf{return}\;\Varid{s})\;\mathrm{0}{}\<[E]%
\\
\>[3]{}(\mathrm{0},\mathrm{1}){}\<[E]%
\ColumnHook
\end{hscode}\resethooks
\noindent
The result is \ensuremath{(\mathrm{0},\mathrm{1})}, with \ensuremath{\mathrm{0}} the resulting value and \ensuremath{\mathrm{1}} the (incremented) state.

% (Op (Put 3 (Op (Get (\s -> return (s + 1)))))) 0

% - - - - - - - - - - - - - - - - - - - - - - - - - - - - - - - - - - - - - - - -
\paragraph{Modularly Combining Effects}

Often, a combination of multiple effects is desired.
Such a composition is commonly modeled by the ``Datatypes \`a la Carte'' \cite{dtc} approach,
using the coproduct of functors, or in this case effect
signatures (denoted by \ensuremath{\mathrel{{+}}})\footnote{
For the sake of readability \cite{dtc}, we have omitted \ensuremath{\Conid{Inl}} and \ensuremath{\Conid{Inr}} in our examples.
}.\begin{hscode}\SaveRestoreHook
\column{B}{@{}>{\hspre}l<{\hspost}@{}}%
\column{3}{@{}>{\hspre}l<{\hspost}@{}}%
\column{E}{@{}>{\hspre}l<{\hspost}@{}}%
\>[3]{}\mathbf{data}\;(\sigma_1\mathrel{{+}}\sigma_2)\;\Varid{a}\mathrel{=}\Conid{Inl}\;(\sigma_1\;\Varid{a})\mid \Conid{Inr}\;(\sigma_2\;\Varid{a}){}\<[E]%
\ColumnHook
\end{hscode}\resethooks
For example, we can combine stateful computations with nondeterminism.
Nondeterministic operations can either fail or branch.

\noindent
\begin{minipage}{0.5\textwidth}
\begin{hscode}\SaveRestoreHook
\column{B}{@{}>{\hspre}l<{\hspost}@{}}%
\column{3}{@{}>{\hspre}l<{\hspost}@{}}%
\column{E}{@{}>{\hspre}l<{\hspost}@{}}%
\>[3]{}\mathbf{data}\;Choice\;\Varid{a}\mathrel{=}\Conid{Fail}\mid \Conid{Or}\;\Varid{a}\;\Varid{a}{}\<[E]%
\ColumnHook
\end{hscode}\resethooks

\end{minipage}%
\begin{minipage}{0.5\textwidth}\begin{hscode}\SaveRestoreHook
\column{B}{@{}>{\hspre}l<{\hspost}@{}}%
\column{3}{@{}>{\hspre}l<{\hspost}@{}}%
\column{E}{@{}>{\hspre}l<{\hspost}@{}}%
\>[3]{}\Conid{Free}\;(State\;\Varid{s}\mathrel{{+}}Choice)\;\Varid{a}{}\<[E]%
\ColumnHook
\end{hscode}\resethooks
\end{minipage}

\noindent
This modular composition allows different effect interactions \cite{justdoit,zhixuan21}, for instance,
achieving local state versus global state semantics \cite{pauwels}, by swapping the handlers
for state and nondeterminism.

One could also combine effectful operations with \emph{unknown} effects, which
are then \emph{forwarded} to other handlers.
A separator \ensuremath{\mathrel{{\#}}} distinguishes between the algebras to interpret and forward
effects, respectively.
%
% < (#) :: (f a -> b) -> (g a -> b) -> (f :+: g) a -> b
% < (alg # fwd) (Inl op) = alg op
% < (alg # fwd) (Inr op) = fwd op
%
We redefine the above state handler to include ``other'' (unknown) effects,
forwarding them to be interpreted by other handlers, and passing the state to these
handlers.
\begin{hscode}\SaveRestoreHook
\column{B}{@{}>{\hspre}l<{\hspost}@{}}%
\column{3}{@{}>{\hspre}l<{\hspost}@{}}%
\column{5}{@{}>{\hspre}l<{\hspost}@{}}%
\column{10}{@{}>{\hspre}l<{\hspost}@{}}%
\column{22}{@{}>{\hspre}l<{\hspost}@{}}%
\column{29}{@{}>{\hspre}c<{\hspost}@{}}%
\column{29E}{@{}l@{}}%
\column{33}{@{}>{\hspre}l<{\hspost}@{}}%
\column{E}{@{}>{\hspre}l<{\hspost}@{}}%
\>[3]{}h_{\textsf{State}}^\prime\mathbin{::}\Conid{Functor}\;\sigma\Rightarrow \Conid{Free}\;(State\;\Varid{s}\mathrel{{+}}\sigma)\;\Varid{a}\to (\Varid{s}\to \Conid{Free}\;\sigma\;(\Varid{a},\Varid{s})){}\<[E]%
\\
\>[3]{}h_{\textsf{State}}^\prime\mathrel{=}\textsf{fold}_{\textsf{Alg}}\;\Varid{gen}\;(\Varid{alg}\mathrel{{\#}}\Varid{fwd})\;\mathbf{where}{}\<[E]%
\\
\>[3]{}\hsindent{2}{}\<[5]%
\>[5]{}\Varid{gen}\;{}\<[10]%
\>[10]{}\Varid{x}{}\<[22]%
\>[22]{}\mathrel{=}\lambda \Varid{s}{}\<[29]%
\>[29]{}\to {}\<[29E]%
\>[33]{}\textsf{return}\;(\Varid{x},\Varid{s}){}\<[E]%
\\
\>[3]{}\hsindent{2}{}\<[5]%
\>[5]{}\Varid{alg}\;{}\<[10]%
\>[10]{}(\Conid{Get}\;\Varid{k}){}\<[22]%
\>[22]{}\mathrel{=}\lambda \Varid{s}{}\<[29]%
\>[29]{}\to {}\<[29E]%
\>[33]{}\Varid{k}\;\Varid{s}\;\Varid{s}{}\<[E]%
\\
\>[3]{}\hsindent{2}{}\<[5]%
\>[5]{}\Varid{alg}\;{}\<[10]%
\>[10]{}(\Conid{Put}\;\Varid{s'}\;\Varid{k}){}\<[22]%
\>[22]{}\mathrel{=}\lambda \anonymous {}\<[29]%
\>[29]{}\to {}\<[29E]%
\>[33]{}\Varid{k}\;\Varid{s'}{}\<[E]%
\\
\>[3]{}\hsindent{2}{}\<[5]%
\>[5]{}\Varid{fwd}\;{}\<[10]%
\>[10]{}\Varid{op}{}\<[22]%
\>[22]{}\mathrel{=}\lambda \Varid{s}{}\<[29]%
\>[29]{}\to {}\<[29E]%
\>[33]{}\Conid{Op}\;(\textsf{fmap}\;(\mathbin{\$}\Varid{s})\;\Varid{op}){}\<[E]%
\ColumnHook
\end{hscode}\resethooks
% In a setting where multiple side-effects are combined, monad transformers \cite{mtl}
% as an alternative approach
% for making a modular composition.
% Algebraic effects \& handlers, however, are more appropriate for modularly composing effects.

%-------------------------------------------------------------------------------
\subsection{Non-Algebraic Effects}
\label{sec:other-effects}

Not all effects are algebraic; some
denote more advanced effectful computations \cite{vandenBerg21,scope14,esop22}.
Effects are algebraic if they satisfy the \emph{algebraicity property}, which
says that algebraic computations commute with sequencing:
\begin{hscode}\SaveRestoreHook
\column{B}{@{}>{\hspre}l<{\hspost}@{}}%
\column{3}{@{}>{\hspre}l<{\hspost}@{}}%
\column{E}{@{}>{\hspre}l<{\hspost}@{}}%
\>[3]{}(\bind \Varid{k})\;\circ\;\Conid{Op}\equiv \Conid{Op}\;\circ\;\textsf{fmap}\;(\bind \Varid{k}){}\<[E]%
\ColumnHook
\end{hscode}\resethooks
Although many effects satisfy this property, not all of them do.

For example,
\emph{scoped effects \& handlers} \cite{lics18,scope14,esop22} distinguish between
a scoped computation, representing the part of the program that is in scope,
and a continuation, referring to the rest of the program, out of scope.
This complication requires a reformulation of the algebraicity property as well
as a more complex free monad to represent these effects.
Examples of scoped effects \& handlers are local variables, catching exceptions,
local pruning of nondeterminism, and more.
Furthermore,
\emph{parallel effects \& handlers} \cite{xie21} parallellize computations with
algebraic effects, using a \ensuremath{\textsf{for}}-loop, for example. These effects contain a computation
to parallellize, as well as a continuation.
% Parallel effects are mainly used to
% improve the efficiency and performance of programs with algebraic effects.
Moreover,
\emph{latent effects \& handlers} \cite{vandenBerg21} delay parts of an effectful
program to be evaluated later in the execution. Their syntax also consists of a computation
to be delayed, and a continuation.
Lazy evaluation strategies in the presence of effects, and staging, among others, can be
modeled by this approach.

Each of these classes of effects\footnote{We discuss them in more detail in \Cref{sec:instances}.}
has been denotationally modeled by a \emph{specialized version}
of the free monad, making their representations ad hoc and inherently distinct.
Moreover, the monads used to model parallel and latent effects have not been shown to be a free structure.

This work builds on the fundamental insight that these effects have something
in common: \textbf{they have an internal computation}.
Indeed, for scoped effects this internal computation is the computation in scope;
for parallel effects the computation to parallellize over; for latent effects the
computation to defer.
We call these effects \emph{higher-order effects}.
With this insight, we argue that we can generalize the framework for algebraic effects
and handlers to model more advanced effects as well.
This generic framework retains the modularity of having separate
syntax (by means of a free monad) and semantics (by means of a fold).
We back the framework by categorical foundations, showing that (1) it is indeed a free monad,
and (2) that it works on a range of examples.
% In what follows we explain the generic framework in more detail and show how
% different effects can be seen as an instantiation of this framework.

%==================================================================
\section{A Generic Free Monad for Modeling Effects}
\label{sec:generic}

Our framework for modeling higher-order effects uses a generic free monad and interpreter.
This representation deliberately does not deviate far from the
\aeh approach (\Cref{sec:aeh}), the modularity of which has already proven
its importance by the adoption of algebraic effects in different programming
languages \cite{effekt,Hillerstrom16,Leijen14,frank}
and libraries \cite{extensible-effects,eff-ocaml,fused-effects}.
In particular, we again use a free monad, with higher-order functors
to include the internal computation in the framework, also featuring a coproduct.

% - - - - - - - - - - - - - - - - - - - - - - - - - - - - - - - - - - - - - - - -
\paragraph{Effect Signatures}

Higher-order effects need access to an internal computation.
In order to reflect this in the signatures, a functor representation is not
sufficiently powerful.
Like Poulsen and
van der Rest \cite{higher-order}, we generalize effect signatures to include
% to include an internal
% computation together with effect operations.
% In particular,
a higher-order functor \ensuremath{\Varid{k}\mathbin{:}(\mathbin{*}\to \mathbin{*})\to (\mathbin{*}\to \mathbin{*})}, which
is a mapping between functors.
Here, the functor argument (of kind \ensuremath{\mathbin{*}\to \mathbin{*}}) represents the internal computation so that
\ensuremath{\Varid{k}\;\Varid{f}} has the familiar form of an algebraic effect signature (a plain functor).
%
% Alternatively, we can view this higher-order bifunctor as a functor between
% functors---an endofunctor in the category of functors.
We model it accordingly \cite{scope14}:
\begin{hscode}\SaveRestoreHook
\column{B}{@{}>{\hspre}l<{\hspost}@{}}%
\column{3}{@{}>{\hspre}l<{\hspost}@{}}%
\column{7}{@{}>{\hspre}l<{\hspost}@{}}%
\column{E}{@{}>{\hspre}l<{\hspost}@{}}%
\>[3]{}\mathbf{class}\;(\forall\kern-2pt\;\Varid{f}\, .\,\Conid{Functor}\;\Varid{f}\Rightarrow \Conid{Functor}\;(\Varid{k}\;\Varid{f}))\Rightarrow HFunctor\;\Varid{k}\;\mathbf{where}{}\<[E]%
\\
\>[3]{}\hsindent{4}{}\<[7]%
\>[7]{}\textsf{hmap}\mathbin{::}(\Conid{Functor}\;\Varid{f},\Conid{Functor}\;\Varid{f'})\Rightarrow \Varid{f}{\leadsto}\Varid{f'}\to \Varid{k}\;\Varid{f}{\leadsto}\Varid{k}\;\Varid{f'}{}\<[E]%
\ColumnHook
\end{hscode}\resethooks

\noindent
Here, \ensuremath{({\leadsto})} represents a natural transformation between two functors\footnote{
\ensuremath{\mathbf{type}\;\Varid{f}{\leadsto}\Varid{g}\mathrel{=}\forall\kern-2pt\;\Varid{a}\, .\,\Varid{f}\;\Varid{a}\to \Varid{g}\;\Varid{a}}}.
For example, a higher-order signature for exceptions is the following:
\begin{hscode}\SaveRestoreHook
\column{B}{@{}>{\hspre}l<{\hspost}@{}}%
\column{3}{@{}>{\hspre}l<{\hspost}@{}}%
\column{E}{@{}>{\hspre}l<{\hspost}@{}}%
\>[3]{}\mathbf{data}\;\Conid{Exc}\;\Varid{f}\;\Varid{r}\mathrel{=}\Conid{Throw}\mid \forall\kern-2pt\;\Varid{a}\, .\,\Conid{Catch}\;(\Varid{f}\;\Varid{a})\;(\Conid{Maybe}\;\Varid{a}\to \Varid{r}){}\<[E]%
\ColumnHook
\end{hscode}\resethooks
The continuation of catching an exception (\ensuremath{\Conid{Maybe}\;\Varid{a}\to \Varid{r}}) depends on whether or not an
exception was thrown in the internal computation \ensuremath{\Varid{f}\;\Varid{a}}.

% - - - - - - - - - - - - - - - - - - - - - - - - - - - - - - - - - - - - - - - -
\paragraph{Computations}

Similar to the free monad for algebraic effects \& handlers, we can now construct a free monad
that uses the above generalization of signatures.
\begin{hscode}\SaveRestoreHook
\column{B}{@{}>{\hspre}l<{\hspost}@{}}%
\column{3}{@{}>{\hspre}l<{\hspost}@{}}%
\column{5}{@{}>{\hspre}l<{\hspost}@{}}%
\column{11}{@{}>{\hspre}l<{\hspost}@{}}%
\column{14}{@{}>{\hspre}l<{\hspost}@{}}%
\column{21}{@{}>{\hspre}l<{\hspost}@{}}%
\column{31}{@{}>{\hspre}l<{\hspost}@{}}%
\column{E}{@{}>{\hspre}l<{\hspost}@{}}%
\>[3]{}\mathbf{data}\;Free_{\textsf{H}}\;\Varid{k}\;\Varid{a}\;\mathbf{where}{}\<[E]%
\\
\>[3]{}\hsindent{2}{}\<[5]%
\>[5]{}Var_{\textsf{H}}{}\<[11]%
\>[11]{}\mathbin{::}\Varid{a}{}\<[31]%
\>[31]{}\to Free_{\textsf{H}}\;\Varid{k}\;\Varid{a}{}\<[E]%
\\
\>[3]{}\hsindent{2}{}\<[5]%
\>[5]{}Op_{\textsf{H}}{}\<[11]%
\>[11]{}\mathbin{::}\Varid{k}\;(Free_{\textsf{H}}\;\Varid{k})\;(Free_{\textsf{H}}\;\Varid{k}\;\Varid{a}){}\<[31]%
\>[31]{}\to Free_{\textsf{H}}\;\Varid{k}\;\Varid{a}{}\<[E]%
\\[\blanklineskip]%
\>[3]{}\mathbf{instance}\;HFunctor\;\Varid{k}\Rightarrow \Conid{Monad}\;(Free_{\textsf{H}}\;\Varid{k})\;\mathbf{where}{}\<[E]%
\\
\>[3]{}\hsindent{2}{}\<[5]%
\>[5]{}\textsf{return}{}\<[21]%
\>[21]{}\mathrel{=}Var_{\textsf{H}}{}\<[E]%
\\
\>[3]{}\hsindent{2}{}\<[5]%
\>[5]{}Var_{\textsf{H}}\;\Varid{x}{}\<[14]%
\>[14]{}\bind \Varid{k}{}\<[21]%
\>[21]{}\mathrel{=}\Varid{k}\;\Varid{x}{}\<[E]%
\\
\>[3]{}\hsindent{2}{}\<[5]%
\>[5]{}Op_{\textsf{H}}\;\Varid{op}{}\<[14]%
\>[14]{}\bind \Varid{k}{}\<[21]%
\>[21]{}\mathrel{=}Op_{\textsf{H}}\;(\textsf{fmap}\;(\bind \Varid{k})\;\Varid{op}){}\<[E]%
\ColumnHook
\end{hscode}\resethooks

\noindent
In \Cref{sec:formalization} we show that this is indeed a free monad.
Again, using the co-yoneda lemma, we rewrite the \ensuremath{Op_{\textsf{H}}} constructor:
it contains an operation, an internal computation and a continuation.
\begin{hscode}\SaveRestoreHook
\column{B}{@{}>{\hspre}l<{\hspost}@{}}%
\column{5}{@{}>{\hspre}l<{\hspost}@{}}%
\column{11}{@{}>{\hspre}l<{\hspost}@{}}%
\column{155}{@{}>{\hspre}l<{\hspost}@{}}%
\column{E}{@{}>{\hspre}l<{\hspost}@{}}%
\>[5]{}Op_{\textsf{H}}{}\<[11]%
\>[11]{}\mathbin{::}\forall\kern-2pt\;\Varid{f}\;\Varid{b}\, .\,\underbrace{(\Varid{k}\;\Varid{f}\;\Varid{b})}_{\Varid{operation}}\to \underbrace{(\Varid{f}{\leadsto}Free_{\textsf{H}}\;\Varid{k})}_{\Varid{internal}\;\Varid{computation}}\to \underbrace{(\Varid{b}\to Free_{\textsf{H}}\;\Varid{k}\;\Varid{a})}_{\Varid{continuation}}{}\<[155]%
\>[155]{}\to Free_{\textsf{H}}\;\Varid{k}\;\Varid{a}{}\<[E]%
\ColumnHook
\end{hscode}\resethooks
% We provide our effect signature |k| with a functor |T k| that represents the internal computation,
% and have a separate continuation.
% In \Cref{sec:formalization} we show that this is indeed a free monad, based
% on solid categorical foundations.
For example, \ensuremath{prog_{\textsf{Exc}}} throws an exception if its argument is smaller than 0.
\begin{hscode}\SaveRestoreHook
\column{B}{@{}>{\hspre}l<{\hspost}@{}}%
\column{3}{@{}>{\hspre}l<{\hspost}@{}}%
\column{9}{@{}>{\hspre}l<{\hspost}@{}}%
\column{16}{@{}>{\hspre}l<{\hspost}@{}}%
\column{27}{@{}>{\hspre}l<{\hspost}@{}}%
\column{E}{@{}>{\hspre}l<{\hspost}@{}}%
\>[3]{}prog_{\textsf{Exc}}\mathbin{::}\Conid{Int}\to Free_{\textsf{H}}\;\Conid{Exc}\;\Conid{String}{}\<[E]%
\\
\>[3]{}prog_{\textsf{Exc}}\;\Varid{x}\mathrel{=}Op_{\textsf{H}}\;(\Conid{Catch}\;(\mathbf{if}\;\Varid{x}\geq \mathrm{0}\;\mathbf{then}\;\textsf{return}\;\Varid{x}\;\mathbf{else}\;Op_{\textsf{H}}\;\Conid{Throw})\;\Varid{k}){}\<[E]%
\\
\>[3]{}\hsindent{6}{}\<[9]%
\>[9]{}\mathbf{where}\;{}\<[16]%
\>[16]{}\Varid{k}\;\Conid{Nothing}{}\<[27]%
\>[27]{}\mathrel{=}\textsf{return}\;\text{\ttfamily \char34 Too~small\char34}{}\<[E]%
\\
\>[16]{}\Varid{k}\;(\Conid{Just}\;\Varid{x})\mathrel{=}\textsf{return}\;(\Varid{show}\;\Varid{x}){}\<[E]%
\ColumnHook
\end{hscode}\resethooks
% - - - - - - - - - - - - - - - - - - - - - - - - - - - - - - - - - - - - - - - -
\paragraph{Interpretation}

We equip this free monad with a fold-style interpreter:
\begin{hscode}\SaveRestoreHook
\column{B}{@{}>{\hspre}l<{\hspost}@{}}%
\column{3}{@{}>{\hspre}l<{\hspost}@{}}%
\column{8}{@{}>{\hspre}l<{\hspost}@{}}%
\column{9}{@{}>{\hspre}l<{\hspost}@{}}%
\column{15}{@{}>{\hspre}l<{\hspost}@{}}%
\column{23}{@{}>{\hspre}l<{\hspost}@{}}%
\column{28}{@{}>{\hspre}l<{\hspost}@{}}%
\column{31}{@{}>{\hspre}l<{\hspost}@{}}%
\column{E}{@{}>{\hspre}l<{\hspost}@{}}%
\>[3]{}\textsf{fold}{}\<[9]%
\>[9]{}\mathbin{::}\forall\kern-2pt\;\Varid{k}\;\Varid{g}\;\Varid{a}\;\Varid{b}\, .\,(HFunctor\;\Varid{k},\Conid{Pointed}\;\Varid{g}){}\<[E]%
\\
\>[9]{}\Rightarrow (\Varid{a}\to \Varid{g}\;\Varid{b})\to (\forall\kern-2pt\;\Varid{x}\, .\,\Varid{k}\;\Varid{g}\;(\Varid{g}\;\Varid{x})\to \Varid{g}\;\Varid{x})\to (Free_{\textsf{H}}\;\Varid{k}\;\Varid{a}\to \Varid{g}\;\Varid{b}){}\<[E]%
\\
\>[3]{}\textsf{fold}\;\Varid{gen}\;\Varid{alg}\;(Var_{\textsf{H}}\;{}\<[23]%
\>[23]{}\Varid{x}){}\<[28]%
\>[28]{}\mathrel{=}\Varid{gen}\;\Varid{x}{}\<[E]%
\\
\>[3]{}\textsf{fold}\;\Varid{gen}\;\Varid{alg}\;(Op_{\textsf{H}}\;{}\<[23]%
\>[23]{}\Varid{op}){}\<[28]%
\>[28]{}\mathrel{=}\Varid{alg}\;(\textsf{hmap}\;\textsf{fold}_{2}\;(\textsf{fmap}\;(\textsf{fold}\;\Varid{gen}\;\Varid{alg})\;\Varid{op})){}\<[E]%
\\
\>[3]{}\hsindent{5}{}\<[8]%
\>[8]{}\mathbf{where}\;{}\<[15]%
\>[15]{}\textsf{fold}_{2}\mathbin{::}Free_{\textsf{H}}\;\Varid{k}{\leadsto}\Varid{g}{}\<[E]%
\\
\>[15]{}\textsf{fold}_{2}\;(Var_{\textsf{H}}\;\Varid{x}){}\<[31]%
\>[31]{}\mathrel{=}\eta\;\Varid{x}{}\<[E]%
\\
\>[15]{}\textsf{fold}_{2}\;(Op_{\textsf{H}}\;\Varid{t}){}\<[31]%
\>[31]{}\mathrel{=}\Varid{alg}\;(\textsf{hmap}\;\textsf{fold}_{2}\;(\textsf{fmap}\;\textsf{fold}_{2}\;\Varid{t})){}\<[E]%
\ColumnHook
\end{hscode}\resethooks
This \ensuremath{\textsf{fold}} consists of two parts: one interpreting \ensuremath{Free_{\textsf{H}}\;\Varid{k}\;\Varid{a}} into semantic domain
\ensuremath{\Varid{g}\;\Varid{b}}, and another interpreting the internal computation: \ensuremath{Free_{\textsf{H}}\;\Varid{k}{\leadsto}\Varid{g}}.
Consequently, from these two \ensuremath{\textsf{fold}}s, one would expect two generators and two algebras.
However, \ensuremath{\textsf{fold}_{2}} relies on the fact that \ensuremath{\Varid{g}} is a pointed functor,
which has an \emph{implicit} generator \ensuremath{\forall\kern-2pt\;\Varid{a}\, .\,\Varid{a}\to \Varid{g}\;\Varid{a}}, so that a single \emph{explicit} generator suffices.
\begin{hscode}\SaveRestoreHook
\column{B}{@{}>{\hspre}l<{\hspost}@{}}%
\column{3}{@{}>{\hspre}l<{\hspost}@{}}%
\column{5}{@{}>{\hspre}l<{\hspost}@{}}%
\column{E}{@{}>{\hspre}l<{\hspost}@{}}%
\>[3]{}\mathbf{class}\;\Conid{Functor}\;\Varid{g}\Rightarrow \Conid{Pointed}\;\Varid{g}\;\mathbf{where}{}\<[E]%
\\
\>[3]{}\hsindent{2}{}\<[5]%
\>[5]{}\eta\mathbin{::}\Varid{a}\to \Varid{g}\;\Varid{a}{}\<[E]%
\ColumnHook
\end{hscode}\resethooks
% Alternatively, we could have chosen to provide a second generator of type |a -> g a|.
Furthermore, to keep things concise, we opt to reuse the same algebra for the two \ensuremath{\textsf{fold}}s (the domain is universally
quantified) to interpret the internal computation and the continuation
consistently. This significantly reduces the handler's complexity but also implies that
some effects are not supported (\Cref{sec:scoped}).
% \todo{be explicit about |g b|.}
For example, our exception handler interprets the result in terms of \ensuremath{\Conid{Maybe}}.

\noindent
\begin{minipage}{0.5\textwidth}
\begin{hscode}\SaveRestoreHook
\column{B}{@{}>{\hspre}l<{\hspost}@{}}%
\column{3}{@{}>{\hspre}l<{\hspost}@{}}%
\column{5}{@{}>{\hspre}l<{\hspost}@{}}%
\column{22}{@{}>{\hspre}l<{\hspost}@{}}%
\column{E}{@{}>{\hspre}l<{\hspost}@{}}%
\>[3]{}h_{\textsf{Exc}}\mathbin{::}Free_{\textsf{H}}\;\Conid{Exc}\;\Varid{a}\to \Conid{Maybe}\;\Varid{a}{}\<[E]%
\\
\>[3]{}h_{\textsf{Exc}}\mathrel{=}\textsf{fold}\;\Conid{Just}\;\Varid{alg}\;\mathbf{where}{}\<[E]%
\\
\>[3]{}\hsindent{2}{}\<[5]%
\>[5]{}\Varid{alg}\;\Conid{Throw}{}\<[22]%
\>[22]{}\mathrel{=}\Conid{Nothing}{}\<[E]%
\\
\>[3]{}\hsindent{2}{}\<[5]%
\>[5]{}\Varid{alg}\;(\Conid{Catch}\;\Varid{c}\;\Varid{k}){}\<[22]%
\>[22]{}\mathrel{=}\Varid{k}\;\Varid{c}{}\<[E]%
\ColumnHook
\end{hscode}\resethooks
\end{minipage}%
\begin{minipage}{0.5\textwidth}
\begin{hscode}\SaveRestoreHook
\column{B}{@{}>{\hspre}l<{\hspost}@{}}%
\column{3}{@{}>{\hspre}l<{\hspost}@{}}%
\column{E}{@{}>{\hspre}l<{\hspost}@{}}%
\>[3]{}\texttt{>>>}\;h_{\textsf{Exc}}\;(prog_{\textsf{Exc}}\;\mathrm{5}){}\<[E]%
\\
\>[3]{}\Conid{Just}\;\text{\ttfamily \char34 5\char34}{}\<[E]%
\\
\>[3]{}\texttt{>>>}\;h_{\textsf{Exc}}\;(prog_{\textsf{Exc}}\;(\mathbin{-}\mathrm{5})){}\<[E]%
\\
\>[3]{}\Conid{Just}\;\text{\ttfamily \char34 Too~small\char34}{}\<[E]%
\ColumnHook
\end{hscode}\resethooks
\end{minipage}

% In what follows, we discuss how different instantiations of the
% bifunctor |K| lead to the representation of different kinds of effects.

% - - - - - - - - - - - - - - - - - - - - - - - - - - - - - - - - - - - - - - - -
\paragraph{Modular Composition}

In order to make a combination of different effects in the style of ``Datatypes
\`a la Carte'' \cite{dtc}, we require a coproduct of higher-order functors
\ensuremath{\Varid{k}_{1}} and \ensuremath{\Varid{k}_{2}}. This coproduct\footnote{
For the sake of readability, we omit \ensuremath{\Conid{In}} and \ensuremath{\Conid{Out}} from our examples.
} (denoted by \ensuremath{\mathrel{\oplus}}, with separator \ensuremath{\kern+2pt\tikz[baseline=(char.base)]{ \node[circle,draw,inner sep=0pt,align=center,scale=0.3] (char) {\Huge{\#}};}\kern+2pt}) works in a similar way as that for functors (\Cref{sec:background}).
\begin{hscode}\SaveRestoreHook
\column{B}{@{}>{\hspre}l<{\hspost}@{}}%
\column{3}{@{}>{\hspre}l<{\hspost}@{}}%
\column{18}{@{}>{\hspre}l<{\hspost}@{}}%
\column{24}{@{}>{\hspre}l<{\hspost}@{}}%
\column{29}{@{}>{\hspre}c<{\hspost}@{}}%
\column{29E}{@{}l@{}}%
\column{32}{@{}>{\hspre}l<{\hspost}@{}}%
\column{E}{@{}>{\hspre}l<{\hspost}@{}}%
\>[3]{}\mathbf{data}\;(\Varid{k}_{1}\mathrel{\oplus}\Varid{k}_{2})\;\Varid{f}\;\Varid{a}\mathrel{=}\Conid{In}\;(\Varid{k}_{1}\;\Varid{f}\;\Varid{a})\mid \Conid{Out}\;(\Varid{k}_{2}\;\Varid{f}\;\Varid{a}){}\<[E]%
\\[\blanklineskip]%
\>[3]{}(\kern+2pt\tikz[baseline=(char.base)]{ \node[circle,draw,inner sep=0pt,align=center,scale=0.3] (char) {\Huge{\#}};}\kern+2pt)\mathbin{::}(\Varid{k}_{1}\;\Varid{f}\;\Varid{a}\to \Varid{g}\;\Varid{b})\to (\Varid{k}_{2}\;\Varid{f}\;\Varid{a}\to \Varid{g}\;\Varid{b})\to (\Varid{k}_{1}\mathrel{\oplus}\Varid{k}_{2})\;\Varid{f}\;\Varid{a}\to \Varid{g}\;\Varid{b}{}\<[E]%
\\
\>[3]{}(\Varid{lft}\kern+2pt\tikz[baseline=(char.base)]{ \node[circle,draw,inner sep=0pt,align=center,scale=0.3] (char) {\Huge{\#}};}\kern+2pt\Varid{rht})\;{}\<[18]%
\>[18]{}(\Conid{In}\;{}\<[24]%
\>[24]{}\Varid{op}){}\<[29]%
\>[29]{}\mathrel{=}{}\<[29E]%
\>[32]{}\Varid{lft}\;\Varid{op}{}\<[E]%
\\
\>[3]{}(\Varid{lft}\kern+2pt\tikz[baseline=(char.base)]{ \node[circle,draw,inner sep=0pt,align=center,scale=0.3] (char) {\Huge{\#}};}\kern+2pt\Varid{rht})\;{}\<[18]%
\>[18]{}(\Conid{Out}\;{}\<[24]%
\>[24]{}\Varid{op}){}\<[29]%
\>[29]{}\mathrel{=}{}\<[29E]%
\>[32]{}\Varid{rht}\;\Varid{op}{}\<[E]%
\ColumnHook
\end{hscode}\resethooks

%==================================================================
\section{Different Effects as Instances of our Framework}
\label{sec:instances}

Different effects (algebraic and higher-order) can be viewed as instances
of our generic framework, defining an appropriate higher-order functor that maps their
signature to the generic setting.
In particular, \emph{instantiating the framework} consists of four steps:

\setlist[itemize]{leftmargin=15mm}
\begin{itemize}
  \item[\textbf{Step 1}] Map the effect signatures to a higher-order representation
  by defining a higher-order functor \ensuremath{\Conid{K}} that maps a functor \ensuremath{\Conid{F}} and type \ensuremath{\Conid{A}} onto a type
  of kind \ensuremath{\mathbin{*}}.
% < K F A = ...
  \item[\textbf{Step 2}] Show that this \ensuremath{\Conid{K}} is indeed a higher-order functor.
% < instance HOFunctor K where ...
  \item[\textbf{Step 3}] Plug it in the generic free monad \ensuremath{Free_{\textsf{H}}} and show that it is isomorphic to the
  specialized effect definition.
% < T K a iso ...
  \item[\textbf{Step 4}] Use the generic \ensuremath{\textsf{fold}} function to write a handler for the effects and show
  that it is isomorphic to the specialized effect handler (if it exists).
\end{itemize}

\noindent
In what follows, we instantiate our framework using these four steps for different classes
of effects.

%- - - - - - - - - - - - - - - - - - - - - - - - - - - - - - - - - - - - - - - -
\subsection{Algebraic Effects \& Handlers}
\label{sec:algebraic}

\noindent
We follow these steps for algebraic effects, showing their specialization of the
framework isomorphic to their definition in \Cref{sec:aeh}.

\noindent \textbf{Step 1}
Our mapping ignores functor argument \ensuremath{\Conid{F}}, since algebraic
effects do not have an internal computation.
\ensuremath{\Sigma} is a functor for algebraic operations.

\noindent
\begin{minipage}{0.4\textwidth}\begin{hscode}\SaveRestoreHook
\column{B}{@{}>{\hspre}l<{\hspost}@{}}%
\column{3}{@{}>{\hspre}l<{\hspost}@{}}%
\column{E}{@{}>{\hspre}l<{\hspost}@{}}%
\>[3]{}K^{\textsf{Alg}}_{\Sigma}\;\Conid{F}\;\Conid{A}\mathrel{=}\Sigma\;\Conid{A}{}\<[E]%
\ColumnHook
\end{hscode}\resethooks
\end{minipage}%
\begin{minipage}{0.6\textwidth}
\begin{hscode}\SaveRestoreHook
\column{B}{@{}>{\hspre}l<{\hspost}@{}}%
\column{3}{@{}>{\hspre}l<{\hspost}@{}}%
\column{5}{@{}>{\hspre}l<{\hspost}@{}}%
\column{E}{@{}>{\hspre}l<{\hspost}@{}}%
\>[3]{}\mathbf{data}\;K^{\textsf{Alg}}\;\sigma\;\Varid{f}\;\Varid{a}\;\mathbf{where}{}\<[E]%
\\
\>[3]{}\hsindent{2}{}\<[5]%
\>[5]{}\textsf{Op}\mathbin{::}\sigma\;\Varid{a}\to K^{\textsf{Alg}}\;\sigma\;\Varid{f}\;\Varid{a}{}\<[E]%
\ColumnHook
\end{hscode}\resethooks
\end{minipage}%

\step{2}
This definition of \ensuremath{K^{\textsf{Alg}}_{\Sigma}} is a higher-order functor:
\begin{hscode}\SaveRestoreHook
\column{B}{@{}>{\hspre}l<{\hspost}@{}}%
\column{3}{@{}>{\hspre}l<{\hspost}@{}}%
\column{5}{@{}>{\hspre}l<{\hspost}@{}}%
\column{E}{@{}>{\hspre}l<{\hspost}@{}}%
\>[3]{}\mathbf{instance}\;\Conid{Functor}\;\sigma\Rightarrow HFunctor\;(K^{\textsf{Alg}}\;\sigma)\;\mathbf{where}{}\<[E]%
\\
\>[3]{}\hsindent{2}{}\<[5]%
\>[5]{}\textsf{hmap}\;\Varid{k}\;(\textsf{Op}\;\Varid{x})\mathrel{=}\textsf{Op}\;\Varid{x}{}\<[E]%
\ColumnHook
\end{hscode}\resethooks
\step{3}
We can show that the following isomorphism holds (\ref{app:iso}):\begin{hscode}\SaveRestoreHook
\column{B}{@{}>{\hspre}l<{\hspost}@{}}%
\column{3}{@{}>{\hspre}l<{\hspost}@{}}%
\column{E}{@{}>{\hspre}l<{\hspost}@{}}%
\>[3]{}\Conid{Free}\;\sigma\;\Varid{a}\;\cong\;Free_{\textsf{H}}\;(K^{\textsf{Alg}}\;\sigma)\;\Varid{a}{}\<[E]%
\ColumnHook
\end{hscode}\resethooks
\step{4}
The generic handler for algebraic effects is defined by \ensuremath{h_{\textsf{Alg}}}.
In \ref{app:iso} we show that \ensuremath{h_{\textsf{Alg}}\;\Varid{gen}\;\Varid{alg}} is isomorphic to \ensuremath{\textsf{fold}_{\textsf{Alg}}\;\Varid{gen}\;(\Varid{alg}\, .\,\textsf{Op})}.
\begin{hscode}\SaveRestoreHook
\column{B}{@{}>{\hspre}l<{\hspost}@{}}%
\column{3}{@{}>{\hspre}l<{\hspost}@{}}%
\column{9}{@{}>{\hspre}l<{\hspost}@{}}%
\column{E}{@{}>{\hspre}l<{\hspost}@{}}%
\>[3]{}h_{\textsf{Alg}}{}\<[9]%
\>[9]{}\mathbin{::}(\Conid{Functor}\;\sigma,\Conid{Pointed}\;\Varid{g}){}\<[E]%
\\
\>[9]{}\Rightarrow (\Varid{a}\to \Varid{g}\;\Varid{b})\to (\forall\kern-2pt\;\Varid{x}\, .\,K^{\textsf{Alg}}\;\sigma\;\Varid{g}\;(\Varid{g}\;\Varid{x})\to \Varid{g}\;\Varid{x})\to Free_{\textsf{H}}\;(K^{\textsf{Alg}}\;\sigma)\;\Varid{a}\to \Varid{g}\;\Varid{b}{}\<[E]%
\\
\>[3]{}h_{\textsf{Alg}}\mathrel{=}\textsf{fold}{}\<[E]%
\ColumnHook
\end{hscode}\resethooks
\paragraph{Example: State}

We have already introduced the state effect as a functor (\Cref{sec:aeh}).
The handler for state is written modularly in terms of \ensuremath{Free_{\textsf{H}}}.
\begin{hscode}\SaveRestoreHook
\column{B}{@{}>{\hspre}l<{\hspost}@{}}%
\column{3}{@{}>{\hspre}l<{\hspost}@{}}%
\column{6}{@{}>{\hspre}l<{\hspost}@{}}%
\column{8}{@{}>{\hspre}l<{\hspost}@{}}%
\column{9}{@{}>{\hspre}l<{\hspost}@{}}%
\column{14}{@{}>{\hspre}l<{\hspost}@{}}%
\column{17}{@{}>{\hspre}l<{\hspost}@{}}%
\column{26}{@{}>{\hspre}l<{\hspost}@{}}%
\column{29}{@{}>{\hspre}l<{\hspost}@{}}%
\column{36}{@{}>{\hspre}l<{\hspost}@{}}%
\column{E}{@{}>{\hspre}l<{\hspost}@{}}%
\>[3]{}h_{\textsf{State}}{}\<[8]%
\>[8]{}\mathbin{::}\Conid{Functor}\;\sigma{}\<[E]%
\\
\>[8]{}\Rightarrow Free_{\textsf{H}}\;(K^{\textsf{Alg}}\;(State\;\Varid{s}\mathrel{{+}}\sigma))\;\Varid{a}\to (\Varid{s}\to Free_{\textsf{H}}\;(K^{\textsf{Alg}}\;\sigma)\;(\Varid{a},\Varid{s})){}\<[E]%
\\
\>[3]{}h_{\textsf{State}}{}\<[8]%
\>[8]{}\mathrel{=}h_{\textsf{Alg}}\;\eta\;alg_{\textsf{Alg}}\;\mathbf{where}{}\<[E]%
\\
\>[3]{}\hsindent{3}{}\<[6]%
\>[6]{}alg_{\textsf{Alg}}\;{}\<[14]%
\>[14]{}(\textsf{Op}\;\Varid{op}){}\<[26]%
\>[26]{}\mathrel{=}(\Varid{alg}\mathrel{{\#}}\Varid{fwd})\;\Varid{op}\;\mathbf{where}{}\<[E]%
\\
\>[6]{}\hsindent{3}{}\<[9]%
\>[9]{}\Varid{alg}\;{}\<[17]%
\>[17]{}(\Conid{Get}\;\Varid{k}){}\<[29]%
\>[29]{}\mathrel{=}\lambda \Varid{s}{}\<[36]%
\>[36]{}\to \Varid{k}\;\Varid{s}\;\Varid{s}{}\<[E]%
\\
\>[6]{}\hsindent{3}{}\<[9]%
\>[9]{}\Varid{alg}\;{}\<[17]%
\>[17]{}(\Conid{Put}\;\Varid{s'}\;\Varid{k}){}\<[29]%
\>[29]{}\mathrel{=}\lambda \anonymous {}\<[36]%
\>[36]{}\to \Varid{k}\;\Varid{s'}{}\<[E]%
\\
\>[6]{}\hsindent{3}{}\<[9]%
\>[9]{}\Varid{fwd}\;{}\<[17]%
\>[17]{}\Varid{op}{}\<[29]%
\>[29]{}\mathrel{=}\lambda \Varid{s}{}\<[36]%
\>[36]{}\to Op_{\textsf{H}}\;(\textsf{Op}\;(\textsf{fmap}\;(\mathbin{\$}\Varid{s})\;\Varid{op}))){}\<[E]%
\ColumnHook
\end{hscode}\resethooks
We define smart constructors \ensuremath{\textsf{get}} and \ensuremath{\textsf{put}} to progam with the state effect
and revisit the example of \Cref{sec:aeh}.
\begin{hscode}\SaveRestoreHook
\column{B}{@{}>{\hspre}l<{\hspost}@{}}%
\column{3}{@{}>{\hspre}l<{\hspost}@{}}%
\column{E}{@{}>{\hspre}l<{\hspost}@{}}%
\>[3]{}\texttt{>>>}\;h_{\textsf{State}}\;(\mathbf{do}\;\Varid{x}\leftarrow \textsf{get};\textsf{put}\;(\Varid{x}\mathbin{+}\mathrm{1});\textsf{return}\;\mathrm{5})\;\mathrm{0}{}\<[E]%
\\
\>[3]{}(\mathrm{5},\mathrm{1}){}\<[E]%
\ColumnHook
\end{hscode}\resethooks

% \noindent
% The result is |(5,1)|, with |5| the resulting value and |1| the (incremented) state.

\paragraph{Example: Nondeterminism}

A handler for nondeterminism interprets computations in terms of a list with all possible results.
\begin{hscode}\SaveRestoreHook
\column{B}{@{}>{\hspre}l<{\hspost}@{}}%
\column{3}{@{}>{\hspre}l<{\hspost}@{}}%
\column{6}{@{}>{\hspre}l<{\hspost}@{}}%
\column{8}{@{}>{\hspre}l<{\hspost}@{}}%
\column{9}{@{}>{\hspre}l<{\hspost}@{}}%
\column{14}{@{}>{\hspre}l<{\hspost}@{}}%
\column{17}{@{}>{\hspre}l<{\hspost}@{}}%
\column{25}{@{}>{\hspre}l<{\hspost}@{}}%
\column{28}{@{}>{\hspre}l<{\hspost}@{}}%
\column{E}{@{}>{\hspre}l<{\hspost}@{}}%
\>[3]{}h_{\textsf{ND}}{}\<[8]%
\>[8]{}\mathbin{::}\Conid{Functor}\;\sigma{}\<[E]%
\\
\>[8]{}\Rightarrow Free_{\textsf{H}}\;(K^{\textsf{Alg}}\;(Choice\mathrel{{+}}\sigma))\;\Varid{a}\to Free_{\textsf{H}}\;(K^{\textsf{Alg}}\;\sigma)\;[\mskip1.5mu \Varid{a}\mskip1.5mu]{}\<[E]%
\\
\>[3]{}h_{\textsf{ND}}{}\<[8]%
\>[8]{}\mathrel{=}h_{\textsf{Alg}}\;\eta\;alg_{\textsf{Alg}}\;\mathbf{where}{}\<[E]%
\\
\>[3]{}\hsindent{3}{}\<[6]%
\>[6]{}alg_{\textsf{Alg}}\;{}\<[14]%
\>[14]{}(\textsf{Op}\;\Varid{op}){}\<[25]%
\>[25]{}\mathrel{=}(\Varid{alg}\mathrel{{\#}}\Varid{fwd})\;\Varid{op}\;\mathbf{where}{}\<[E]%
\\
\>[6]{}\hsindent{3}{}\<[9]%
\>[9]{}\Varid{alg}\;{}\<[17]%
\>[17]{}\Conid{Fail}{}\<[28]%
\>[28]{}\mathrel{=}\textsf{return}\;[\mskip1.5mu \mskip1.5mu]{}\<[E]%
\\
\>[6]{}\hsindent{3}{}\<[9]%
\>[9]{}\Varid{alg}\;{}\<[17]%
\>[17]{}(\Conid{Or}\;\Varid{p}\;\Varid{q}){}\<[28]%
\>[28]{}\mathrel{=}(\plus )\mathrel{{\langle\kern-1pt}{\$}{\kern-1pt\rangle}}\Varid{p}\mathrel{{\langle\kern-1pt}{\ast}{\kern-1pt\rangle}}\Varid{q}{}\<[E]%
\\
\>[6]{}\hsindent{3}{}\<[9]%
\>[9]{}\Varid{fwd}{}\<[28]%
\>[28]{}\mathrel{=}Op_{\textsf{H}}\, .\,\textsf{Op}{}\<[E]%
\ColumnHook
\end{hscode}\resethooks

\noindent
Constructors \ensuremath{\textsf{fail}} and \ensuremath{\textsf{or}} allow programming with nondeterminism as an effect
(similar to \ensuremath{\textsf{get}} and \ensuremath{\textsf{put}}).
For instance, consider the following example.
\begin{hscode}\SaveRestoreHook
\column{B}{@{}>{\hspre}l<{\hspost}@{}}%
\column{3}{@{}>{\hspre}l<{\hspost}@{}}%
\column{E}{@{}>{\hspre}l<{\hspost}@{}}%
\>[3]{}\texttt{>>>}\;h_{\textsf{ND}}\;(\textsf{or}\;(\textsf{return}\;\mathrm{1})\;(\textsf{or}\;(\textsf{or}\;(\textsf{return}\;\mathrm{2})\;(\textsf{return}\;\mathrm{3}))\;\textsf{fail})){}\<[E]%
\\
\>[3]{}[\mskip1.5mu \mathrm{1},\mathrm{2},\mathrm{3}\mskip1.5mu]{}\<[E]%
\ColumnHook
\end{hscode}\resethooks

%-------------------------------------------------------------------------------
\subsection{Scoped Effects \& Handlers}
\label{sec:scoped}

%- - - - - - - - - - - - - - - - - - - - - - - - - - - - - - - - - - - - - - - -
\paragraph{Definition}
Scoped effects \& handlers model effects that delimit a certain scope, with as
most prominent examples exceptions, nondeterminism with once,
and local state variables.
In order to retain the separation between syntax and semantics, and to achieve
the same modularity as for algebraic effects, the literature \cite{lics18,scope14,esop22}
proposes to model scoped effects by a recursive
datatype \ensuremath{\Conid{Prog}} that captures both algebraic and scoped effects.
Algebraic operations are represented by a functor \ensuremath{\sigma}, whereas another functor
\ensuremath{\gamma} is used for scoped operations.
\begin{hscode}\SaveRestoreHook
\column{B}{@{}>{\hspre}l<{\hspost}@{}}%
\column{3}{@{}>{\hspre}l<{\hspost}@{}}%
\column{5}{@{}>{\hspre}l<{\hspost}@{}}%
\column{14}{@{}>{\hspre}l<{\hspost}@{}}%
\column{54}{@{}>{\hspre}l<{\hspost}@{}}%
\column{E}{@{}>{\hspre}l<{\hspost}@{}}%
\>[3]{}\mathbf{data}\;\Conid{Prog}\;\sigma\;\gamma\;\Varid{a}\;\mathbf{where}{}\<[E]%
\\
\>[3]{}\hsindent{2}{}\<[5]%
\>[5]{}Var{}\<[14]%
\>[14]{}\mathbin{::}\Varid{a}{}\<[54]%
\>[54]{}\to \Conid{Prog}\;\sigma\;\gamma\;\Varid{a}{}\<[E]%
\\
\>[3]{}\hsindent{2}{}\<[5]%
\>[5]{}Op{}\<[14]%
\>[14]{}\mathbin{::}\sigma\;(\Conid{Prog}\;\sigma\;\gamma\;\Varid{a}){}\<[54]%
\>[54]{}\to \Conid{Prog}\;\sigma\;\gamma\;\Varid{a}{}\<[E]%
\\
\>[3]{}\hsindent{2}{}\<[5]%
\>[5]{}Enter{}\<[14]%
\>[14]{}\mathbin{::}\gamma\;(\Conid{Prog}\;\sigma\;\gamma\;(\Conid{Prog}\;\sigma\;\gamma\;\Varid{a})){}\<[54]%
\>[54]{}\to \Conid{Prog}\;\sigma\;\gamma\;\Varid{a}{}\<[E]%
\ColumnHook
\end{hscode}\resethooks

\noindent
Here, \ensuremath{\Conid{Op}} corresponds to algebraic effects, and \ensuremath{\Conid{Enter}} enters a scope, representing scoped effects.
\ensuremath{\Conid{Enter}} can be rewritten (using the co-yoneda lemma) as a program with
(1) a \emph{scoped computation} that represents the program in scope, and (2)
a \emph{continuation}, outside the scope.
\begin{hscode}\SaveRestoreHook
\column{B}{@{}>{\hspre}l<{\hspost}@{}}%
\column{3}{@{}>{\hspre}l<{\hspost}@{}}%
\column{E}{@{}>{\hspre}l<{\hspost}@{}}%
\>[3]{}\Conid{Enter}\mathbin{::}\forall\kern-2pt\;\Varid{b}\;\Varid{c}\, .\,\gamma\;\Varid{b}\to \underbrace{(\Varid{b}\to \Conid{Prog}\;\sigma\;\gamma\;\Varid{c})}_{\Varid{scoped}\;\Varid{computation}}\to \underbrace{(\Varid{c}\to \Conid{Prog}\;\sigma\;\gamma\;\Varid{a})}_{\Varid{continuation}}\to \Conid{Prog}\;\sigma\;\gamma\;\Varid{a}{}\<[E]%
\ColumnHook
\end{hscode}\resethooks
We zoom in on scoped effects only,
to later compose them again with algebraic effects.
\ensuremath{Free_{\textsf{Sc}}} is isomorphic to \ensuremath{\Conid{Prog}} without \ensuremath{Op} for algebraic effects.
\begin{hscode}\SaveRestoreHook
\column{B}{@{}>{\hspre}l<{\hspost}@{}}%
\column{3}{@{}>{\hspre}l<{\hspost}@{}}%
\column{5}{@{}>{\hspre}l<{\hspost}@{}}%
\column{13}{@{}>{\hspre}l<{\hspost}@{}}%
\column{49}{@{}>{\hspre}l<{\hspost}@{}}%
\column{E}{@{}>{\hspre}l<{\hspost}@{}}%
\>[3]{}\mathbf{data}\;Free_{\textsf{Sc}}\;\gamma\;\Varid{a}\;\mathbf{where}{}\<[E]%
\\
\>[3]{}\hsindent{2}{}\<[5]%
\>[5]{}Var{}\<[13]%
\>[13]{}\mathbin{::}\Varid{a}{}\<[49]%
\>[49]{}\to Free_{\textsf{Sc}}\;\gamma\;\Varid{a}{}\<[E]%
\\
\>[3]{}\hsindent{2}{}\<[5]%
\>[5]{}\Conid{Enter}{}\<[13]%
\>[13]{}\mathbin{::}\gamma\;(Free_{\textsf{Sc}}\;\gamma\;(Free_{\textsf{Sc}}\;\gamma\;\Varid{a})){}\<[49]%
\>[49]{}\to Free_{\textsf{Sc}}\;\gamma\;\Varid{a}{}\<[E]%
\ColumnHook
\end{hscode}\resethooks

%- - - - - - - - - - - - - - - - - - - - - - - - - - - - - - - - - - - - - - - -
\paragraph{Interpretation}
To interpret scoped effects, Yang et al. \cite{esop22}
have proposed functorial algebras as a structured way of
handling scoped effects, in contrast with the more tedious approach of Pir\'og et al.
\cite{lics18} in terms of indexed algebras.
A functorial algebra \cite{esop22} consists of two parts:
an \emph{endo-algebra} (\ensuremath{Alg_{E}}) interprets the part of the program in scope,
whereas a \emph{base algebra} (\ensuremath{Alg_{B}}) interprets the continuation.
% In particular, a functorial algebra is a quadruple | \< f, b, ealg, balg \> |,
% with |f| and |ealg| the endofunctor carrier and endofunctor algebra, respectively,
% and |b| and |balg| the base carrier and base algebra, respectively.

\noindent
\begin{minipage}[t]{0.5\textwidth}
\begin{hscode}\SaveRestoreHook
\column{B}{@{}>{\hspre}l<{\hspost}@{}}%
\column{3}{@{}>{\hspre}l<{\hspost}@{}}%
\column{5}{@{}>{\hspre}l<{\hspost}@{}}%
\column{14}{@{}>{\hspre}l<{\hspost}@{}}%
\column{29}{@{}>{\hspre}l<{\hspost}@{}}%
\column{E}{@{}>{\hspre}l<{\hspost}@{}}%
\>[3]{}\mathbf{data}\;Alg_{E}\;\gamma\;\Varid{f}\mathrel{=}Alg_{E}\;\{\mskip1.5mu {}\<[E]%
\\
\>[3]{}\hsindent{2}{}\<[5]%
\>[5]{}return_E{}\<[14]%
\>[14]{}\mathbin{::}\forall\kern-2pt\;\Varid{x}\, .\,{}\<[29]%
\>[29]{}\Varid{x}\to \Varid{f}\;\Varid{x},{}\<[E]%
\\
\>[3]{}\hsindent{2}{}\<[5]%
\>[5]{}enter_E{}\<[14]%
\>[14]{}\mathbin{::}\forall\kern-2pt\;\Varid{x}\, .\,{}\<[29]%
\>[29]{}\gamma\;(\Varid{f}\;(\Varid{f}\;\Varid{x}))\to \Varid{f}\;\Varid{x}\mskip1.5mu\}{}\<[E]%
\ColumnHook
\end{hscode}\resethooks
\end{minipage}%
\begin{minipage}[t]{0.5\textwidth}
\begin{hscode}\SaveRestoreHook
\column{B}{@{}>{\hspre}l<{\hspost}@{}}%
\column{3}{@{}>{\hspre}l<{\hspost}@{}}%
\column{5}{@{}>{\hspre}l<{\hspost}@{}}%
\column{14}{@{}>{\hspre}l<{\hspost}@{}}%
\column{E}{@{}>{\hspre}l<{\hspost}@{}}%
\>[3]{}\mathbf{data}\;Alg_{B}\;\gamma\;\Varid{f}\;\Varid{a}\mathrel{=}Alg_{B}\;\{\mskip1.5mu {}\<[E]%
\\
\>[3]{}\hsindent{2}{}\<[5]%
\>[5]{}enter_B{}\<[14]%
\>[14]{}\mathbin{::}\gamma\;(\Varid{f}\;\Varid{a})\to \Varid{a}\mskip1.5mu\}{}\<[E]%
\ColumnHook
\end{hscode}\resethooks
\end{minipage}%

\noindent
Nearly always, the two algebras (endo- and base algebra) have the
same implementation, simplifying the code significantly but
disallowing a different interpretation inside and outside the scope.
A different interpretation is desirable, for example, when abstracting over
evaluation strategies such as depth-first search, breadth-first search or
depth-bound search, the latter of which can be modeled as a scoped effect \cite{esop22}.
In our generic framework, we require the two algebras to be the same for simplicity reasons.
The structural recursion scheme \ensuremath{\textsf{fold}_{\textsf{Sc}}} is specialized to interpreting \ensuremath{Free_{\textsf{Sc}}}, a
free monad with only scoped effects.
% |foldSc| is defined in terms of |hcata|:
\begin{hscode}\SaveRestoreHook
\column{B}{@{}>{\hspre}l<{\hspost}@{}}%
\column{3}{@{}>{\hspre}l<{\hspost}@{}}%
\column{7}{@{}>{\hspre}l<{\hspost}@{}}%
\column{11}{@{}>{\hspre}l<{\hspost}@{}}%
\column{14}{@{}>{\hspre}l<{\hspost}@{}}%
\column{33}{@{}>{\hspre}l<{\hspost}@{}}%
\column{37}{@{}>{\hspre}l<{\hspost}@{}}%
\column{E}{@{}>{\hspre}l<{\hspost}@{}}%
\>[3]{}\textsf{fold}_{\textsf{Sc}}{}\<[11]%
\>[11]{}\mathbin{::}\Conid{Functor}\;\gamma{}\<[E]%
\\
\>[11]{}\Rightarrow (\Varid{a}\to \Varid{b})\to Alg_{E}\;\gamma\;\Varid{f}\to Alg_{B}\;\gamma\;\Varid{f}\;\Varid{b}\to Free_{\textsf{Sc}}\;\gamma\;\Varid{a}\to \Varid{b}{}\<[E]%
\\
\>[3]{}\textsf{fold}_{\textsf{Sc}}\;\Varid{gen}\;alg_{E}\;alg_{B}\;(Var\;\Varid{x}){}\<[37]%
\>[37]{}\mathrel{=}\Varid{gen}\;\Varid{x}{}\<[E]%
\\
\>[3]{}\textsf{fold}_{\textsf{Sc}}\;\Varid{gen}\;alg_{E}\;alg_{B}\;(\Conid{Enter}\;\Varid{sc}){}\<[37]%
\>[37]{}\mathrel{=}enter_B\;alg_{B}\;(\textsf{fmap}\;\Varid{endo}\;\Varid{sc}){}\<[E]%
\\
\>[3]{}\hsindent{4}{}\<[7]%
\>[7]{}\mathbf{where}\;{}\<[14]%
\>[14]{}\Varid{endo}{}\<[33]%
\>[33]{}\mathrel{=}h_{cata}\, .\,\textsf{fmap}\;(\textsf{fold}_{\textsf{Sc}}\;\Varid{gen}\;alg_{E}\;alg_{B}){}\<[E]%
\\
\>[14]{}h_{cata}\;(Var\;\Varid{x}){}\<[33]%
\>[33]{}\mathrel{=}return_E\;alg_{E}\;\Varid{x}{}\<[E]%
\\
\>[14]{}h_{cata}\;(\Conid{Enter}\;\Varid{sc}){}\<[33]%
\>[33]{}\mathrel{=}(enter_E\;alg_{E}\, .\,\textsf{fmap}\;(h_{cata}\, .\,\textsf{fmap}\;h_{cata}))\;\Varid{sc}{}\<[E]%
\ColumnHook
\end{hscode}\resethooks

%- - - - - - - - - - - - - - - - - - - - - - - - - - - - - - - - - - - - - - - -
\paragraph{Generic Framework}
Scoped effects \& handlers fit our generic framework.

\par\noindent\textbf{Step 1}
We define a mapping using a functor \ensuremath{\Gamma} to represent scoped operations.

\noindent
\begin{minipage}{0.4\textwidth}\begin{hscode}\SaveRestoreHook
\column{B}{@{}>{\hspre}l<{\hspost}@{}}%
\column{3}{@{}>{\hspre}l<{\hspost}@{}}%
\column{E}{@{}>{\hspre}l<{\hspost}@{}}%
\>[3]{}K^{\textsf{Sc}}_{\Gamma}\;\Conid{F}\;\Conid{A}\mathrel{=}\Gamma\;(\Conid{F}\;\Conid{A}){}\<[E]%
\ColumnHook
\end{hscode}\resethooks
\end{minipage}%
\begin{minipage}{0.6\textwidth}
\begin{hscode}\SaveRestoreHook
\column{B}{@{}>{\hspre}l<{\hspost}@{}}%
\column{3}{@{}>{\hspre}l<{\hspost}@{}}%
\column{5}{@{}>{\hspre}l<{\hspost}@{}}%
\column{13}{@{}>{\hspre}l<{\hspost}@{}}%
\column{E}{@{}>{\hspre}l<{\hspost}@{}}%
\>[3]{}\mathbf{data}\;K^{\textsf{Sc}}\;\gamma\;\Varid{f}\;\Varid{a}\;\mathbf{where}{}\<[E]%
\\
\>[3]{}\hsindent{2}{}\<[5]%
\>[5]{}\textsf{Enter}{}\<[13]%
\>[13]{}\mathbin{::}\gamma\;(\Varid{f}\;\Varid{a})\to K^{\textsf{Sc}}\;\gamma\;\Varid{f}\;\Varid{a}{}\<[E]%
\ColumnHook
\end{hscode}\resethooks
\end{minipage}%

\step{2}
This mapping \ensuremath{K^{\textsf{Sc}}_{\Gamma}} is a higher-order functor.
\begin{hscode}\SaveRestoreHook
\column{B}{@{}>{\hspre}l<{\hspost}@{}}%
\column{3}{@{}>{\hspre}l<{\hspost}@{}}%
\column{5}{@{}>{\hspre}l<{\hspost}@{}}%
\column{21}{@{}>{\hspre}l<{\hspost}@{}}%
\column{26}{@{}>{\hspre}l<{\hspost}@{}}%
\column{E}{@{}>{\hspre}l<{\hspost}@{}}%
\>[3]{}\mathbf{instance}\;\Conid{Functor}\;\gamma\Rightarrow HFunctor\;(K^{\textsf{Sc}}\;\gamma)\;\mathbf{where}{}\<[E]%
\\
\>[3]{}\hsindent{2}{}\<[5]%
\>[5]{}\textsf{hmap}\;\Varid{k}\;(\textsf{Enter}\;{}\<[21]%
\>[21]{}\Varid{sc}){}\<[26]%
\>[26]{}\mathrel{=}\textsf{Enter}\;(\textsf{fmap}\;\Varid{k}\;\Varid{sc}){}\<[E]%
\ColumnHook
\end{hscode}\resethooks
\step{3}
We say that the following isomorphism holds (\ref{app:iso}):\begin{hscode}\SaveRestoreHook
\column{B}{@{}>{\hspre}l<{\hspost}@{}}%
\column{3}{@{}>{\hspre}l<{\hspost}@{}}%
\column{E}{@{}>{\hspre}l<{\hspost}@{}}%
\>[3]{}Free_{\textsf{Sc}}\;\gamma\;\Varid{a}\;\cong\;Free_{\textsf{H}}\;(K^{\textsf{Sc}}\;\gamma)\;\Varid{a}{}\<[E]%
\ColumnHook
\end{hscode}\resethooks
\step{4}
We write a handler in terms of the generic recursion scheme.
In \ref{app:iso} we show that
\ensuremath{\textsf{fold}_{\textsf{Sc}}\;\Varid{gen}\;\Varid{alg}\;\Varid{alg}}
is isomorphic to \ensuremath{h_{\textsf{Sc}}\;\Varid{gen}\;(\lambda (\textsf{Enter}\;\Varid{sc})\to enter_E\;\Varid{alg}\;\Varid{sc})} with \ensuremath{return_E\;\Varid{alg}\mathrel{=}\eta}.
Notice that we use the same implementation for endo-algebra and base-algebra.
\begin{hscode}\SaveRestoreHook
\column{B}{@{}>{\hspre}l<{\hspost}@{}}%
\column{3}{@{}>{\hspre}l<{\hspost}@{}}%
\column{8}{@{}>{\hspre}l<{\hspost}@{}}%
\column{E}{@{}>{\hspre}l<{\hspost}@{}}%
\>[3]{}h_{\textsf{Sc}}{}\<[8]%
\>[8]{}\mathbin{::}(\Conid{Functor}\;\gamma,\Conid{Pointed}\;\Varid{g}){}\<[E]%
\\
\>[8]{}\Rightarrow (\Varid{a}\to \Varid{g}\;\Varid{b})\to (\forall\kern-2pt\;\Varid{x}\, .\,K^{\textsf{Sc}}\;\gamma\;\Varid{g}\;(\Varid{g}\;\Varid{x})\to \Varid{g}\;\Varid{x})\to Free_{\textsf{H}}\;(K^{\textsf{Sc}}\;\gamma)\;\Varid{a}\to \Varid{g}\;\Varid{b}{}\<[E]%
\\
\>[3]{}h_{\textsf{Sc}}{}\<[8]%
\>[8]{}\mathrel{=}\textsf{fold}{}\<[E]%
\ColumnHook
\end{hscode}\resethooks
% %- - - - - - - - - - - - - - - - - - - - - - - - - - - - - - - - - - - - - - - -
% \subsection{Coproducts}
%
We reconstruct the \ensuremath{\Conid{Prog}} datatype using the coproduct
of the higher-order functors for algebraic and scoped effects, and define a
corresponding handler.
\begin{hscode}\SaveRestoreHook
\column{B}{@{}>{\hspre}l<{\hspost}@{}}%
\column{3}{@{}>{\hspre}l<{\hspost}@{}}%
\column{10}{@{}>{\hspre}l<{\hspost}@{}}%
\column{41}{@{}>{\hspre}l<{\hspost}@{}}%
\column{E}{@{}>{\hspre}l<{\hspost}@{}}%
\>[3]{}\Conid{Prog}\;\sigma\;\gamma\;\Varid{a}\;\cong\;Free_{\textsf{H}}\;(K^{\textsf{Alg}}\;\sigma\mathbin{+}K^{\textsf{Sc}}\;\gamma)\;\Varid{a}{}\<[E]%
\\[\blanklineskip]%
\>[3]{}h_{\textsf{Prog}}{}\<[10]%
\>[10]{}\mathbin{::}(\Conid{Functor}\;\sigma,\Conid{Functor}\;\gamma,\Conid{Pointed}\;\Varid{g}){}\<[E]%
\\
\>[10]{}\Rightarrow (\Varid{a}\to \Varid{g}\;\Varid{b})\to (\forall\kern-2pt\;\Varid{x}\, .\,(K^{\textsf{Alg}}\;\sigma\mathrel{\oplus}K^{\textsf{Sc}}\;\gamma)\;\Varid{g}\;(\Varid{g}\;\Varid{x})\to \Varid{g}\;\Varid{x}){}\<[E]%
\\
\>[10]{}\to Free_{\textsf{H}}\;(K^{\textsf{Alg}}\;\sigma\mathrel{\oplus}K^{\textsf{Sc}}\;\gamma)\;\Varid{a}{}\<[41]%
\>[41]{}\to \Varid{g}\;\Varid{b}{}\<[E]%
\\
\>[3]{}h_{\textsf{Prog}}{}\<[10]%
\>[10]{}\mathrel{=}\textsf{fold}{}\<[E]%
\ColumnHook
\end{hscode}\resethooks

%- - - - - - - - - - - - - - - - - - - - - - - - - - - - - - - - - - - - - - - -
\paragraph{Example: Nondeterminism with Once}

To exemplify this approach, we use nondeterminism with \ensuremath{\textsf{once}}, an operation that
only returns the first result of a nondeterministic program.
We distinguish between the nondeterministic algebraic operations \ensuremath{\Conid{Fail}} and \ensuremath{\Conid{Or}},
and the scoped operation \ensuremath{\Conid{Once}}.
\begin{hscode}\SaveRestoreHook
\column{B}{@{}>{\hspre}l<{\hspost}@{}}%
\column{3}{@{}>{\hspre}l<{\hspost}@{}}%
\column{18}{@{}>{\hspre}l<{\hspost}@{}}%
\column{E}{@{}>{\hspre}l<{\hspost}@{}}%
\>[3]{}\mathbf{data}\;\Conid{Once}\;\Varid{a}{}\<[18]%
\>[18]{}\mathrel{=}\Conid{Once}\;\Varid{a}{}\<[E]%
\ColumnHook
\end{hscode}\resethooks

\noindent
Again, we interpret nondeterministic programs in terms of a list, that retains all
found results.
The handler for nondeterminism with once features both algebraic (\ensuremath{\Conid{Or}} and \ensuremath{\Conid{Fail}})
and scoped (\ensuremath{\Conid{Once}}) operations.
\begin{hscode}\SaveRestoreHook
\column{B}{@{}>{\hspre}l<{\hspost}@{}}%
\column{3}{@{}>{\hspre}l<{\hspost}@{}}%
\column{6}{@{}>{\hspre}l<{\hspost}@{}}%
\column{10}{@{}>{\hspre}l<{\hspost}@{}}%
\column{14}{@{}>{\hspre}l<{\hspost}@{}}%
\column{18}{@{}>{\hspre}l<{\hspost}@{}}%
\column{22}{@{}>{\hspre}l<{\hspost}@{}}%
\column{31}{@{}>{\hspre}l<{\hspost}@{}}%
\column{35}{@{}>{\hspre}l<{\hspost}@{}}%
\column{45}{@{}>{\hspre}c<{\hspost}@{}}%
\column{45E}{@{}l@{}}%
\column{48}{@{}>{\hspre}l<{\hspost}@{}}%
\column{59}{@{}>{\hspre}l<{\hspost}@{}}%
\column{62}{@{}>{\hspre}l<{\hspost}@{}}%
\column{E}{@{}>{\hspre}l<{\hspost}@{}}%
\>[3]{}h_{\textsf{Once}}{}\<[10]%
\>[10]{}\mathbin{::}(\Conid{Functor}\;\sigma,\Conid{Functor}\;\gamma){}\<[E]%
\\
\>[10]{}\Rightarrow Free_{\textsf{H}}\;(K^{\textsf{Alg}}\;(Choice\mathrel{{+}}\sigma)\mathrel{\oplus}K^{\textsf{Sc}}\;(\Conid{Once}\mathrel{{+}}\gamma))\;\Varid{a}{}\<[E]%
\\
\>[10]{}\to Free_{\textsf{H}}\;(K^{\textsf{Alg}}\;\sigma\mathrel{\oplus}K^{\textsf{Sc}}\;\gamma)\;[\mskip1.5mu \Varid{a}\mskip1.5mu]{}\<[E]%
\\
\>[3]{}h_{\textsf{Once}}{}\<[10]%
\>[10]{}\mathrel{=}\textsf{fold}\;\Varid{gen}\;(alg_{\textsf{Alg}}\kern+2pt\tikz[baseline=(char.base)]{ \node[circle,draw,inner sep=0pt,align=center,scale=0.3] (char) {\Huge{\#}};}\kern+2ptalg_{\textsf{Sc}})\;\mathbf{where}{}\<[E]%
\\
\>[3]{}\hsindent{3}{}\<[6]%
\>[6]{}\Varid{gen}\;{}\<[18]%
\>[18]{}\Varid{x}{}\<[31]%
\>[31]{}\mathrel{=}\textsf{return}\;[\mskip1.5mu \Varid{x}\mskip1.5mu]{}\<[E]%
\\
\>[3]{}\hsindent{3}{}\<[6]%
\>[6]{}alg_{\textsf{Alg}}\;{}\<[18]%
\>[18]{}(\textsf{Op}\;\Varid{op}){}\<[31]%
\>[31]{}\mathrel{=}(alg_{\textsf{Choice}}{}\<[45]%
\>[45]{}\mathrel{{\#}}{}\<[45E]%
\>[48]{}fwd_{\textsf{Choice}}{}\<[59]%
\>[59]{})\;{}\<[62]%
\>[62]{}\Varid{op}\;\mathbf{where}\mathbin{...}{}\<[E]%
\\
\>[3]{}\hsindent{3}{}\<[6]%
\>[6]{}alg_{\textsf{Sc}}\;{}\<[18]%
\>[18]{}(\textsf{Enter}\;\Varid{sc}){}\<[31]%
\>[31]{}\mathrel{=}(alg_{\textsf{Once}}{}\<[45]%
\>[45]{}\mathrel{{\#}}{}\<[45E]%
\>[48]{}fwd_{\textsf{Once}}{}\<[59]%
\>[59]{})\;{}\<[62]%
\>[62]{}\Varid{sc}\;\mathbf{where}{}\<[E]%
\\
\>[6]{}\hsindent{4}{}\<[10]%
\>[10]{}alg_{\textsf{Once}}\;{}\<[22]%
\>[22]{}(\Conid{Once}\;\Varid{y}){}\<[35]%
\>[35]{}\mathrel{=}\textsf{join}\;(\textsf{fmap}\;\textsf{head}\;\Varid{y}){}\<[E]%
\\
\>[6]{}\hsindent{4}{}\<[10]%
\>[10]{}fwd_{\textsf{Once}}{}\<[35]%
\>[35]{}\mathrel{=}Op_{\textsf{H}}\, .\,\textsf{Enter}\, .\,\textsf{fmap}\;(\textsf{fmap}\;\textsf{lift}_{\textsf{Once}}){}\<[E]%
\\
\>[10]{}\hsindent{4}{}\<[14]%
\>[14]{}\mathbf{where}\;{}\<[22]%
\>[22]{}\textsf{lift}_{\textsf{Once}}\mathrel{=}\Varid{foldr}\;(\lambda \Varid{x}\;\Varid{xs}\to (\plus )\mathrel{{\langle\kern-1pt}{\$}{\kern-1pt\rangle}}\Varid{x}\mathrel{{\langle\kern-1pt}{\ast}{\kern-1pt\rangle}}\Varid{xs})\;(\textsf{return}\;[\mskip1.5mu \mskip1.5mu]){}\<[E]%
\ColumnHook
\end{hscode}\resethooks
% <
% <    algND       Fail         = return []
% <    algND       (Or x y)     = (++) <$> x <*> y
% <    fwdND                    = OpT . Op_
% <

\noindent
We define smart constructors \ensuremath{\textsf{fail}}, \ensuremath{\textsf{or}} and \ensuremath{\textsf{once}} in the usual way to write
nondeterministic programs.
For example, consider the difference between the following program with scoped operation \ensuremath{\textsf{once}},
which effectively continues the rest of the program after the first result, and the equivalent
program without \ensuremath{\textsf{once}}.
% |once(or(1,5)) >>= \ x. or(x,x+1) = [1,2]|

% > exOnce :: [Int]
\begin{hscode}\SaveRestoreHook
\column{B}{@{}>{\hspre}l<{\hspost}@{}}%
\column{3}{@{}>{\hspre}l<{\hspost}@{}}%
\column{9}{@{}>{\hspre}l<{\hspost}@{}}%
\column{E}{@{}>{\hspre}l<{\hspost}@{}}%
\>[3]{}\texttt{>>>}\;{}\<[9]%
\>[9]{}h_{\textsf{Once}}\;(\colorbox{lightgray}{$\textsf{once}$}\;(\textsf{or}\;(\textsf{return}\;\mathrm{1})\;(\textsf{return}\;\mathrm{5}))\bind {}\<[E]%
\\
\>[9]{}\lambda \Varid{x}\to \textsf{or}\;(\textsf{return}\;\Varid{x})\;(\textsf{return}\;(\Varid{x}\mathbin{+}\mathrm{1}))){}\<[E]%
\\
\>[3]{}[\mskip1.5mu \mathrm{1},\mathrm{2}\mskip1.5mu]{}\<[E]%
\\
\>[3]{}\texttt{>>>}\;{}\<[9]%
\>[9]{}h_{\textsf{Once}}\;(\textsf{or}\;(\textsf{return}\;\mathrm{1})\;(\textsf{return}\;\mathrm{5})\bind {}\<[E]%
\\
\>[9]{}\lambda \Varid{x}\to \textsf{or}\;(\textsf{return}\;\Varid{x})\;(\textsf{return}\;(\Varid{x}\mathbin{+}\mathrm{1}))){}\<[E]%
\\
\>[3]{}[\mskip1.5mu \mathrm{1},\mathrm{2},\mathrm{5},\mathrm{6}\mskip1.5mu]{}\<[E]%
\ColumnHook
\end{hscode}\resethooks

% \noindent
% In contrast, the following program only has nondeterminism (omitting |once|).

% < repl hOnce (or (return 1) (return 5) >>= \x -> or (return x) (return (x + 1)))
% < [1,2,5,6]

%-------------------------------------------------------------------------------
\subsection{Parallel Effects \& Handlers}
\label{sec:parallel}

%- - - - - - - - - - - - - - - - - - - - - - - - - - - - - - - - - - - - - - - -
\paragraph{Definition}
In general, algebraic effects are executed sequentially. However, in several cases,
for example for performance reasons, a parallel execution of effects might be
desired. Xie et al. \cite{xie21} define parallel effects \& handlers, with an operational semantics that
supports both algebraic and parallel effects.
In particular, they represent parallel effects by means of a \ensuremath{\textsf{for}} keyword to iterate
over parallellizable computations.
Their work comes with an implementation in Dex \cite{dex} and in Haskell. %\footnote{
% \url{https://github.com/google-research/dex-lang/tree/parallel-effects-exploration}}.
%
They represent parallel effects by \ensuremath{Free_{\textsf{Par}}\;\rho\;\Varid{a}}, where \ensuremath{\rho} is an
iterable functor to represent parallel operations\footnote{This differs from the implementation
of Xie et al. \cite{xie21} in two ways: it omits algebraic effects and uses a generic functor
\ensuremath{\rho} instead of \ensuremath{\Conid{List}}s to iterate over computations.}.
\begin{hscode}\SaveRestoreHook
\column{B}{@{}>{\hspre}l<{\hspost}@{}}%
\column{3}{@{}>{\hspre}l<{\hspost}@{}}%
\column{5}{@{}>{\hspre}l<{\hspost}@{}}%
\column{11}{@{}>{\hspre}l<{\hspost}@{}}%
\column{63}{@{}>{\hspre}l<{\hspost}@{}}%
\column{E}{@{}>{\hspre}l<{\hspost}@{}}%
\>[3]{}\mathbf{data}\;Free_{\textsf{Par}}\;\rho\;\Varid{a}\;\mathbf{where}{}\<[E]%
\\
\>[3]{}\hsindent{2}{}\<[5]%
\>[5]{}Var{}\<[11]%
\>[11]{}\mathbin{::}\Varid{a}{}\<[63]%
\>[63]{}\to Free_{\textsf{Par}}\;\rho\;\Varid{a}{}\<[E]%
\\
\>[3]{}\hsindent{2}{}\<[5]%
\>[5]{}\Conid{For}{}\<[11]%
\>[11]{}\mathbin{::}\rho\;(Free_{\textsf{Par}}\;\rho\;\Varid{b})\to (\rho\;\Varid{b}\to Free_{\textsf{Par}}\;\rho\;\Varid{a}){}\<[63]%
\>[63]{}\to Free_{\textsf{Par}}\;\rho\;\Varid{a}{}\<[E]%
\ColumnHook
\end{hscode}\resethooks

\noindent
Here, \ensuremath{Var} is a pure computation and \ensuremath{\Conid{For}} represents a parallellizable computation.
\ensuremath{\Conid{For}} takes two arguments: an iterable structure of computations
(indicated by functor \ensuremath{\rho}), and a continuation.

Xie et al. \cite{xie21} do not show that this representation of parallel effects
is a free monad.

% %- - - - - - - - - - - - - - - - - - - - - - - - - - - - - - - - - - - - - - - -
% \paragraph{Interpretation}
%
% To interpret parallel effects, we need an algebra that can handle
% parallel computations.
% The algebra has the same structure as the |For| constructor of |FreePar|:
% it takes an iterable structure of computations and a continuation.
%
% > data AlgPar rho f a = AlgPar { hFor :: forall x . rho (f x) -> (rho x -> a) -> a }
%
% The corresponding handler for parallel effects is defined as follows:
%
% > foldPar :: (a -> b) -> AlgPar rho (FreePar rho) b -> FreePar rho a -> b
% > foldPar gen alg (Pure x)       = gen x
% > foldPar gen alg (For iters k)  = hFor alg iters (foldPar gen alg . k)

%- - - - - - - - - - - - - - - - - - - - - - - - - - - - - - - - - - - - - - - -
\paragraph{Generic Framework}

Parallel effects can be expressed in terms of our framework.

\par\noindent\textbf{Step 1}
We define a higher-order functor \ensuremath{K^{\textsf{Par}}_{\mathrm{P}}} for mapping effect signatures:
\begin{hscode}\SaveRestoreHook
\column{B}{@{}>{\hspre}l<{\hspost}@{}}%
\column{3}{@{}>{\hspre}l<{\hspost}@{}}%
\column{E}{@{}>{\hspre}l<{\hspost}@{}}%
\>[3]{}K^{\textsf{Par}}_{\mathrm{P}}\;\Conid{F}\;\Conid{A}\mathrel{=}\mathrm{P}\;(\Conid{F}\;\Conid{B})\times(\mathrm{P}\;\Conid{B}\Rightarrow \Conid{A}){}\<[E]%
\ColumnHook
\end{hscode}\resethooks
\begin{hscode}\SaveRestoreHook
\column{B}{@{}>{\hspre}l<{\hspost}@{}}%
\column{3}{@{}>{\hspre}l<{\hspost}@{}}%
\column{5}{@{}>{\hspre}l<{\hspost}@{}}%
\column{14}{@{}>{\hspre}l<{\hspost}@{}}%
\column{E}{@{}>{\hspre}l<{\hspost}@{}}%
\>[3]{}\mathbf{data}\;K^{\textsf{Par}}\;\rho\;\Varid{f}\;\Varid{a}\;\mathbf{where}{}\<[E]%
\\
\>[3]{}\hsindent{2}{}\<[5]%
\>[5]{}\textsf{For}{}\<[14]%
\>[14]{}\mathbin{::}\rho\;(\Varid{f}\;\Varid{b})\to (\rho\;\Varid{b}\to \Varid{a})\to K^{\textsf{Par}}\;\rho\;\Varid{f}\;\Varid{a}{}\<[E]%
\ColumnHook
\end{hscode}\resethooks

\step{2}
Indeed, \ensuremath{K^{\textsf{Par}}_{\mathrm{P}}} is a higher-order functor:
\begin{hscode}\SaveRestoreHook
\column{B}{@{}>{\hspre}l<{\hspost}@{}}%
\column{3}{@{}>{\hspre}l<{\hspost}@{}}%
\column{5}{@{}>{\hspre}l<{\hspost}@{}}%
\column{E}{@{}>{\hspre}l<{\hspost}@{}}%
\>[3]{}\mathbf{instance}\;\Conid{Functor}\;\rho\Rightarrow HFunctor\;(K^{\textsf{Par}}\;\rho)\;\mathbf{where}{}\<[E]%
\\
\>[3]{}\hsindent{2}{}\<[5]%
\>[5]{}\textsf{hmap}\;\Varid{k}\;(\textsf{For}\;\Varid{iters}\;\Varid{c})\mathrel{=}\textsf{For}\;(\textsf{fmap}\;\Varid{k}\;\Varid{iters})\;\Varid{c}{}\<[E]%
\ColumnHook
\end{hscode}\resethooks
\step{3}
Now, the following isomorphism holds (\ref{app:iso}):
\begin{hscode}\SaveRestoreHook
\column{B}{@{}>{\hspre}l<{\hspost}@{}}%
\column{3}{@{}>{\hspre}l<{\hspost}@{}}%
\column{E}{@{}>{\hspre}l<{\hspost}@{}}%
\>[3]{}Free_{\textsf{Par}}\;\rho\;\Varid{a}\;\cong\;Free_{\textsf{H}}\;(K^{\textsf{Par}}\;\rho)\;\Varid{a}{}\<[E]%
\ColumnHook
\end{hscode}\resethooks
This implies that also \ensuremath{Free_{\textsf{Par}}} is a free monad.

\step{4}
A handler for parallel effects can now be defined generically.
In \ref{app:iso}, we show that the handler of \cite{xie21} (in our adapted version)
is isomorphic to \ensuremath{h_{\textsf{Par}}\;\Varid{gen}\;(\lambda (\textsf{For}\;\Varid{iters}\;\Varid{k})\to h_{\textsf{For}}\;\Varid{alg}\;\Varid{iters}\;\Varid{k})}.
\begin{hscode}\SaveRestoreHook
\column{B}{@{}>{\hspre}l<{\hspost}@{}}%
\column{3}{@{}>{\hspre}l<{\hspost}@{}}%
\column{10}{@{}>{\hspre}l<{\hspost}@{}}%
\column{E}{@{}>{\hspre}l<{\hspost}@{}}%
\>[3]{}h_{\textsf{Par}}{}\<[10]%
\>[10]{}\mathbin{::}(\Conid{Functor}\;\rho,\Conid{Pointed}\;\Varid{g}){}\<[E]%
\\
\>[10]{}\Rightarrow (\Varid{a}\to \Varid{g}\;\Varid{b})\to (\forall\kern-2pt\;\Varid{x}\, .\,K^{\textsf{Par}}\;\rho\;\Varid{g}\;(\Varid{g}\;\Varid{x})\to \Varid{g}\;\Varid{x})\to Free_{\textsf{H}}\;(K^{\textsf{Par}}\;\rho)\;\Varid{a}\to \Varid{g}\;\Varid{b}{}\<[E]%
\\
\>[3]{}h_{\textsf{Par}}{}\<[10]%
\>[10]{}\mathrel{=}\textsf{fold}{}\<[E]%
\ColumnHook
\end{hscode}\resethooks

%- - - - - - - - - - - - - - - - - - - - - - - - - - - - - - - - - - - - - - - -
\paragraph{Example: Parallel Accumulation}

We revisit the example of Xie et al. \cite{xie21} that imitates Dex's accumulation
effect, which is similar to state but can only increment and is implicitly
initialized with the identity of this increment.
The accumulation is represented by
an algebraic operation \ensuremath{\Conid{Accum}\;\Varid{m}}, where \ensuremath{\Varid{m}} is a monoid\footnote{A monoid is defined
as a set equipped with an associative binary operation \ensuremath{\diamond}, and an identity \ensuremath{\epsilon}.}.
\begin{hscode}\SaveRestoreHook
\column{B}{@{}>{\hspre}l<{\hspost}@{}}%
\column{3}{@{}>{\hspre}l<{\hspost}@{}}%
\column{E}{@{}>{\hspre}l<{\hspost}@{}}%
\>[3]{}\mathbf{data}\;\Conid{Accum}\;\Varid{m}\;\Varid{a}\mathrel{=}\Conid{Accum}\;\Varid{m}\;\Varid{a}{}\<[E]%
\ColumnHook
\end{hscode}\resethooks

\noindent
The handler for parallel accumulation features algebraic and parallel
effects.
% It is defined in terms of our generic fold as follows:
\begin{hscode}\SaveRestoreHook
\column{B}{@{}>{\hspre}l<{\hspost}@{}}%
\column{3}{@{}>{\hspre}l<{\hspost}@{}}%
\column{6}{@{}>{\hspre}l<{\hspost}@{}}%
\column{9}{@{}>{\hspre}l<{\hspost}@{}}%
\column{11}{@{}>{\hspre}l<{\hspost}@{}}%
\column{15}{@{}>{\hspre}l<{\hspost}@{}}%
\column{16}{@{}>{\hspre}l<{\hspost}@{}}%
\column{19}{@{}>{\hspre}l<{\hspost}@{}}%
\column{34}{@{}>{\hspre}l<{\hspost}@{}}%
\column{37}{@{}>{\hspre}l<{\hspost}@{}}%
\column{43}{@{}>{\hspre}l<{\hspost}@{}}%
\column{E}{@{}>{\hspre}l<{\hspost}@{}}%
\>[3]{}h_{\textsf{Accum}}{}\<[11]%
\>[11]{}\mathbin{::}(\Conid{Monoid}\;\Varid{m},\Conid{Functor}\;\sigma){}\<[E]%
\\
\>[11]{}\Rightarrow Free_{\textsf{H}}\;(K^{\textsf{Alg}}\;(\Conid{Accum}\;\Varid{m}\mathrel{{+}}\sigma)\mathrel{\oplus}K^{\textsf{Par}}\;[\mskip1.5mu \mskip1.5mu]\;\Varid{a}{}\<[E]%
\\
\>[11]{}\to Free_{\textsf{H}}\;(K^{\textsf{Alg}}\;\sigma)\;(\Varid{m},\Varid{a}){}\<[E]%
\\
\>[3]{}h_{\textsf{Accum}}{}\<[11]%
\>[11]{}\mathrel{=}\textsf{fold}\;\Varid{gen}\;(alg_{\textsf{Alg}}\kern+2pt\tikz[baseline=(char.base)]{ \node[circle,draw,inner sep=0pt,align=center,scale=0.3] (char) {\Huge{\#}};}\kern+2ptalg_{\textsf{Par}})\;{}\<[43]%
\>[43]{}\mathbf{where}{}\<[E]%
\\
\>[3]{}\hsindent{3}{}\<[6]%
\>[6]{}\Varid{gen}\;{}\<[16]%
\>[16]{}\Varid{x}{}\<[34]%
\>[34]{}\mathrel{=}\textsf{return}\;(\epsilon,\Varid{x}){}\<[E]%
\\
\>[3]{}\hsindent{3}{}\<[6]%
\>[6]{}alg_{\textsf{Alg}}\;{}\<[16]%
\>[16]{}(\textsf{Op}\;\Varid{op}){}\<[34]%
\>[34]{}\mathrel{=}(alg_{\textsf{Accum}}\mathrel{{\#}}fwd_{\textsf{Accum}})\;\Varid{op}\;\mathbf{where}{}\<[E]%
\\
\>[6]{}\hsindent{3}{}\<[9]%
\>[9]{}alg_{\textsf{Accum}}\;{}\<[19]%
\>[19]{}(\Conid{Accum}\;\Varid{m'}\;\Varid{k}){}\<[37]%
\>[37]{}\mathrel{=}\textsf{fmap}\;(\lambda (\Varid{m},\Varid{x})\to (\Varid{m'}\diamond\Varid{m},\Varid{x}))\;\Varid{k}{}\<[E]%
\\
\>[6]{}\hsindent{3}{}\<[9]%
\>[9]{}fwd_{\textsf{Accum}}{}\<[37]%
\>[37]{}\mathrel{=}Op_{\textsf{H}}\, .\,\textsf{Op}{}\<[E]%
\\
\>[3]{}\hsindent{3}{}\<[6]%
\>[6]{}alg_{\textsf{Par}}\;{}\<[16]%
\>[16]{}(\textsf{For}\;\Varid{iters}\;\Varid{k}){}\<[34]%
\>[34]{}\mathrel{=}\mathbf{do}{}\<[E]%
\\
\>[6]{}\hsindent{9}{}\<[15]%
\>[15]{}(\Varid{ms},\Varid{xs})\leftarrow \textsf{fmap}\;\Varid{unzip}\;(\Varid{sequence}\;\Varid{iters}){}\<[E]%
\\
\>[6]{}\hsindent{9}{}\<[15]%
\>[15]{}\mathbf{let}\;\Varid{append}\;(\Varid{m},\Varid{x}){}\<[34]%
\>[34]{}\mathrel{=}(\Varid{foldr}\;(\diamond)\;\Varid{m}\;\Varid{ms},\Varid{x})\;\mathbf{in}\;\textsf{fmap}\;\Varid{append}\;(\Varid{k}\;\Varid{xs}){}\<[E]%
\ColumnHook
\end{hscode}\resethooks
% <    algPar    (For_ op k)       = OpT (For_ op (k . fmap snd))

\noindent
We define constructors \ensuremath{\textsf{accum}} and \ensuremath{\textsf{for}} to accumulate
and iterate over computations, respectively.
Consider the following example that computes the sum of a list of
integers. We use the \ensuremath{\Conid{Sum}} monoid for our accumulator.
\begin{hscode}\SaveRestoreHook
\column{B}{@{}>{\hspre}l<{\hspost}@{}}%
\column{3}{@{}>{\hspre}l<{\hspost}@{}}%
\column{E}{@{}>{\hspre}l<{\hspost}@{}}%
\>[3]{}\texttt{>>>}\;h_{\textsf{Accum}}\;(\textsf{for}\;(\textsf{fmap}\;(\textsf{accum}\, .\,\Conid{Sum})\;[\mskip1.5mu \mathrm{1},\mathrm{2},\mathrm{10},\mathrm{4}\mskip1.5mu])){}\<[E]%
\\
\>[3]{}\mathrm{17}{}\<[E]%
\ColumnHook
\end{hscode}\resethooks

%-------------------------------------------------------------------------------
\subsection{Writer Effect \& Handler}
\label{sec:inner}

To define writer effects \& handlers, we use the writer monad as a running example.
The writer monad keeps track of both return values and output messages, for example in a log file.
The minimal complete definition of the writer effect, according to the MonadWriter
library\footnote{\url{https://hackage.haskell.org/package/mtl-2.2.2/docs/Control-Monad-Writer-Class.html}},
requires three operations: \ensuremath{\textsf{tell}}, \ensuremath{\textsf{listen}} and \ensuremath{\textsf{pass}}.
The former is an algebraic operation, producing output messages of type \ensuremath{\Varid{w}}.
\begin{hscode}\SaveRestoreHook
\column{B}{@{}>{\hspre}l<{\hspost}@{}}%
\column{3}{@{}>{\hspre}l<{\hspost}@{}}%
\column{E}{@{}>{\hspre}l<{\hspost}@{}}%
\>[3]{}\textsf{tell}\mathbin{::}\Varid{w}\to \Varid{m}\;(){}\<[E]%
\ColumnHook
\end{hscode}\resethooks
Both \ensuremath{\textsf{listen}} and \ensuremath{\textsf{pass}} do not fit any of the previously defined effects.
We reformulate their definitions to make them fit our generic framework.
\ensuremath{\textsf{listen}} executes a computation and returns a tuple of the resulting
value and the output message.
It is used to \emph{inspect} what a subcomputation has written to the output.
% By equational reasoning, we restructure the type of |listen|.
\begin{hscode}\SaveRestoreHook
\column{B}{@{}>{\hspre}l<{\hspost}@{}}%
\column{4}{@{}>{\hspre}l<{\hspost}@{}}%
\column{9}{@{}>{\hspre}l<{\hspost}@{}}%
\column{17}{@{}>{\hspre}c<{\hspost}@{}}%
\column{17E}{@{}l@{}}%
\column{22}{@{}>{\hspre}l<{\hspost}@{}}%
\column{46}{@{}>{\hspre}l<{\hspost}@{}}%
\column{56}{@{}>{\hspre}l<{\hspost}@{}}%
\column{E}{@{}>{\hspre}l<{\hspost}@{}}%
\>[9]{}\textsf{listen}{}\<[17]%
\>[17]{}\mathbin{::}{}\<[17E]%
\>[22]{}\Varid{m}\;\Varid{a}\to \Varid{m}\;(\Varid{a},\Varid{w}){}\<[E]%
\\
\>[4]{}\cong\;{}\<[9]%
\>[9]{}\textsf{listen}{}\<[17]%
\>[17]{}\mathbin{::}{}\<[17E]%
\>[22]{}\Varid{m}\;\Varid{a}\to ((\Varid{a},\Varid{w})\to \Varid{m}\;\Varid{b}){}\<[46]%
\>[46]{}\to \Varid{m}\;\Varid{b}{}\<[E]%
\\
\>[4]{}\cong\;{}\<[9]%
\>[9]{}\textsf{listen}{}\<[17]%
\>[17]{}\mathbin{::}{}\<[17E]%
\>[22]{}\Varid{m}\;(\Varid{w}\to \Varid{m}\;\Varid{b}){}\<[46]%
\>[46]{}\to \Varid{m}\;\Varid{b}{}\<[E]%
\\
\>[4]{}\cong\;{}\<[9]%
\>[9]{}\textsf{listen}{}\<[17]%
\>[17]{}\mathbin{::}{}\<[17E]%
\>[22]{}\Varid{m}\;(\varphi_{\textsf{listen}}\;(\Varid{m}\;\Varid{b})){}\<[46]%
\>[46]{}\to \Varid{m}\;\Varid{b}\;{}\<[56]%
\>[56]{}\hspace{15pt}\textsf{with}\;\varphi_{\textsf{listen}}\;\Varid{w}\mathrel{=}((\to )\;\Varid{w}){}\<[E]%
\ColumnHook
\end{hscode}\resethooks
Furthermore, \ensuremath{\textsf{pass}} executes a computation, resulting in a value and
a function, the latter of which is applied to the output message.
It is used to \emph{modify} what is written to the output.
\begin{hscode}\SaveRestoreHook
\column{B}{@{}>{\hspre}l<{\hspost}@{}}%
\column{4}{@{}>{\hspre}l<{\hspost}@{}}%
\column{9}{@{}>{\hspre}l<{\hspost}@{}}%
\column{17}{@{}>{\hspre}c<{\hspost}@{}}%
\column{17E}{@{}l@{}}%
\column{22}{@{}>{\hspre}l<{\hspost}@{}}%
\column{37}{@{}>{\hspre}l<{\hspost}@{}}%
\column{55}{@{}>{\hspre}l<{\hspost}@{}}%
\column{E}{@{}>{\hspre}l<{\hspost}@{}}%
\>[9]{}\textsf{pass}{}\<[17]%
\>[17]{}\mathbin{::}{}\<[17E]%
\>[22]{}\Varid{m}\;(\Varid{a},\Varid{w}\to \Varid{w}){}\<[37]%
\>[37]{}\to \Varid{m}\;\Varid{a}{}\<[E]%
\\
\>[4]{}\cong\;{}\<[9]%
\>[9]{}\textsf{pass}{}\<[17]%
\>[17]{}\mathbin{::}{}\<[17E]%
\>[22]{}\Varid{m}\;(\varphi_{\textsf{pass}}\;\Varid{a}){}\<[37]%
\>[37]{}\to \Varid{m}\;\Varid{a}\;{}\<[55]%
\>[55]{}\hspace{15pt}\textsf{with}\;\varphi_{\textsf{pass}}\;\Varid{w}\mathrel{=}((,)\;(\Varid{w}\to \Varid{w})){}\<[E]%
\ColumnHook
\end{hscode}\resethooks
We now abstract over these two definitions of \ensuremath{\textsf{listen}} and \ensuremath{\textsf{pass}},
defining a novel kind of effects and showing that they are a special case of
our generic framework.

%- - - - - - - - - - - - - - - - - - - - - - - - - - - - - - - - - - - - - - - -
\paragraph{Definition}

We dub these effects \emph{\ensuremath{\text{writer}} effects} and denote them
by \ensuremath{Free_{\textsf{Write}}\;\varphi\;\Varid{a}}, where \ensuremath{\varphi} is a functor for \ensuremath{\text{writer}} effect operations.
\begin{hscode}\SaveRestoreHook
\column{B}{@{}>{\hspre}l<{\hspost}@{}}%
\column{3}{@{}>{\hspre}l<{\hspost}@{}}%
\column{5}{@{}>{\hspre}l<{\hspost}@{}}%
\column{12}{@{}>{\hspre}l<{\hspost}@{}}%
\column{54}{@{}>{\hspre}l<{\hspost}@{}}%
\column{E}{@{}>{\hspre}l<{\hspost}@{}}%
\>[3]{}\mathbf{data}\;Free_{\textsf{Write}}\;\varphi\;\Varid{a}\;\mathbf{where}{}\<[E]%
\\
\>[3]{}\hsindent{2}{}\<[5]%
\>[5]{}Var{}\<[12]%
\>[12]{}\mathbin{::}\Varid{a}{}\<[54]%
\>[54]{}\to Free_{\textsf{Write}}\;\varphi\;\Varid{a}{}\<[E]%
\\
\>[3]{}\hsindent{2}{}\<[5]%
\>[5]{}\Conid{Exec}{}\<[12]%
\>[12]{}\mathbin{::}Free_{\textsf{Write}}\;\varphi\;(\varphi\;(Free_{\textsf{Write}}\;\varphi\;\Varid{a})){}\<[54]%
\>[54]{}\to Free_{\textsf{Write}}\;\varphi\;\Varid{a}{}\<[E]%
\ColumnHook
\end{hscode}\resethooks
% Both |listen| and |pass| do not fit in any of the previous categories of effects.
% We design a free monad |FreeWrite phi a|
% that captures these two operations, where |sig| refers to the functor for
% algebraic operations, and |phi| is a functor to denote |inner| effects.
%
% > data Action sig phi a where
% >   Ret'      :: a -> Action sig phi a
% >   Produce'  :: sig (Action sig phi a) -> Action sig phi a
% >   Exec'     :: Action sig phi (phi (Action sig phi a)) -> Action sig phi a
%
Here, \ensuremath{Var} represents a pure computation and \ensuremath{\Conid{Exec}} is a \ensuremath{\text{writer}} computation.
% In what follows, we isolate the novel class of effects from the algebraic part
% and work with a free monad |FreeWrite|:
In fact, one can rewrite (using the co-yoneda lemma) \ensuremath{\Conid{Exec}} so that it consists of a \ensuremath{\text{writer}} action,
of which the result is decorated by some functor \ensuremath{\varphi}, and a continuation.
\begin{hscode}\SaveRestoreHook
\column{B}{@{}>{\hspre}l<{\hspost}@{}}%
\column{3}{@{}>{\hspre}l<{\hspost}@{}}%
\column{E}{@{}>{\hspre}l<{\hspost}@{}}%
\>[3]{}\Conid{Exec}\mathbin{::}\forall\kern-2pt\;\Varid{b}\, .\,\underbrace{Free_{\textsf{Write}}\;\varphi\;(\varphi\;\Varid{b})}_{\text{writer}\;\Varid{computation}}\to \underbrace{(\Varid{b}\to Free_{\textsf{Write}}\;\varphi\;\Varid{a})}_{\Varid{continuation}}\to Free_{\textsf{Write}}\;\varphi\;\Varid{a}{}\<[E]%
\ColumnHook
\end{hscode}\resethooks
In particular, for \ensuremath{\textsf{listen}} this functor \ensuremath{\varphi} is \ensuremath{((\to )\;\Varid{w})} as the result type of the
\ensuremath{\text{writer}} action is \ensuremath{\Varid{w}\to \Varid{m}\;\Varid{a}}.
Similarly, \ensuremath{\textsf{pass}} is decorated by the functor \ensuremath{((,)\;(\Varid{w}\to \Varid{w}))} as its inner action
has result type \ensuremath{(\Varid{a},\Varid{w}\to \Varid{w})}.

% %- - - - - - - - - - - - - - - - - - - - - - - - - - - - - - - - - - - - - - - -
% \paragraph{Interpretation}
%
% To interpret |inner| effects, we design an algebra for |inner| effects.
% This algebra has a structure similar to |Exec| and a corresponding
% handler.
%
% > data AlgWrite phi f a = AlgWrite { hExec  :: f (phi a) -> a }
% >
% > foldWrite :: (Functor phi) => (a -> b) -> AlgWrite phi (FreeWrite phi) b -> FreeWrite phi a -> b
% > foldWrite gen alg (Ret x)   = gen x
% > foldWrite gen alg (Exec k)  = hExec alg (fmap (fmap (foldWrite gen alg)) k)
%
% %if False
%
% > instance Functor phi => Functor (FreeWrite phi) where
% >   fmap f (Ret x)  = Ret (f x)
% >   fmap f (Exec k) = Exec (fmap (fmap (fmap f)) k)
%
% %endif

%- - - - - - - - - - - - - - - - - - - - - - - - - - - - - - - - - - - - - - - -
\paragraph{Generic Framework}
We show that the writer effect fits our generic framework.

\par\noindent\textbf{Step 1}
We choose an appropriate mapping \ensuremath{K^{\textsf{Write}}_{\Phi}} to express \ensuremath{\text{writer}} effects
in terms of our theoretical model.
Notice the similarity with scoped effects \& handlers, with the order
of the functors reversed.

\noindent
\begin{minipage}{0.5\textwidth}\begin{hscode}\SaveRestoreHook
\column{B}{@{}>{\hspre}l<{\hspost}@{}}%
\column{3}{@{}>{\hspre}l<{\hspost}@{}}%
\column{E}{@{}>{\hspre}l<{\hspost}@{}}%
\>[3]{}K^{\textsf{Write}}_{\Phi}\;\Conid{F}\;\Conid{A}\mathrel{=}\Conid{F}\;(\Phi\;\Conid{A}){}\<[E]%
\ColumnHook
\end{hscode}\resethooks
\end{minipage}%
\begin{minipage}{0.5\textwidth}
\begin{hscode}\SaveRestoreHook
\column{B}{@{}>{\hspre}l<{\hspost}@{}}%
\column{3}{@{}>{\hspre}l<{\hspost}@{}}%
\column{5}{@{}>{\hspre}l<{\hspost}@{}}%
\column{13}{@{}>{\hspre}l<{\hspost}@{}}%
\column{E}{@{}>{\hspre}l<{\hspost}@{}}%
\>[3]{}\mathbf{data}\;K^{\textsf{Write}}\;\varphi\;\Varid{f}\;\Varid{a}\;\mathbf{where}{}\<[E]%
\\
\>[3]{}\hsindent{2}{}\<[5]%
\>[5]{}\textsf{Exec}{}\<[13]%
\>[13]{}\mathbin{::}\Varid{f}\;(\varphi\;\Varid{a})\to K^{\textsf{Write}}\;\varphi\;\Varid{f}\;\Varid{a}{}\<[E]%
\ColumnHook
\end{hscode}\resethooks
\end{minipage}

% < KWrite_Phi F A = intg Phi G >< F (G A)
% > data KWrite phi f a where
% >   Exec_   :: Functor g => phi g -> f (g a) -> KWrite phi f a

\step{2}
\ensuremath{K^{\textsf{Write}}_{\Phi}} is a higher-order functor:

\begin{hscode}\SaveRestoreHook
\column{B}{@{}>{\hspre}l<{\hspost}@{}}%
\column{3}{@{}>{\hspre}l<{\hspost}@{}}%
\column{5}{@{}>{\hspre}l<{\hspost}@{}}%
\column{E}{@{}>{\hspre}l<{\hspost}@{}}%
\>[3]{}\mathbf{instance}\;\Conid{Functor}\;\varphi\Rightarrow HFunctor\;(K^{\textsf{Write}}\;\varphi)\;\mathbf{where}{}\<[E]%
\\
\>[3]{}\hsindent{2}{}\<[5]%
\>[5]{}\textsf{hmap}\;\Varid{k}\;(\textsf{Exec}\;\Varid{x})\mathrel{=}\textsf{Exec}\;(\Varid{k}\;\Varid{x}){}\<[E]%
\ColumnHook
\end{hscode}\resethooks
\step{3}
Furthermore, the following isomorphism holds (\ref{app:iso}):
\begin{hscode}\SaveRestoreHook
\column{B}{@{}>{\hspre}l<{\hspost}@{}}%
\column{3}{@{}>{\hspre}l<{\hspost}@{}}%
\column{E}{@{}>{\hspre}l<{\hspost}@{}}%
\>[3]{}Free_{\textsf{Write}}\;\varphi\;\Varid{a}\;\cong\;Free_{\textsf{H}}\;(K^{\textsf{Write}}\;\varphi)\;\Varid{a}{}\<[E]%
\ColumnHook
\end{hscode}\resethooks
This implies that also \ensuremath{Free_{\textsf{Write}}} is a free monad.

\step{4}
A handler for \ensuremath{\text{writer}} effects can now be defined in terms of our framework:
\begin{hscode}\SaveRestoreHook
\column{B}{@{}>{\hspre}l<{\hspost}@{}}%
\column{3}{@{}>{\hspre}l<{\hspost}@{}}%
\column{10}{@{}>{\hspre}l<{\hspost}@{}}%
\column{E}{@{}>{\hspre}l<{\hspost}@{}}%
\>[3]{}h_{\textsf{Wr}}{}\<[10]%
\>[10]{}\mathbin{::}(\Conid{Functor}\;\varphi,\Conid{Pointed}\;\Varid{g}){}\<[E]%
\\
\>[10]{}\Rightarrow (\Varid{a}\to \Varid{g}\;\Varid{b})\to (\forall\kern-2pt\;\Varid{x}\, .\,K^{\textsf{Write}}\;\varphi\;\Varid{g}\;(\Varid{g}\;\Varid{x})\to \Varid{g}\;\Varid{x})\to Free_{\textsf{H}}\;(K^{\textsf{Write}}\;\varphi)\;\Varid{a}\to \Varid{g}\;\Varid{b}{}\<[E]%
\\
\>[3]{}h_{\textsf{Wr}}{}\<[10]%
\>[10]{}\mathrel{=}\textsf{fold}{}\<[E]%
\ColumnHook
\end{hscode}\resethooks
Writing an isomorphism for this handler is not meaningful as no specialized handler
for \ensuremath{Free_{\textsf{Write}}} exists.

%- - - - - - - - - - - - - - - - - - - - - - - - - - - - - - - - - - - - - - - -
\paragraph{Example: Resetting the Log}

The writer monad uses both algebraic and \ensuremath{\text{writer}} effects.
\ensuremath{\Conid{Tell}} constructs output messages, where \ensuremath{\Varid{w}} is a monoid that
keeps track of the output message.
\begin{hscode}\SaveRestoreHook
\column{B}{@{}>{\hspre}l<{\hspost}@{}}%
\column{3}{@{}>{\hspre}l<{\hspost}@{}}%
\column{E}{@{}>{\hspre}l<{\hspost}@{}}%
\>[3]{}\mathbf{data}\;\Conid{Tell}\;\Varid{w}\;\Varid{a}\mathrel{=}\Conid{Tell}\;\Varid{w}\;\Varid{a}{}\<[E]%
\ColumnHook
\end{hscode}\resethooks

\noindent
\ensuremath{\Conid{Listen}} sees what a (sub-)computation wrote to the output.
\ensuremath{\Conid{Pass}} is able to adapt the message that is written to the output.
Both operations are represented by a functor corresponding to the argument type
of \ensuremath{\textsf{listen}} and \ensuremath{\textsf{pass}}, respectively.

\noindent
\begin{minipage}{0.5\textwidth}
\begin{hscode}\SaveRestoreHook
\column{B}{@{}>{\hspre}l<{\hspost}@{}}%
\column{3}{@{}>{\hspre}l<{\hspost}@{}}%
\column{E}{@{}>{\hspre}l<{\hspost}@{}}%
\>[3]{}\mathbf{type}\;\Conid{Listen}\;\Varid{w}\mathrel{=}((\to )\;\Varid{w}){}\<[E]%
\ColumnHook
\end{hscode}\resethooks
\end{minipage}%
\begin{minipage}{0.5\textwidth}
\begin{hscode}\SaveRestoreHook
\column{B}{@{}>{\hspre}l<{\hspost}@{}}%
\column{3}{@{}>{\hspre}l<{\hspost}@{}}%
\column{E}{@{}>{\hspre}l<{\hspost}@{}}%
\>[3]{}\mathbf{type}\;\Conid{Pass}\;\Varid{w}\mathrel{=}((,)\;(\Varid{w}\to \Varid{w})){}\<[E]%
\ColumnHook
\end{hscode}\resethooks
\end{minipage}

% These two operations are similar. Indeed, we can capture them in a datatype |Write w|,
% which takes a functor to distinguish between the |listen| and |pass| functionality.

% > data Write w :: (* -> *) -> * where
% >   Listen  :: Write w  ((->)  w)
% >   Pass    :: Write w  ((,)  (w -> w))

\noindent
The handler for writing is defined in terms of our generic framework:
\begin{hscode}\SaveRestoreHook
\column{B}{@{}>{\hspre}l<{\hspost}@{}}%
\column{3}{@{}>{\hspre}l<{\hspost}@{}}%
\column{6}{@{}>{\hspre}l<{\hspost}@{}}%
\column{11}{@{}>{\hspre}l<{\hspost}@{}}%
\column{12}{@{}>{\hspre}l<{\hspost}@{}}%
\column{16}{@{}>{\hspre}l<{\hspost}@{}}%
\column{27}{@{}>{\hspre}l<{\hspost}@{}}%
\column{28}{@{}>{\hspre}l<{\hspost}@{}}%
\column{31}{@{}>{\hspre}l<{\hspost}@{}}%
\column{34}{@{}>{\hspre}l<{\hspost}@{}}%
\column{E}{@{}>{\hspre}l<{\hspost}@{}}%
\>[3]{}h_{\textsf{Write}}{}\<[12]%
\>[12]{}\mathbin{::}(\Conid{Functor}\;\sigma,\Conid{Monoid}\;\Varid{w}){}\<[E]%
\\
\>[12]{}\Rightarrow Free_{\textsf{H}}\;(K^{\textsf{Alg}}\;(\Conid{Tell}\;\Varid{w}\mathrel{{+}}\sigma)\mathrel{\oplus}K^{\textsf{Write}}\;(\Conid{Listen}\;\Varid{w}\mathrel{{+}}\Conid{Pass}\;\Varid{w}\mathrel{{+}}\varphi))\;\Varid{a}{}\<[E]%
\\
\>[12]{}\to Free_{\textsf{H}}\;(K^{\textsf{Alg}}\;\sigma\mathrel{\oplus}K^{\textsf{Write}}\;\varphi)\;(\Varid{a},\Varid{w}){}\<[E]%
\\
\>[3]{}h_{\textsf{Write}}{}\<[12]%
\>[12]{}\mathrel{=}\textsf{fold}\;\Varid{gen}\;(alg_{\textsf{Alg}}\kern+2pt\tikz[baseline=(char.base)]{ \node[circle,draw,inner sep=0pt,align=center,scale=0.3] (char) {\Huge{\#}};}\kern+2ptalg_{\textsf{Write}})\;\mathbf{where}{}\<[E]%
\\
\>[3]{}\hsindent{3}{}\<[6]%
\>[6]{}\Varid{gen}\;\Varid{x}{}\<[28]%
\>[28]{}\mathrel{=}\textsf{return}\;(\Varid{x},\epsilon){}\<[E]%
\\
\>[3]{}\hsindent{3}{}\<[6]%
\>[6]{}alg_{\textsf{Alg}}\;(\textsf{Op}\;\Varid{op}){}\<[28]%
\>[28]{}\mathrel{=}(alg_{\textsf{Tell}}\mathrel{{\#}}fwd_{\textsf{Tell}})\;\Varid{op}\;\mathbf{where}{}\<[E]%
\\
\>[6]{}\hsindent{5}{}\<[11]%
\>[11]{}alg_{\textsf{Tell}}\;(\Conid{Tell}\;\Varid{w}\;\Varid{k}){}\<[31]%
\>[31]{}\mathrel{=}\mathbf{do}\;(\Varid{x},\Varid{w'})\leftarrow \Varid{k};\textsf{return}\;(\Varid{x},\Varid{w}\diamond\Varid{w'}){}\<[E]%
\\
\>[6]{}\hsindent{5}{}\<[11]%
\>[11]{}fwd_{\textsf{Tell}}{}\<[31]%
\>[31]{}\mathrel{=}Op_{\textsf{H}}\, .\,\textsf{Op}{}\<[E]%
\\
\>[3]{}\hsindent{3}{}\<[6]%
\>[6]{}alg_{\textsf{Write}}\;(\textsf{Exec}\;\Varid{k}){}\<[28]%
\>[28]{}\mathrel{=}\Varid{k}\bind \lambda \mathbf{case}{}\<[E]%
\\
\>[6]{}\hsindent{10}{}\<[16]%
\>[16]{}(\Varid{f}{}\<[27]%
\>[27]{},\Varid{w}){}\<[34]%
\>[34]{}\to \Varid{f}\;\Varid{w}{}\<[E]%
\\
\>[6]{}\hsindent{10}{}\<[16]%
\>[16]{}((\Varid{f},\Varid{mx})){}\<[27]%
\>[27]{},\anonymous ){}\<[34]%
\>[34]{}\to \textsf{fmap}\;(\textsf{fmap}\;\Varid{f})\;\Varid{mx}{}\<[E]%
\\
\>[6]{}\hsindent{10}{}\<[16]%
\>[16]{}(\Varid{op}{}\<[27]%
\>[27]{},\anonymous ){}\<[34]%
\>[34]{}\to Op_{\textsf{H}}\;(\textsf{Exec}\;(\textsf{return}\;\Varid{op})){}\<[E]%
\ColumnHook
\end{hscode}\resethooks
\noindent
We define constructors \ensuremath{\textsf{tell}}, \ensuremath{\textsf{listen}} and \ensuremath{\textsf{pass}} to define a function \ensuremath{\Varid{reset}} which
resets the log to empty it.
\begin{hscode}\SaveRestoreHook
\column{B}{@{}>{\hspre}l<{\hspost}@{}}%
\column{3}{@{}>{\hspre}l<{\hspost}@{}}%
\column{10}{@{}>{\hspre}l<{\hspost}@{}}%
\column{E}{@{}>{\hspre}l<{\hspost}@{}}%
\>[3]{}\Varid{reset}{}\<[10]%
\>[10]{}\mathbin{::}(\Conid{Functor}\;\sigma,\Conid{Functor}\;\varphi,\Conid{Monoid}\;\Varid{w}){}\<[E]%
\\
\>[10]{}\Rightarrow Free_{\textsf{H}}\;(K^{\textsf{Alg}}\;(\Conid{Tell}\;\Varid{w}\mathrel{{+}}\sigma)\mathrel{\oplus}K^{\textsf{Write}}\;(\Conid{Listen}\;\Varid{w}\mathrel{{+}}(\Conid{Pass}\;\Varid{w}\mathrel{{+}}\varphi)))\;(){}\<[E]%
\\
\>[3]{}\Varid{reset}{}\<[10]%
\>[10]{}\mathrel{=}\textsf{pass}\;(\textsf{return}\;((),\Varid{const}\;\epsilon)){}\<[E]%
\ColumnHook
\end{hscode}\resethooks
For example, consider the following program which first logs \ensuremath{\text{\ttfamily \char34 pre\char34}}, then
resets the log and then logs \ensuremath{\text{\ttfamily \char34 post\char34}}.
\begin{hscode}\SaveRestoreHook
\column{B}{@{}>{\hspre}l<{\hspost}@{}}%
\column{3}{@{}>{\hspre}l<{\hspost}@{}}%
\column{E}{@{}>{\hspre}l<{\hspost}@{}}%
\>[3]{}\texttt{>>>}\;h_{\textsf{Write}}\;(\textsf{tell}\;\text{\ttfamily \char34 post\char34}\sequ \Varid{reset}\sequ \textsf{tell}\;\text{\ttfamily \char34 pre\char34}){}\<[E]%
\\
\>[3]{}((),\text{\ttfamily \char34 post\char34}){}\<[E]%
\ColumnHook
\end{hscode}\resethooks
Alternatively, this \ensuremath{\Varid{reset}} function can be written in terms of \ensuremath{\textsf{censor}}.
a derived method in the MonadWriter library, which
takes a function \ensuremath{\Varid{w}\to \Varid{w}} and a computation \ensuremath{\Varid{m}\;\Varid{a}} and modifies the writer after
the computation has taken place, leaving the return value untouched.
\ensuremath{\textsf{censor}} can either be defined as a scoped effect,
or as a special case of \ensuremath{\textsf{pass}} (\ref{app:censor}).

%-------------------------------------------------------------------------------
\subsection{Latent Effects \& Handlers}
\label{sec:latent}

%- - - - - - - - - - - - - - - - - - - - - - - - - - - - - - - - - - - - - - - -
\paragraph{Definition}
Latent effects \& handlers \cite{vandenBerg21} represent those effects that have
a control-flow in which some computations are deferred for evaluation
at a later point in the program interpretation.
Examples of latent effects are function abstractions with effectful bodies,
call-by-name or call-by-need evaluation strategies and staging.
Latent effects are represented by \ensuremath{Free_{\textsf{Lat}}\;\zeta\;\ell\;\Varid{a}} where
\ensuremath{\zeta} is the regular effect functor, and \ensuremath{\ell} is the \emph{latent} effect functor,
representing the effects that are deferred\footnote{This datatype is equivalent
to the \ensuremath{\Conid{Tree}} datatype of van den Berg et al. \cite{vandenBerg21}.}.
\begin{hscode}\SaveRestoreHook
\column{B}{@{}>{\hspre}l<{\hspost}@{}}%
\column{3}{@{}>{\hspre}l<{\hspost}@{}}%
\column{5}{@{}>{\hspre}l<{\hspost}@{}}%
\column{11}{@{}>{\hspre}l<{\hspost}@{}}%
\column{44}{@{}>{\hspre}l<{\hspost}@{}}%
\column{E}{@{}>{\hspre}l<{\hspost}@{}}%
\>[3]{}\mathbf{data}\;Free_{\textsf{Lat}}\;\zeta\;\ell\;\Varid{a}\;\mathbf{where}{}\<[E]%
\\
\>[3]{}\hsindent{2}{}\<[5]%
\>[5]{}\Conid{Leaf}{}\<[11]%
\>[11]{}\mathbin{::}\Varid{a}\to Free_{\textsf{Lat}}\;\zeta\;\ell\;\Varid{a}{}\<[E]%
\\
\>[3]{}\hsindent{2}{}\<[5]%
\>[5]{}\Conid{Node}{}\<[11]%
\>[11]{}\mathbin{::}\zeta\;\Varid{p}\;\Varid{c}\to \ell\;()\to (\forall\kern-2pt\;\Varid{x}\, .\,\Varid{c}\;\Varid{x}\to \ell\;()\to Free_{\textsf{Lat}}\;\zeta\;\ell\;(\ell\;\Varid{x})){}\<[E]%
\\
\>[11]{}\to (\ell\;\Varid{p}\to Free_{\textsf{Lat}}\;\zeta\;\ell\;\Varid{a}){}\<[44]%
\>[44]{}\to Free_{\textsf{Lat}}\;\zeta\;\ell\;\Varid{a}{}\<[E]%
\ColumnHook
\end{hscode}\resethooks
Here, \ensuremath{\Conid{Leaf}} is a pure computation, and \ensuremath{\Conid{Node}} is an internal node that
contains
(1) an operation \ensuremath{\zeta\;\Varid{p}\;\Varid{c}} with result type \ensuremath{\Varid{p}} and a number of subcomputations \ensuremath{\Varid{c}};
(2) the effect state \ensuremath{\ell\;()} in the node;
(3) a function \ensuremath{(\forall\kern-2pt\;\Varid{x}\, .\,\Varid{c}\;\Varid{x}\to \ell\;()\to Free_{\textsf{Lat}}\;\zeta\;\ell\;(\ell\;\Varid{x}))} to interpret a subcomputation \ensuremath{\Varid{c}}
with result type \ensuremath{\Varid{x}}, given the current effect state; and
(4) a continuation \ensuremath{(\ell\;\Varid{p}\to Free_{\textsf{Lat}}\;\zeta\;\ell\;\Varid{a})} to interpret the remainder of the program.

Van den Berg et al. \cite{vandenBerg21} do not show that \ensuremath{Free_{\textsf{Lat}}\;\zeta\;\ell\;\Varid{a}} is a free monad.

% %- - - - - - - - - - - - - - - - - - - - - - - - - - - - - - - - - - - - - - - -
% \paragraph{Interpretation}
% To interpret latent effects, van den Berg et al. \cite{vandenBerg21} use a specialized handler for each
% latent effect.
% A generic handler function as a structural recursion scheme is---as far as we
% know---currently missing from the literature.
% An ad hoc solution would be to use an algebra that has the same structure as
% our node.
%
% > data AlgLat zet l f a = AlgLat {
% >   hNode  :: forall p c . zet p c -> l () -> (forall x . c x -> l () -> f (l x)) -> (l p -> a) -> a }
% >
% > foldLat :: (a -> b) -> AlgLat zet l (FreeLat zet l) b -> FreeLat zet l a -> b
% > foldLat gen alg (Leaf x)          = gen x
% > foldLat gen alg (Node op l st k)  = (hNode alg) op l
% >              (\c -> foldLat return (AlgLat Node) . st c) (foldLat gen alg . k)
%
% We can also use our generic framework to define a handler using |fold|.

%- - - - - - - - - - - - - - - - - - - - - - - - - - - - - - - - - - - - - - - -
\paragraph{Generic Framework}
Latent effects are an instance of our generic free monad.

\par\noindent\textbf{Step 1}
We choose a mapping\footnote{Using ends and coends from the Algebra of Types,
corresponding to Haskell's universal and existential quantification, respectively.}
\ensuremath{K^{\textsf{Lat}}_{\zeta, \mathcal{L}}} to represent latent effects \& handlers with
effect functor \ensuremath{\zeta} and \emph{latent} effect functor \ensuremath{\mathcal{L}}.
It contains a \ensuremath{\textsf{Node}} constructor to represent internal nodes, similar to the \ensuremath{\Conid{Node}}
constructor of \ensuremath{Free_{\textsf{Lat}}}.
\begin{hscode}\SaveRestoreHook
\column{B}{@{}>{\hspre}l<{\hspost}@{}}%
\column{3}{@{}>{\hspre}l<{\hspost}@{}}%
\column{5}{@{}>{\hspre}l<{\hspost}@{}}%
\column{12}{@{}>{\hspre}l<{\hspost}@{}}%
\column{E}{@{}>{\hspre}l<{\hspost}@{}}%
\>[3]{}K^{\textsf{Lat}}_{\zeta, \mathcal{L}}\;\Conid{F}\;\Conid{A}\mathrel{=}\int^{P, C}\kern-3pt\;\zeta\;\Conid{P}\;\Conid{C}\times\mathcal{L}\;\mathrm{1}\times(\int_{X}\kern-3pt\;\Conid{C}\;\Conid{X}\times\mathcal{L}\;\mathrm{1}\Rightarrow \Conid{F}\;(\mathcal{L}\;\Conid{X}))\times(\mathcal{L}\;\Conid{P}\Rightarrow \Conid{A}){}\<[E]%
\\[\blanklineskip]%
\>[3]{}\mathbf{data}\;K^{\textsf{Lat}}\;\zeta\;\ell\;\Varid{f}\;\Varid{a}\;\mathbf{where}{}\<[E]%
\\
\>[3]{}\hsindent{2}{}\<[5]%
\>[5]{}\textsf{Node}{}\<[12]%
\>[12]{}\mathbin{::}\zeta\;\Varid{p}\;\Varid{c}\to \ell\;()\to (\forall\kern-2pt\;\Varid{x}\, .\,\Varid{c}\;\Varid{x}\to \ell\;()\to \Varid{f}\;(\ell\;\Varid{x})){}\<[E]%
\\
\>[12]{}\to (\ell\;\Varid{p}\to \Varid{a})\to K^{\textsf{Lat}}\;\zeta\;\ell\;\Varid{f}\;\Varid{a}{}\<[E]%
\ColumnHook
\end{hscode}\resethooks

\step{2}
It has a corresponding higher-order functor instance.
\begin{hscode}\SaveRestoreHook
\column{B}{@{}>{\hspre}l<{\hspost}@{}}%
\column{3}{@{}>{\hspre}l<{\hspost}@{}}%
\column{5}{@{}>{\hspre}l<{\hspost}@{}}%
\column{E}{@{}>{\hspre}l<{\hspost}@{}}%
\>[3]{}\mathbf{instance}\;HFunctor\;(K^{\textsf{Lat}}\;\zeta\;\ell)\;\mathbf{where}{}\<[E]%
\\
\>[3]{}\hsindent{2}{}\<[5]%
\>[5]{}\textsf{hmap}\;\Varid{k}\;(\textsf{Node}\;\Varid{sub}\;\ell\;\Varid{st}\;\Varid{c})\mathrel{=}\textsf{Node}\;\Varid{sub}\;\ell\;(\textsf{fmap}\;\Varid{k}\, .\,\Varid{st})\;\Varid{c}{}\<[E]%
\ColumnHook
\end{hscode}\resethooks
\step{3}
From this, the following isomorphism holds (\ref{app:iso}):
\begin{hscode}\SaveRestoreHook
\column{B}{@{}>{\hspre}l<{\hspost}@{}}%
\column{3}{@{}>{\hspre}l<{\hspost}@{}}%
\column{E}{@{}>{\hspre}l<{\hspost}@{}}%
\>[3]{}Free_{\textsf{Lat}}\;\zeta\;\ell\;\Varid{a}\;\cong\;Free_{\textsf{H}}\;(K^{\textsf{Lat}}\;\zeta\;\ell)\;\Varid{a}{}\<[E]%
\ColumnHook
\end{hscode}\resethooks
This implies that also \ensuremath{Free_{\textsf{Lat}}} is a free monad.

\step{4}
We can now write a generic handler for latent effects.
\begin{hscode}\SaveRestoreHook
\column{B}{@{}>{\hspre}l<{\hspost}@{}}%
\column{3}{@{}>{\hspre}l<{\hspost}@{}}%
\column{9}{@{}>{\hspre}l<{\hspost}@{}}%
\column{E}{@{}>{\hspre}l<{\hspost}@{}}%
\>[3]{}h_{\textsf{Lat}}{}\<[9]%
\>[9]{}\mathbin{::}\Conid{Pointed}\;\Varid{g}{}\<[E]%
\\
\>[9]{}\Rightarrow (\Varid{a}\to \Varid{g}\;\Varid{b})\to (\forall\kern-2pt\;\Varid{x}\, .\,K^{\textsf{Lat}}\;\zeta\;\ell\;\Varid{g}\;(\Varid{g}\;\Varid{x})\to \Varid{g}\;\Varid{x})\to Free_{\textsf{H}}\;(K^{\textsf{Lat}}\;\zeta\;\ell)\;\Varid{a}\to \Varid{g}\;\Varid{b}{}\<[E]%
\\
\>[3]{}h_{\textsf{Lat}}{}\<[9]%
\>[9]{}\mathrel{=}\textsf{fold}{}\<[E]%
\ColumnHook
\end{hscode}\resethooks
Writing an isomorphism for this handler is not meaningful as no specialized handler
for \ensuremath{Free_{\textsf{Lat}}} exists in the literature.

%- - - - - - - - - - - - - - - - - - - - - - - - - - - - - - - - - - - - - - - -
\paragraph{Example: Lazy Evaluation with Memoization}

% We require latent effects, for example, when we want to implement different evaluation
% strategies, and for lazy evaluation in particular.
We implement call-by-need with memoization and call-by-value.
This example is an adapted version of the lazy evaluation example
of van den Berg et al.\cite{vandenBerg21}.
Consider the following program:
\begin{hscode}\SaveRestoreHook
\column{B}{@{}>{\hspre}l<{\hspost}@{}}%
\column{3}{@{}>{\hspre}l<{\hspost}@{}}%
\column{E}{@{}>{\hspre}l<{\hspost}@{}}%
\>[3]{}prog_{\textsf{Lazy}}\mathrel{=}\textsf{app}\;(\textsf{abs}\;(\textsf{return}\;\mathrm{3}))\;(\mathbf{do}\;\textsf{put}\;\mathrm{42};\textsf{return}\;\mathrm{5}){}\<[E]%
\ColumnHook
\end{hscode}\resethooks
It applies a constant function to an argument that
changes the state of the program.
With a call-by-need evaluation strategy, we expect the state to remain unchanged,
as the argument is never needed and thus never executed. With an eager strategy,
we expect the state of the program to be 42.

We require algebraic, scoped and latent effects to model this example.
First, the algebraic state effect (\Cref{sec:algebraic}) can \ensuremath{\textsf{get}} the state and \ensuremath{\textsf{put}} a new state.
Second, we require an environment to know which variables are in scope. We model this
functionality with the scoped reader effect, which has two operations: \ensuremath{\textsf{ask}},
which is algebraic, for getting the current environment, and \ensuremath{\textsf{local}}, which is scoped,
for modifying the environment before executing a computation in the modified environment.
For more details on \ensuremath{\textsf{ask}} and \ensuremath{\textsf{local}}, we refer to Haskell's
MonadReader\footnote{\url{https://hackage.haskell.org/package/mtl-2.3/docs/Control-Monad-Reader.html}}
library.
Finally, the latent effect for lazy evaluation with memoization has two operations:
\ensuremath{\textsf{thunk}} for deferring the evaluation of an operation, and \ensuremath{\textsf{force}} for forcing it to
evaluate.
\begin{hscode}\SaveRestoreHook
\column{B}{@{}>{\hspre}l<{\hspost}@{}}%
\column{3}{@{}>{\hspre}l<{\hspost}@{}}%
\column{5}{@{}>{\hspre}l<{\hspost}@{}}%
\column{12}{@{}>{\hspre}l<{\hspost}@{}}%
\column{18}{@{}>{\hspre}l<{\hspost}@{}}%
\column{23}{@{}>{\hspre}l<{\hspost}@{}}%
\column{39}{@{}>{\hspre}l<{\hspost}@{}}%
\column{E}{@{}>{\hspre}l<{\hspost}@{}}%
\>[3]{}\mathbf{data}\;\Conid{Thunking}\;\Varid{v}\mathbin{::}\mathbin{*}\to (\mathbin{*}\to \mathbin{*})\to \mathbin{*}\mathbf{where}{}\<[E]%
\\
\>[3]{}\hsindent{2}{}\<[5]%
\>[5]{}\Conid{Thunk}{}\<[12]%
\>[12]{}\mathbin{::}{}\<[23]%
\>[23]{}\Conid{Thunking}\;\Varid{v}\;\Conid{Ptr}\;{}\<[39]%
\>[39]{}(\Conid{OneSub}\;\Varid{v}){}\<[E]%
\\
\>[3]{}\hsindent{2}{}\<[5]%
\>[5]{}\Conid{Force}{}\<[12]%
\>[12]{}\mathbin{::}\Conid{Ptr}\to {}\<[23]%
\>[23]{}\Conid{Thunking}\;\Varid{v}\;\Varid{v}\;{}\<[39]%
\>[39]{}\Conid{NoSub}{}\<[E]%
\\[\blanklineskip]%
\>[3]{}\mathbf{data}\;\Conid{NoSub}\mathbin{::}\mathbin{*}\to \mathbin{*}\mathbf{where}{}\<[E]%
\\[\blanklineskip]%
\>[3]{}\mathbf{data}\;\Conid{OneSub}\;\Varid{v}{}\<[18]%
\>[18]{}\mathbin{::}\mathbin{*}\to \mathbin{*}{}\<[E]%
\\
\>[3]{}\hsindent{2}{}\<[5]%
\>[5]{}\mathbf{where}\;\Conid{One}\mathbin{::}\Conid{OneSub}\;\Varid{v}\;\Varid{v}{}\<[E]%
\ColumnHook
\end{hscode}\resethooks
\noindent
Here, \ensuremath{\Conid{Ptr}} is a pointer (\ensuremath{\Conid{Int}}) to the environment that keeps track of thunks and
memoized values.
\ensuremath{\Conid{NoSub}} and \ensuremath{\Conid{OneSub}} indicate that \ensuremath{\Conid{Thunk}} and \ensuremath{\Conid{Force}} have no or one latent subcomputations, respectively.

For brevity, we omit forwarding of
unknown effects and define an expression datatype that contains the above
algebraic, scoped and latent effects.
\begin{hscode}\SaveRestoreHook
\column{B}{@{}>{\hspre}l<{\hspost}@{}}%
\column{3}{@{}>{\hspre}l<{\hspost}@{}}%
\column{14}{@{}>{\hspre}l<{\hspost}@{}}%
\column{19}{@{}>{\hspre}l<{\hspost}@{}}%
\column{25}{@{}>{\hspre}c<{\hspost}@{}}%
\column{25E}{@{}l@{}}%
\column{31}{@{}>{\hspre}l<{\hspost}@{}}%
\column{37}{@{}>{\hspre}l<{\hspost}@{}}%
\column{50}{@{}>{\hspre}c<{\hspost}@{}}%
\column{50E}{@{}l@{}}%
\column{56}{@{}>{\hspre}l<{\hspost}@{}}%
\column{E}{@{}>{\hspre}l<{\hspost}@{}}%
\>[3]{}\mathbf{type}\;\Conid{Expr}\;{}\<[14]%
\>[14]{}\Varid{v}\;\Varid{a}{}\<[19]%
\>[19]{}\mathrel{=}Free_{\textsf{H}}\;{}\<[25]%
\>[25]{}({}\<[25E]%
\>[31]{}K^{\textsf{Alg}}\;{}\<[37]%
\>[37]{}(State\;\Varid{v}{}\<[50]%
\>[50]{}\mathrel{{+}}{}\<[50E]%
\>[56]{}\Conid{Ask}\;[\mskip1.5mu \Varid{v}\mskip1.5mu]){}\<[E]%
\\
\>[25]{}\mathrel{\oplus}{}\<[25E]%
\>[31]{}K^{\textsf{Sc}}\;{}\<[37]%
\>[37]{}(\Conid{Local}\;[\mskip1.5mu \Varid{v}\mskip1.5mu]){}\<[E]%
\\
\>[25]{}\mathrel{\oplus}{}\<[25E]%
\>[31]{}K^{\textsf{Lat}}\;{}\<[37]%
\>[37]{}(\Conid{Thunking}\;\Varid{v})\;\mathit{Id})\;\Varid{a}{}\<[E]%
\ColumnHook
\end{hscode}\resethooks
The semantic domain of our handler is \ensuremath{State_{L}\;\Varid{s}\;\Varid{l}\;\Varid{a}}, where in this case,
state \ensuremath{\Varid{s}} consists of a tuple \ensuremath{(\Varid{v},[\mskip1.5mu \Conid{Thunk}\;\Varid{v}\mskip1.5mu])} and \ensuremath{\Varid{l}} is the identity functor.
\begin{hscode}\SaveRestoreHook
\column{B}{@{}>{\hspre}l<{\hspost}@{}}%
\column{3}{@{}>{\hspre}l<{\hspost}@{}}%
\column{25}{@{}>{\hspre}l<{\hspost}@{}}%
\column{E}{@{}>{\hspre}l<{\hspost}@{}}%
\>[3]{}\mathbf{newtype}\;State_{L}\;\Varid{s}\;\Varid{l}\;\Varid{a}{}\<[25]%
\>[25]{}\mathrel{=}State_{L}\;\{\mskip1.5mu \Varid{unStateL}\mathbin{::}(\Varid{s},\Varid{l}\;\Varid{a})\mskip1.5mu\}{}\<[E]%
\ColumnHook
\end{hscode}\resethooks
\noindent
A handler for lazy evaluation has three environments:
(1) the \emph{state} of type \ensuremath{\Varid{v}};
(2) the \emph{evaluation environment} of type \ensuremath{[\mskip1.5mu \Varid{v}\mskip1.5mu]} to know which variables are in scope; and
(3) the \emph{environment of thunks and memoized values} of type \ensuremath{[\mskip1.5mu \Conid{Thunk}\;\Varid{v}\mskip1.5mu]}.
  A thunk is an unevaluated operation.\begin{hscode}\SaveRestoreHook
\column{B}{@{}>{\hspre}l<{\hspost}@{}}%
\column{3}{@{}>{\hspre}l<{\hspost}@{}}%
\column{27}{@{}>{\hspre}l<{\hspost}@{}}%
\column{64}{@{}>{\hspre}l<{\hspost}@{}}%
\column{E}{@{}>{\hspre}l<{\hspost}@{}}%
\>[3]{}\mathbf{type}\;\Conid{Thunk}\;\Varid{v}\mathrel{=}\Conid{Either}\;{}\<[27]%
\>[27]{}(\mathit{Id}\;()\to \Varid{v}\to [\mskip1.5mu \Varid{v}\mskip1.5mu]\to [\mskip1.5mu \Conid{Thunk}\;\Varid{v}\mskip1.5mu]{}\<[E]%
\\
\>[27]{}\to State_{L}\;(\Varid{v},[\mskip1.5mu \Conid{Thunk}\;\Varid{v}\mskip1.5mu])\;\mathit{Id}\;\Varid{v})\;{}\<[64]%
\>[64]{}\Varid{v}{}\<[E]%
\ColumnHook
\end{hscode}\resethooks
Environments (1) and (3) are part of the result of the interpreted expression.
With this in place we can define \ensuremath{h_{\textsf{Lazy}}}, which lazily interprets a stateful
program\footnote{We use a compact datatype \ensuremath{\mathbf{data}\;\mathit{Id}\;\Varid{a}\mathrel{=}\mathit{Id}\;\{\mskip1.5mu \Varid{unId}\mathbin{::}\Varid{a}\mskip1.5mu\}} for
the identity functor.}.
\begin{hscode}\SaveRestoreHook
\column{B}{@{}>{\hspre}l<{\hspost}@{}}%
\column{3}{@{}>{\hspre}l<{\hspost}@{}}%
\column{7}{@{}>{\hspre}l<{\hspost}@{}}%
\column{10}{@{}>{\hspre}l<{\hspost}@{}}%
\column{11}{@{}>{\hspre}l<{\hspost}@{}}%
\column{14}{@{}>{\hspre}l<{\hspost}@{}}%
\column{15}{@{}>{\hspre}l<{\hspost}@{}}%
\column{20}{@{}>{\hspre}l<{\hspost}@{}}%
\column{28}{@{}>{\hspre}l<{\hspost}@{}}%
\column{30}{@{}>{\hspre}l<{\hspost}@{}}%
\column{32}{@{}>{\hspre}l<{\hspost}@{}}%
\column{33}{@{}>{\hspre}l<{\hspost}@{}}%
\column{35}{@{}>{\hspre}l<{\hspost}@{}}%
\column{37}{@{}>{\hspre}l<{\hspost}@{}}%
\column{39}{@{}>{\hspre}l<{\hspost}@{}}%
\column{45}{@{}>{\hspre}l<{\hspost}@{}}%
\column{48}{@{}>{\hspre}l<{\hspost}@{}}%
\column{50}{@{}>{\hspre}l<{\hspost}@{}}%
\column{52}{@{}>{\hspre}l<{\hspost}@{}}%
\column{59}{@{}>{\hspre}c<{\hspost}@{}}%
\column{59E}{@{}l@{}}%
\column{E}{@{}>{\hspre}l<{\hspost}@{}}%
\>[3]{}h_{\textsf{Lazy}}{}\<[10]%
\>[10]{}\mathbin{::}\Conid{Expr}\;\Varid{v}\;\Varid{a}\to \Varid{v}\to [\mskip1.5mu \Varid{v}\mskip1.5mu]\to [\mskip1.5mu \Conid{Thunk}\;\Varid{v}\mskip1.5mu]\to State_{L}\;(\Varid{v},[\mskip1.5mu \Conid{Thunk}\;\Varid{v}\mskip1.5mu])\;\mathit{Id}\;\Varid{a}{}\<[E]%
\\
\>[3]{}h_{\textsf{Lazy}}\;\Varid{prog}\;\Varid{s}\;\Varid{nv}\;\Varid{th}\mathrel{=}\textsf{fold}\;\Varid{gen}\;(alg_{\textsf{Alg}}\kern+2pt\tikz[baseline=(char.base)]{ \node[circle,draw,inner sep=0pt,align=center,scale=0.3] (char) {\Huge{\#}};}\kern+2ptalg_{\textsf{Sc}}\kern+2pt\tikz[baseline=(char.base)]{ \node[circle,draw,inner sep=0pt,align=center,scale=0.3] (char) {\Huge{\#}};}\kern+2ptalg_{\textsf{Lat}})\;\Varid{prog}\;\mathbf{where}{}\<[E]%
\\[\blanklineskip]%
\>[3]{}\hsindent{4}{}\<[7]%
\>[7]{}\Varid{gen}\;{}\<[15]%
\>[15]{}\Varid{x}{}\<[28]%
\>[28]{}\mathrel{=}State_{L}\;((\Varid{s},\Varid{th}),\mathit{Id}\;\Varid{x}){}\<[E]%
\\[\blanklineskip]%
\>[3]{}\hsindent{4}{}\<[7]%
\>[7]{}alg_{\textsf{Alg}}\;{}\<[15]%
\>[15]{}(\textsf{Op}\;\Varid{op}){}\<[28]%
\>[28]{}\mathrel{=}(alg_{\textsf{St}}\mathrel{{\#}}alg_{\textsf{Ask}})\;{}\<[48]%
\>[48]{}\Varid{op}\;{}\<[52]%
\>[52]{}\mathbf{where}{}\<[59]%
\>[59]{}\mathbin{...}{}\<[59E]%
\\
\>[3]{}\hsindent{4}{}\<[7]%
\>[7]{}alg_{\textsf{Sc}}\;{}\<[15]%
\>[15]{}(\textsf{Enter}\;\Varid{sc}){}\<[28]%
\>[28]{}\mathrel{=}alg_{\textsf{Local}}\;{}\<[48]%
\>[48]{}\Varid{sc}\;{}\<[52]%
\>[52]{}\mathbf{where}{}\<[59]%
\>[59]{}\mathbin{...}{}\<[59E]%
\\[\blanklineskip]%
\>[3]{}\hsindent{4}{}\<[7]%
\>[7]{}alg_{\textsf{Lat}}\;(\textsf{Node}\;\Conid{Thunk}\;{}\<[30]%
\>[30]{}\Varid{l}\;{}\<[33]%
\>[33]{}\Varid{st}\;{}\<[37]%
\>[37]{}\Varid{k}){}\<[45]%
\>[45]{}\mathrel{=}\Varid{k}\;{}\<[50]%
\>[50]{}(\textsf{length}\;\Varid{th}\mathrel{{<\kern-1pt}{\$}}\Varid{l})\;\Varid{s}\;\Varid{nv}\;{}\<[E]%
\\
\>[50]{}(\Varid{th}\plus [\mskip1.5mu \Conid{Left}\;(\Varid{st}\;\Conid{One})\mskip1.5mu]){}\<[E]%
\\
\>[3]{}\hsindent{4}{}\<[7]%
\>[7]{}alg_{\textsf{Lat}}\;(\textsf{Node}\;(\Conid{Force}\;\Varid{p})\;{}\<[32]%
\>[32]{}\Varid{l}\;{}\<[35]%
\>[35]{}\Varid{st}\;{}\<[39]%
\>[39]{}\Varid{k}){}\<[45]%
\>[45]{}\mathrel{=}\mathbf{case}\;(\Varid{th}\mathbin{!!}\Varid{p})\;\mathbf{of}{}\<[E]%
\\
\>[7]{}\hsindent{4}{}\<[11]%
\>[11]{}\Conid{Left}\;\Varid{t}{}\<[20]%
\>[20]{}\to \mathbf{let}\;State_{L}\;((\Varid{s'},\Varid{th'}),\mathit{Id}\;\Varid{lv})\mathrel{=}(\Varid{t}\;\Varid{l})\;\Varid{s}\;\Varid{nv}\;\Varid{th}{}\<[E]%
\\
\>[11]{}\hsindent{3}{}\<[14]%
\>[14]{}\mathbf{in}\;\Varid{unId}\mathbin{\$}\textsf{fmap}\;(\lambda \Varid{v}\to (\Varid{k}\;\Varid{lv})\;\Varid{s'}\;\Varid{nv}\;(\Varid{replace}\;\Varid{p}\;(\Conid{Right}\;\Varid{v})\;\Varid{th'}))\;\Varid{lv}{}\<[E]%
\\
\>[7]{}\hsindent{4}{}\<[11]%
\>[11]{}\Conid{Right}\;\Varid{v}{}\<[20]%
\>[20]{}\to (\Varid{k}\;(\Varid{v}\mathrel{{<\kern-1pt}{\$}}\Varid{l}))\;\Varid{s}\;\Varid{nv}\;\Varid{th}{}\<[E]%
\ColumnHook
\end{hscode}\resethooks
The algebras for algebraic and scoped effects hold little surprises.
When thunking an operation, we call the continuation and add the current operation to the
list of thunks.
When forcing an operation, we get the thunk or value from the list of thunks.
In case of a thunk we evaluate it, then replace it with its value in the list of thunks
in order to memoize it, and call the continuation.
In case of a value, we immediately call the continuation
with this value.

An eager handler has the same type signature and implementation for the algebraic and scoped effects,
but swaps the behaviour of \ensuremath{\Conid{Thunk}} and \ensuremath{\Conid{Force}}:
\begin{hscode}\SaveRestoreHook
\column{B}{@{}>{\hspre}l<{\hspost}@{}}%
\column{3}{@{}>{\hspre}l<{\hspost}@{}}%
\column{7}{@{}>{\hspre}l<{\hspost}@{}}%
\column{11}{@{}>{\hspre}l<{\hspost}@{}}%
\column{13}{@{}>{\hspre}l<{\hspost}@{}}%
\column{33}{@{}>{\hspre}l<{\hspost}@{}}%
\column{36}{@{}>{\hspre}l<{\hspost}@{}}%
\column{38}{@{}>{\hspre}l<{\hspost}@{}}%
\column{45}{@{}>{\hspre}c<{\hspost}@{}}%
\column{45E}{@{}l@{}}%
\column{47}{@{}>{\hspre}l<{\hspost}@{}}%
\column{E}{@{}>{\hspre}l<{\hspost}@{}}%
\>[3]{}h_{\textsf{Eager}}\;\Varid{prog}\;\Varid{s}\;\Varid{nv}\;\Varid{th}\mathrel{=}\mathbin{...}{}\<[33]%
\>[33]{}\mathbf{where}{}\<[E]%
\\
\>[3]{}\hsindent{4}{}\<[7]%
\>[7]{}alg_{\textsf{Lat}}\;(\textsf{Node}\;\colorbox{lightgray}{$\Conid{Thunk}$}\;{}\<[36]%
\>[36]{}\Varid{l}\;\Varid{st}\;\Varid{k}){}\<[45]%
\>[45]{}\mathrel{=}{}\<[45E]%
\\
\>[7]{}\hsindent{4}{}\<[11]%
\>[11]{}\mathbf{let}\;State_{L}\;((\Varid{s'},\Varid{th'}),\Varid{lv})\mathrel{=}(\Varid{st}\;\Conid{One}\;\Varid{l})\;\Varid{s}\;\Varid{nv}\;\Varid{th}{}\<[E]%
\\
\>[7]{}\hsindent{4}{}\<[11]%
\>[11]{}\mathbf{in}\;\Varid{unId}\mathbin{\$}{}\<[E]%
\\
\>[11]{}\hsindent{2}{}\<[13]%
\>[13]{}\textsf{fmap}\;(\lambda \Varid{v}\to \Varid{k}\;(\textsf{length}\;\Varid{th'}\mathrel{{<\kern-1pt}{\$}}\Varid{l})\;\Varid{s'}\;\Varid{nv}\;(\Varid{th'}\plus [\mskip1.5mu \Conid{Right}\;\Varid{v}\mskip1.5mu]))\;(\Varid{unId}\;\Varid{lv}){}\<[E]%
\\
\>[3]{}\hsindent{4}{}\<[7]%
\>[7]{}alg_{\textsf{Lat}}\;(\textsf{Node}\;\colorbox{lightgray}{$(\Conid{Force}\;\Varid{p})$}\;{}\<[38]%
\>[38]{}\Varid{l}\;\Varid{st}\;\Varid{k}){}\<[47]%
\>[47]{}\mathrel{=}\mathbf{case}\;(\Varid{th}\mathbin{!!}\Varid{p})\;\mathbf{of}{}\<[E]%
\\
\>[7]{}\hsindent{4}{}\<[11]%
\>[11]{}\Conid{Right}\;\Varid{v}\to \Varid{k}\;(\Varid{v}\mathrel{{<\kern-1pt}{\$}}\Varid{l})\;\Varid{s}\;\Varid{nv}\;\Varid{th}{}\<[E]%
\ColumnHook
\end{hscode}\resethooks
We define constructors \ensuremath{\textsf{get}}, \ensuremath{\textsf{put}}, \ensuremath{\textsf{ask}}, \ensuremath{\textsf{local}}, \ensuremath{\textsf{thunk}} and \ensuremath{\textsf{force}}
and go back to our example program to define the standard \ensuremath{\lambda}-calculus functionality
in terms of these constructors.
Values can be either integers or function abstractions:
\begin{hscode}\SaveRestoreHook
\column{B}{@{}>{\hspre}l<{\hspost}@{}}%
\column{3}{@{}>{\hspre}l<{\hspost}@{}}%
\column{E}{@{}>{\hspre}l<{\hspost}@{}}%
\>[3]{}\mathbf{data}\;V\mathrel{=}\Conid{Val}\;\Conid{Int}\mid \Conid{Abs}\;(\Conid{Expr}\;V\;V){}\<[E]%
\ColumnHook
\end{hscode}\resethooks
We define variables, function abstraction and function application as follows.
We thunk the argument of function application and evaluate it
\begin{hscode}\SaveRestoreHook
\column{B}{@{}>{\hspre}l<{\hspost}@{}}%
\column{3}{@{}>{\hspre}l<{\hspost}@{}}%
\column{8}{@{}>{\hspre}l<{\hspost}@{}}%
\column{19}{@{}>{\hspre}l<{\hspost}@{}}%
\column{E}{@{}>{\hspre}l<{\hspost}@{}}%
\>[3]{}\textsf{var}{}\<[8]%
\>[8]{}\mathbin{::}\Conid{Ptr}\to \Conid{Expr}\;V\;V{}\<[E]%
\\
\>[3]{}\textsf{var}\;{}\<[8]%
\>[8]{}\Varid{x}\mathrel{=}\mathbf{do}\;\Varid{nv}\leftarrow \textsf{ask};\textsf{local}\;([\mskip1.5mu \Varid{nv}\mathbin{!!}\Varid{x}\mskip1.5mu])\;(\textsf{force}\;\mathrm{0}){}\<[E]%
\\[\blanklineskip]%
\>[3]{}\textsf{abs}{}\<[8]%
\>[8]{}\mathbin{::}\Conid{Expr}\;V\;V\to \Conid{Expr}\;V\;V{}\<[E]%
\\
\>[3]{}\textsf{abs}\;\Varid{body}\mathrel{=}\textsf{return}\;(\Conid{Abs}\;\Varid{body}){}\<[E]%
\\[\blanklineskip]%
\>[3]{}\textsf{app}{}\<[8]%
\>[8]{}\mathbin{::}\Conid{Expr}\;V\;V\to \Conid{Expr}\;V\;V\to \Conid{Expr}\;V\;V{}\<[E]%
\\
\>[3]{}\textsf{app}\;\Varid{e}_{1}\;\Varid{e}_{2}\mathrel{=}\mathbf{do}\;{}\<[19]%
\>[19]{}\Varid{vf}\leftarrow \Varid{e}_{1};\Varid{nv}\leftarrow \textsf{ask};\Varid{p}\leftarrow \textsf{thunk}\;\Varid{e}_{2}{}\<[E]%
\\
\>[19]{}\mathbf{case}\;\Varid{vf}\;\mathbf{of}\;\Conid{Abs}\;\Varid{body}\to \textsf{local}\;([\mskip1.5mu \Varid{nv}\mathbin{!!}\Varid{p}\mskip1.5mu])\;\Varid{body}{}\<[E]%
\ColumnHook
\end{hscode}\resethooks
We now lazily and eagerly evaluate the above program \ensuremath{\Varid{prog}}.
We start with state \ensuremath{\mathrm{0}}, an empty environment and an empty list of thunks.

\noindent
\begin{minipage}{0.5\textwidth}\begin{hscode}\SaveRestoreHook
\column{B}{@{}>{\hspre}l<{\hspost}@{}}%
\column{3}{@{}>{\hspre}l<{\hspost}@{}}%
\column{9}{@{}>{\hspre}l<{\hspost}@{}}%
\column{17}{@{}>{\hspre}l<{\hspost}@{}}%
\column{26}{@{}>{\hspre}c<{\hspost}@{}}%
\column{26E}{@{}l@{}}%
\column{27}{@{}>{\hspre}l<{\hspost}@{}}%
\column{36}{@{}>{\hspre}l<{\hspost}@{}}%
\column{40}{@{}>{\hspre}l<{\hspost}@{}}%
\column{E}{@{}>{\hspre}l<{\hspost}@{}}%
\>[3]{}\texttt{>>>}\;{}\<[9]%
\>[9]{}h_{\textsf{Lazy}}\;{}\<[17]%
\>[17]{}prog_{\textsf{Lazy}}\;{}\<[27]%
\>[27]{}(\Conid{Val}\;\mathrm{0})\;{}\<[36]%
\>[36]{}[\mskip1.5mu \mskip1.5mu]\;{}\<[40]%
\>[40]{}[\mskip1.5mu \mskip1.5mu]{}\<[E]%
\\
\>[3]{}(\mathrm{0},{}\<[9]%
\>[9]{}[\mskip1.5mu \Conid{Left}\;\text{\ttfamily \char34 thunk\char34}\mskip1.5mu],{}\<[26]%
\>[26]{}\mathrm{3}){}\<[26E]%
\ColumnHook
\end{hscode}\resethooks
\end{minipage}%
\begin{minipage}{0.5\textwidth}\begin{hscode}\SaveRestoreHook
\column{B}{@{}>{\hspre}l<{\hspost}@{}}%
\column{3}{@{}>{\hspre}l<{\hspost}@{}}%
\column{9}{@{}>{\hspre}l<{\hspost}@{}}%
\column{17}{@{}>{\hspre}l<{\hspost}@{}}%
\column{26}{@{}>{\hspre}c<{\hspost}@{}}%
\column{26E}{@{}l@{}}%
\column{27}{@{}>{\hspre}l<{\hspost}@{}}%
\column{36}{@{}>{\hspre}l<{\hspost}@{}}%
\column{40}{@{}>{\hspre}l<{\hspost}@{}}%
\column{E}{@{}>{\hspre}l<{\hspost}@{}}%
\>[3]{}\texttt{>>>}\;{}\<[9]%
\>[9]{}h_{\textsf{Eager}}\;{}\<[17]%
\>[17]{}prog_{\textsf{Lazy}}\;{}\<[27]%
\>[27]{}(\Conid{Val}\;\mathrm{0})\;{}\<[36]%
\>[36]{}[\mskip1.5mu \mskip1.5mu]\;{}\<[40]%
\>[40]{}[\mskip1.5mu \mskip1.5mu]{}\<[E]%
\\
\>[3]{}(\mathrm{42},{}\<[9]%
\>[9]{}[\mskip1.5mu \Conid{Right}\;\mathrm{5}\mskip1.5mu],{}\<[26]%
\>[26]{}\mathrm{3}){}\<[26E]%
\ColumnHook
\end{hscode}\resethooks
\end{minipage}

\noindent
The first element of the triple is the current state, the second element
is the list of thunks or memoized values, and the third element is the program result.

%-------------------------------------------------------------------------------
\subsection{Bracketing Effect \& Handler}
\label{sec:resource}

Consider the situation where a (built-in) exception is raised while a resource (such as a
file) was being used. It is not always straightforward what has to happen in this
scenario, but in each case we want to avoid that our resource is unreachable by code.
The function \ensuremath{\textsf{bracket}} \cite{bracket} acquires a resource, does something with it and then
safely releases the resource again.
Also if an exception occurs, \ensuremath{\textsf{bracket}} correctly releases the resource and re-raises the exception.
This is reflected in its type: \ensuremath{\Varid{m}\;\Varid{r}} acquires the resource \ensuremath{\Varid{r}},
the function \ensuremath{\Varid{r}\to \Varid{m}\;\Varid{b}} releases the resource,
the function \ensuremath{\Varid{r}\to \Varid{m}\;\Varid{a}} uses the resource, and \ensuremath{\Varid{m}\;\Varid{a}} is the result.
\begin{hscode}\SaveRestoreHook
\column{B}{@{}>{\hspre}l<{\hspost}@{}}%
\column{3}{@{}>{\hspre}l<{\hspost}@{}}%
\column{9}{@{}>{\hspre}l<{\hspost}@{}}%
\column{E}{@{}>{\hspre}l<{\hspost}@{}}%
\>[9]{}\textsf{bracket}\mathbin{::}\Varid{m}\;\Varid{r}\to (\Varid{r}\to \Varid{m}\;\Varid{b})\to (\Varid{r}\to \Varid{m}\;\Varid{a})\to \Varid{m}\;\Varid{a}{}\<[E]%
\\
\>[3]{}\cong\;{}\<[9]%
\>[9]{}\textsf{bracket}\mathbin{::}\Varid{m}\;\Varid{r}\to (\Varid{r}\to (\Varid{m}\;\Varid{b},\Varid{m}\;\Varid{a}))\to \Varid{m}\;\Varid{a}{}\<[E]%
\\
\>[3]{}\cong\;{}\<[9]%
\>[9]{}\textsf{bracket}\mathbin{::}\Varid{m}\;(\Varid{m}\;\Varid{b},\Varid{m}\;\Varid{a})\to \Varid{m}\;\Varid{a}{}\<[E]%
\ColumnHook
\end{hscode}\resethooks
Typically, \ensuremath{\Varid{b}} is the unit type \ensuremath{()}\footnote{In what follows, we replace this type \ensuremath{\Varid{b}} by \ensuremath{()}.}
and the monad \ensuremath{\Varid{m}} is the \ensuremath{\textsf{IO}}-monad.
For example, consider the program \ensuremath{\textsf{firstTwo}} that opens a file \ensuremath{\text{\ttfamily \char34 foo.txt\char34}} and
reads and prints its first two charachters. In case an exception occurs, the file
is released (indicated by printing \ensuremath{\text{\ttfamily \char34 released\char34}}).
\begin{hscode}\SaveRestoreHook
\column{B}{@{}>{\hspre}l<{\hspost}@{}}%
\column{3}{@{}>{\hspre}l<{\hspost}@{}}%
\column{24}{@{}>{\hspre}l<{\hspost}@{}}%
\column{E}{@{}>{\hspre}l<{\hspost}@{}}%
\>[3]{}\textsf{firstTwo}\mathrel{=}\textsf{bracket}\;{}\<[24]%
\>[24]{}(\textsf{openFile}\;\text{\ttfamily \char34 foo.txt\char34}\;\Conid{ReadMode})\;{}\<[E]%
\\
\>[24]{}(\lambda \anonymous \to \textsf{print}\;\text{\ttfamily \char34 released\char34})\;{}\<[E]%
\\
\>[24]{}(\lambda \Varid{h}\to \mathbf{do}\;\Varid{x}\leftarrow \textsf{hGetChar}\;\Varid{h};\Varid{y}\leftarrow \textsf{hGetChar}\;\Varid{h};\textsf{print}\;(\Varid{x},\Varid{y})){}\<[E]%
\ColumnHook
\end{hscode}\resethooks
The following prints the contents of \ensuremath{\text{\ttfamily \char34 foo.txt\char34}} and then executes \ensuremath{\textsf{firstTwo}}.
\begin{hscode}\SaveRestoreHook
\column{B}{@{}>{\hspre}l<{\hspost}@{}}%
\column{3}{@{}>{\hspre}l<{\hspost}@{}}%
\column{9}{@{}>{\hspre}l<{\hspost}@{}}%
\column{27}{@{}>{\hspre}l<{\hspost}@{}}%
\column{37}{@{}>{\hspre}l<{\hspost}@{}}%
\column{E}{@{}>{\hspre}l<{\hspost}@{}}%
\>[3]{}\texttt{>>>}\;{}\<[9]%
\>[9]{}\textsf{readFile}\;\text{\ttfamily \char34 foo.txt\char34}\bind \textsf{print}\sequ \textsf{firstTwo}{}\<[E]%
\\
\>[3]{}\text{\ttfamily \char34 HELLO,~WORLD!\char34}\;{}\<[27]%
\>[27]{}\hspace{10pt}\;{}\<[37]%
\>[37]{}\text{\ttfamily \char34 H\char34}{}\<[E]%
\\
\>[3]{}(\text{\ttfamily 'H'},\text{\ttfamily 'E'})\;{}\<[27]%
\>[27]{}\hspace{10pt}\;{}\<[37]%
\>[37]{}\text{\ttfamily \char34 released\char34}{}\<[E]%
\\
\>[3]{}\text{\ttfamily \char34 released\char34}\;{}\<[27]%
\>[27]{}\hspace{10pt}\;{}\<[37]%
\>[37]{}\text{\ttfamily \char34 ***Exception:~foo.txt~hGetChar~end~of~file\char34}{}\<[E]%
\ColumnHook
\end{hscode}\resethooks
There are two possible situations: either \ensuremath{\text{\ttfamily \char34 foo.txt\char34}} contains enough characters (left)
or an exception occurs as no more characters can be read (right). Notice that the
resource is released in both cases, but in case of an error, the exception is re-raised
after release.

%- - - - - - - - - - - - - - - - - - - - - - - - - - - - - - - - - - - - - - - -
\paragraph{Definition}

\ensuremath{\textsf{bracket}} does not fit any of the previous higher-order effects.
For that reason, we define \ensuremath{Free_{\textsf{Res}}\;\Varid{a}} to represent the \emph{bracketing effect \& handler}.
\begin{hscode}\SaveRestoreHook
\column{B}{@{}>{\hspre}l<{\hspost}@{}}%
\column{3}{@{}>{\hspre}l<{\hspost}@{}}%
\column{5}{@{}>{\hspre}l<{\hspost}@{}}%
\column{14}{@{}>{\hspre}l<{\hspost}@{}}%
\column{50}{@{}>{\hspre}l<{\hspost}@{}}%
\column{E}{@{}>{\hspre}l<{\hspost}@{}}%
\>[3]{}\mathbf{data}\;Free_{\textsf{Res}}\;\Varid{a}\;\mathbf{where}{}\<[E]%
\\
\>[3]{}\hsindent{2}{}\<[5]%
\>[5]{}Var{}\<[14]%
\>[14]{}\mathbin{::}\Varid{a}{}\<[50]%
\>[50]{}\to Free_{\textsf{Res}}\;\Varid{a}{}\<[E]%
\\
\>[3]{}\hsindent{2}{}\<[5]%
\>[5]{}\Conid{Bracket}{}\<[14]%
\>[14]{}\mathbin{::}Free_{\textsf{Res}}\;(Free_{\textsf{Res}}\;(),Free_{\textsf{Res}}\;\Varid{a}){}\<[50]%
\>[50]{}\to Free_{\textsf{Res}}\;\Varid{a}{}\<[E]%
\ColumnHook
\end{hscode}\resethooks
Here, \ensuremath{Var} represents a pure computation and \ensuremath{\Conid{Bracket}} has the same structure as
\ensuremath{\textsf{bracket}}, rewritten above.

%- - - - - - - - - - - - - - - - - - - - - - - - - - - - - - - - - - - - - - - -
\paragraph{Generic Framework}

We show that the bracketing effect fits our framework.

\par\noindent\textbf{Step 1}
We choose an appropriate mapping \ensuremath{K^{\textsf{Res}}} to express bracketing.

\noindent
\begin{minipage}{0.4\textwidth}\begin{hscode}\SaveRestoreHook
\column{B}{@{}>{\hspre}l<{\hspost}@{}}%
\column{3}{@{}>{\hspre}l<{\hspost}@{}}%
\column{E}{@{}>{\hspre}l<{\hspost}@{}}%
\>[3]{}K^{\textsf{Res}}\;\Conid{F}\;\Conid{A}\mathrel{=}\Conid{F}\;(\Conid{F}\;(),\Conid{F}\;\Conid{A}){}\<[E]%
\ColumnHook
\end{hscode}\resethooks
\end{minipage}%
\begin{minipage}{0.6\textwidth}
\begin{hscode}\SaveRestoreHook
\column{B}{@{}>{\hspre}l<{\hspost}@{}}%
\column{3}{@{}>{\hspre}l<{\hspost}@{}}%
\column{5}{@{}>{\hspre}l<{\hspost}@{}}%
\column{16}{@{}>{\hspre}l<{\hspost}@{}}%
\column{E}{@{}>{\hspre}l<{\hspost}@{}}%
\>[3]{}\mathbf{data}\;K^{\textsf{Res}}\;\Varid{f}\;\Varid{a}\;\mathbf{where}{}\<[E]%
\\
\>[3]{}\hsindent{2}{}\<[5]%
\>[5]{}\textsf{Bracket}{}\<[16]%
\>[16]{}\mathbin{::}\Varid{f}\;(\Varid{f}\;(),\Varid{f}\;\Varid{a})\to K^{\textsf{Res}}\;\Varid{f}\;\Varid{a}{}\<[E]%
\ColumnHook
\end{hscode}\resethooks
\end{minipage}

\step{2}
\ensuremath{K^{\textsf{Res}}} is a higher-order functor:
\begin{hscode}\SaveRestoreHook
\column{B}{@{}>{\hspre}l<{\hspost}@{}}%
\column{3}{@{}>{\hspre}l<{\hspost}@{}}%
\column{5}{@{}>{\hspre}l<{\hspost}@{}}%
\column{E}{@{}>{\hspre}l<{\hspost}@{}}%
\>[3]{}\mathbf{instance}\;HFunctor\;K^{\textsf{Res}}\;\mathbf{where}{}\<[E]%
\\
\>[3]{}\hsindent{2}{}\<[5]%
\>[5]{}\textsf{hmap}\;\Varid{k}\;(\textsf{Bracket}\;\Varid{res})\mathrel{=}\textsf{Bracket}\;(\Varid{k}\;(\textsf{fmap}\;(\lambda (\Varid{rel},\Varid{use})\to (\Varid{k}\;\Varid{rel},\Varid{k}\;\Varid{use}))\;\Varid{res})){}\<[E]%
\ColumnHook
\end{hscode}\resethooks

\step{3}
Furthermore, the following isomorphism holds (\ref{app:iso}):
\begin{hscode}\SaveRestoreHook
\column{B}{@{}>{\hspre}l<{\hspost}@{}}%
\column{3}{@{}>{\hspre}l<{\hspost}@{}}%
\column{E}{@{}>{\hspre}l<{\hspost}@{}}%
\>[3]{}Free_{\textsf{Res}}\;\Varid{a}\;\cong\;Free_{\textsf{H}}\;K^{\textsf{Res}}\;\Varid{a}{}\<[E]%
\ColumnHook
\end{hscode}\resethooks
\step{4}
A handler for the bracketing effect can be defined in terms of our framework:
\begin{hscode}\SaveRestoreHook
\column{B}{@{}>{\hspre}l<{\hspost}@{}}%
\column{3}{@{}>{\hspre}l<{\hspost}@{}}%
\column{9}{@{}>{\hspre}l<{\hspost}@{}}%
\column{10}{@{}>{\hspre}l<{\hspost}@{}}%
\column{E}{@{}>{\hspre}l<{\hspost}@{}}%
\>[3]{}h_{\textsf{Res}}{}\<[9]%
\>[9]{}\mathbin{::}(\Conid{Pointed}\;\Varid{g}){}\<[E]%
\\
\>[9]{}\hsindent{1}{}\<[10]%
\>[10]{}\Rightarrow (\Varid{a}\to \Varid{g}\;\Varid{b})\to (\forall\kern-2pt\;\Varid{x}\, .\,K^{\textsf{Res}}\;\Varid{g}\;(\Varid{g}\;\Varid{x})\to \Varid{g}\;\Varid{x})\to Free_{\textsf{H}}\;K^{\textsf{Res}}\;\Varid{a}\to \Varid{g}\;\Varid{b}{}\<[E]%
\\
\>[3]{}h_{\textsf{Res}}{}\<[9]%
\>[9]{}\mathrel{=}\textsf{fold}{}\<[E]%
\ColumnHook
\end{hscode}\resethooks
Writing an isomorphism for this handler is not meaningful as no specialized handler
for \ensuremath{Free_{\textsf{Res}}} exists.

%- - - - - - - - - - - - - - - - - - - - - - - - - - - - - - - - - - - - - - - -
\paragraph{Example: Print First Two Characters of File}
We revisit the above example that prints the first two characters of a file, if
possible.
To model this example with our framework, we require algebraic effects to open (\ensuremath{\Conid{OpenFile}})
and read a file (\ensuremath{\Conid{ReadFile}}), to read a character from a file (\ensuremath{\Conid{HGetChar}}),
and to print it to the standard output (\ensuremath{\Conid{Print}}).
Furthermore, we require the bracketing effect to encode the bracketing behaviour for
safely releasing resources.
\begin{hscode}\SaveRestoreHook
\column{B}{@{}>{\hspre}l<{\hspost}@{}}%
\column{3}{@{}>{\hspre}l<{\hspost}@{}}%
\column{20}{@{}>{\hspre}l<{\hspost}@{}}%
\column{23}{@{}>{\hspre}c<{\hspost}@{}}%
\column{23E}{@{}l@{}}%
\column{26}{@{}>{\hspre}l<{\hspost}@{}}%
\column{37}{@{}>{\hspre}l<{\hspost}@{}}%
\column{47}{@{}>{\hspre}l<{\hspost}@{}}%
\column{E}{@{}>{\hspre}l<{\hspost}@{}}%
\>[3]{}\mathbf{data}\;\Conid{Teletype}\;{}\<[20]%
\>[20]{}\Varid{a}{}\<[23]%
\>[23]{}\mathrel{=}{}\<[23E]%
\>[26]{}\Conid{HGetChar}\;{}\<[37]%
\>[37]{}\Conid{Handle}\;{}\<[47]%
\>[47]{}(\Conid{Char}\to \Varid{a}){}\<[E]%
\\
\>[23]{}\mid {}\<[23E]%
\>[26]{}\Conid{Print}\;{}\<[37]%
\>[37]{}\Conid{String}\;{}\<[47]%
\>[47]{}\Varid{a}{}\<[E]%
\\
\>[23]{}\mid {}\<[23E]%
\>[26]{}\Conid{ReadFile}\;{}\<[37]%
\>[37]{}\Conid{FilePath}\;{}\<[47]%
\>[47]{}(\Conid{String}\to \Varid{a}){}\<[E]%
\\
\>[23]{}\mid {}\<[23E]%
\>[26]{}\Conid{OpenFile}\;{}\<[37]%
\>[37]{}\Conid{FilePath}\;{}\<[47]%
\>[47]{}\Conid{IOMode}\;(\Conid{Handle}\to \Varid{a}){}\<[E]%
\ColumnHook
\end{hscode}\resethooks
The handler for bracketing is defined in terms of our generic framework:
\begin{hscode}\SaveRestoreHook
\column{B}{@{}>{\hspre}l<{\hspost}@{}}%
\column{3}{@{}>{\hspre}l<{\hspost}@{}}%
\column{5}{@{}>{\hspre}l<{\hspost}@{}}%
\column{29}{@{}>{\hspre}l<{\hspost}@{}}%
\column{33}{@{}>{\hspre}l<{\hspost}@{}}%
\column{38}{@{}>{\hspre}l<{\hspost}@{}}%
\column{62}{@{}>{\hspre}c<{\hspost}@{}}%
\column{62E}{@{}l@{}}%
\column{67}{@{}>{\hspre}l<{\hspost}@{}}%
\column{E}{@{}>{\hspre}l<{\hspost}@{}}%
\>[3]{}h_{\textsf{Bracket}}\mathbin{::}Free_{\textsf{H}}\;(K^{\textsf{Alg}}\;\Conid{Teletype}\mathrel{\oplus}K^{\textsf{Res}})\;\Varid{a}\to \textsf{IO}\;\Varid{a}{}\<[E]%
\\
\>[3]{}h_{\textsf{Bracket}}\mathrel{=}\textsf{fold}\;\eta\;(alg_{\textsf{Tele}}\kern+2pt\tikz[baseline=(char.base)]{ \node[circle,draw,inner sep=0pt,align=center,scale=0.3] (char) {\Huge{\#}};}\kern+2ptalg_{\textsf{Res}})\;\mathbf{where}{}\<[E]%
\\[\blanklineskip]%
\>[3]{}\hsindent{2}{}\<[5]%
\>[5]{}alg_{\textsf{Tele}}\;(\textsf{Op}\;(\Conid{HGetChar}\;{}\<[29]%
\>[29]{}\Varid{h}\;{}\<[38]%
\>[38]{}\Varid{k}))\mathrel{=}\textsf{hGetChar}\;\Varid{h}{}\<[62]%
\>[62]{}\bind {}\<[62E]%
\>[67]{}\Varid{k}{}\<[E]%
\\
\>[3]{}\hsindent{2}{}\<[5]%
\>[5]{}alg_{\textsf{Tele}}\;(\textsf{Op}\;(\Conid{Print}\;{}\<[29]%
\>[29]{}\Varid{s}\;{}\<[38]%
\>[38]{}\Varid{k}))\mathrel{=}\textsf{print}\;\Varid{s}{}\<[62]%
\>[62]{}\sequ {}\<[62E]%
\>[67]{}\Varid{k}{}\<[E]%
\\
\>[3]{}\hsindent{2}{}\<[5]%
\>[5]{}alg_{\textsf{Tele}}\;(\textsf{Op}\;(\Conid{ReadFile}\;{}\<[29]%
\>[29]{}\Varid{fp}\;{}\<[38]%
\>[38]{}\Varid{k}))\mathrel{=}\textsf{readFile}\;\Varid{fp}{}\<[62]%
\>[62]{}\bind {}\<[62E]%
\>[67]{}\Varid{k}{}\<[E]%
\\
\>[3]{}\hsindent{2}{}\<[5]%
\>[5]{}alg_{\textsf{Tele}}\;(\textsf{Op}\;(\Conid{OpenFile}\;{}\<[29]%
\>[29]{}\Varid{fp}\;\Varid{mode}\;{}\<[38]%
\>[38]{}\Varid{k}))\mathrel{=}\textsf{openFile}\;\Varid{fp}\;\Varid{mode}{}\<[62]%
\>[62]{}\bind {}\<[62E]%
\>[67]{}\Varid{k}{}\<[E]%
\\[\blanklineskip]%
\>[3]{}\hsindent{2}{}\<[5]%
\>[5]{}alg_{\textsf{Res}}\;(\textsf{Bracket}\;\Varid{res})\mathrel{=}\mathbf{do}\;{}\<[33]%
\>[33]{}(\Varid{rel},\Varid{use})\leftarrow \Varid{res};{}\<[E]%
\\
\>[33]{}\textsf{bracket}\;(\textsf{return}\;())\;(\Varid{const}\;\Varid{rel})\;(\Varid{const}\;(\textsf{join}\;\Varid{use})){}\<[E]%
\ColumnHook
\end{hscode}\resethooks
We define constructors \ensuremath{\textsf{hGetC}}, \ensuremath{\textsf{prnt}}, \ensuremath{\textsf{brckt}}, \ensuremath{\textsf{readF}} and \ensuremath{\textsf{openF}} to
revisit the example.
\begin{hscode}\SaveRestoreHook
\column{B}{@{}>{\hspre}l<{\hspost}@{}}%
\column{3}{@{}>{\hspre}l<{\hspost}@{}}%
\column{20}{@{}>{\hspre}l<{\hspost}@{}}%
\column{27}{@{}>{\hspre}l<{\hspost}@{}}%
\column{E}{@{}>{\hspre}l<{\hspost}@{}}%
\>[3]{}\texttt{>>>}\;h_{\textsf{Bracket}}\;(\textsf{readF}\;\text{\ttfamily \char34 foo.txt\char34}\bind \textsf{prnt}\sequ \textsf{firstTwo}){}\<[E]%
\\
\>[3]{}\text{\ttfamily \char34 HELLO,~WORLD!\char34}\;{}\<[20]%
\>[20]{}\hspace{10pt}\;{}\<[27]%
\>[27]{}\text{\ttfamily \char34 H\char34}{}\<[E]%
\\
\>[3]{}\text{\ttfamily \char34 ('H','E')\char34}\;{}\<[20]%
\>[20]{}\hspace{10pt}\;{}\<[27]%
\>[27]{}\text{\ttfamily \char34 released\char34}{}\<[E]%
\\
\>[3]{}\text{\ttfamily \char34 released\char34}\;{}\<[20]%
\>[20]{}\hspace{10pt}\;{}\<[27]%
\>[27]{}\text{\ttfamily \char34 ***Exception:~foo.txt~hGetChar~end~of~file\char34}{}\<[E]%
\ColumnHook
\end{hscode}\resethooks

%==================================================================
\section{Formalizing the Free Monad}
\label{sec:formalization}

In what follows we present the categorical foundations of our generic free monad.
For the reader who is not familiar with categories, adjunctions and free structures,
we refer to \ref{app:ct-background}.

Summarized, our framework is based on the \ensuremath{\overline{\mathit{Free}}\;\circ\;(\mathit{Id}\rtimes\mathbin{-})\dashv \mathit{App}\kern+2pt\;\circ\;\overline{\mathit{U}}} adjunction,
which is a composition of two simpler adjunctions \ensuremath{\overline{\mathit{Free}}\dashv \overline{\mathit{U}}} and \ensuremath{\mathit{Id}\rtimes\mathbin{-}\dashv \mathit{App}\kern+2pt}.
This composed adjunction gives rise to our monad \ensuremath{Free_{\textsf{H}}} in \ensuremath{\mathbb{C}}.

\begin{center}
\begin{tikzcd}
{\ensuremath{H\mathit{-Alg}(\mathbb{P}\rtimes\mathbb{C})}} & {\ensuremath{\bot}} & {\ensuremath{\mathbb{P}\rtimes\mathbb{C}}} & {\ensuremath{\bot}} & {\ensuremath{\mathbb{C}}}
\arrow["{\ensuremath{\overline{\mathit{Free}}}}"', from=1-3, to=1-1, shift right=3]
\arrow["{\ensuremath{\overline{\mathit{U}}}}"', from=1-1, to=1-3, shift right=3]
\arrow["{\ensuremath{\mathit{App}\kern+2pt}}"', from=1-3, to=1-5, shift right=3]
\arrow["{\ensuremath{\mathit{Id}\rtimes\mathbin{-}}}"', from=1-5, to=1-3, shift right=3]
\end{tikzcd}
\end{center}

\noindent
We want a monad in \ensuremath{\mathbb{C}} that represents higher-order effects.
However, our operation signatures are no longer endofunctors in \ensuremath{\mathbb{C}} (as
for algebraic effects), but rather higher-order endofuctors.
For that reason, it is necessary to make an intermediate step from \ensuremath{\mathbb{C}} to a
custom product category before constructing the free-forgetful adjunction.

In what follows we explain the two adjunctions with their corresponding categories and
functors before putting them together.

%-------------------------------------------------------------------------------
\subsection{The \ensuremath{\mathit{Id}\rtimes\mathbin{-}\dashv \mathit{App}\kern+2pt} Adjunction}

We first focus on the rightmost adjunction in the overview diagram.
%
% \begin{center}
% \begin{tikzcd}
% {|pointed ></ catC|} & {|bot|} & {|catC|}
% \arrow["{|App|}"', from=1-1, to=1-3, shift right=3]
% \arrow["{|Id ></ -|}"', from=1-3, to=1-1, shift right=3]
% \end{tikzcd}
% \end{center}
%
\noindent
It connects
the base category \ensuremath{\mathbb{C}} with the category \ensuremath{\mathbb{P}\rtimes\mathbb{C}}, the latter of which
%
% \paragraph{The |pointed ></ catC|-Category}
%
% The category |pointed ></ catC|
is defined as a variation on the
product of the category \ensuremath{\mathbb{P}} of pointed endofunctors on \ensuremath{\mathbb{C}}
and the base category \ensuremath{\mathbb{C}}.
This category \ensuremath{\mathbb{P}\rtimes\mathbb{C}} represents the higher-order signatures of effect
operations, consisting of an internal computation (in \ensuremath{\mathbb{P}}) and a continuation
(in \ensuremath{\mathbb{C}}).
% An endofunctor |H : pointed ></ catC -> pointed ></ catC| in this category is
% defined in terms of a higher-order bifunctor |K : pointed ></ catC -> catC|.
%
The following table summarizes the objects and morphisms of the different categories.
% The bottom row shows the category |pointed ></ catC| we are working with.

\begin{center}
\begin{tabular}{|l|l|l|}
\hline
\textbf{Symbol}    & \textbf{Objects}     & \textbf{Morphisms}            \\
\hline
\ensuremath{\mathbb{C}}             & \ensuremath{\Conid{A}}, \ensuremath{\Conid{B}}             & \ensuremath{\Varid{f}\mathbin{:}\Conid{A}\to \Conid{B}}                  \\
\ensuremath{\mathbb{P}}          & \ensuremath{\Conid{F}}, \ensuremath{\Conid{G}}             & \ensuremath{\alpha\mathbin{:}\Conid{F}\to \Conid{G}}               \\
\ensuremath{\mathbb{P}\times\mathbb{C}}  & \ensuremath{\langle \Conid{F},\Conid{A}\rangle }, \ensuremath{\langle \Conid{G},\Conid{B}\rangle }  & \ensuremath{\langle \alpha\mathbin{:}\Conid{F}\to \Conid{G},\Varid{f}\mathbin{:}\Conid{A}\to \Conid{B}\rangle } \\
\hline
\ensuremath{\mathbb{P}\rtimes\mathbb{C}} & \ensuremath{\langle \Conid{F},\Conid{A}\rangle }, \ensuremath{\langle \Conid{G},\Conid{B}\rangle }  & \ensuremath{\langle \alpha\mathbin{:}\Conid{F}\to \Conid{G},\Varid{f}\mathbin{:}\Conid{F}\;\Conid{A}\to \Conid{G}\;\Conid{B}\rangle } \\
\hline
\end{tabular}
\end{center}

% \noindent
% \begin{tabular}{||l||lll||}
% \hline
% \textbf{Symbol}    & \textbf{Description}                       & \textbf{Objects}     & \textbf{Morphisms}            \\
% \hline
% |catC|             & base category                              & |A|, |B|             & |f : A -> B|                  \\
% |pointed|          & category of pointed endofunctors on |catC| & |F|, |G|             & |alph : F -> G|               \\
% |pointed >< catC|  & product category of |pointed| and |catC|   & |\<F, A \>|, |\<G, B \>|  & |\<alph : F -> G, f : A -> B \>| \\
% \hline
% |pointed ></ catC| & our variation on the product category      & |\<F, A \>|, |\<G, B \>|  & |\<alph : F -> G, f : F A -> G B \>| \\
% \hline
% \end{tabular}

\noindent
Objects in \ensuremath{\mathbb{P}\rtimes\mathbb{C}} have the form \ensuremath{\langle \Conid{F},\Conid{A}\rangle }, where \ensuremath{\Conid{F}} is a pointed endofunctor on \ensuremath{\mathbb{C}} and \ensuremath{\Conid{A}} is an object of \ensuremath{\mathbb{C}}.
Morphisms are represented by pairs \ensuremath{\langle \alpha,\Varid{f}\rangle } of a structure-preserving natural transformation \ensuremath{\alpha\mathbin{:}\Conid{F}\to \Conid{G}}
(\ensuremath{\alpha\;\circ\;\eta^F\mathrel{=}\eta^G}), and a morphism \ensuremath{\Varid{f}\mathbin{:}\Conid{F}\;\Conid{A}\to \Conid{G}\;\Conid{B}}.
Composition and identity morphisms are defined componentwise:

\begin{minipage}{0.5\textwidth}\begin{hscode}\SaveRestoreHook
\column{B}{@{}>{\hspre}l<{\hspost}@{}}%
\column{3}{@{}>{\hspre}l<{\hspost}@{}}%
\column{E}{@{}>{\hspre}l<{\hspost}@{}}%
\>[3]{}\langle \beta,\Varid{g}\rangle \circ\langle \alpha,\Varid{f}\rangle \mathrel{=}\langle \beta\;\circ\;\alpha,\Varid{g}\;\circ\;\Varid{f}\rangle {}\<[E]%
\ColumnHook
\end{hscode}\resethooks
\end{minipage}%
\begin{minipage}{0.5\textwidth}\begin{hscode}\SaveRestoreHook
\column{B}{@{}>{\hspre}l<{\hspost}@{}}%
\column{3}{@{}>{\hspre}l<{\hspost}@{}}%
\column{E}{@{}>{\hspre}l<{\hspost}@{}}%
\>[3]{}id_{ \langle \Conid{F},\Conid{A}\rangle  }\mathrel{=}\langle id_{ \Conid{F} },id_{ \Conid{F}\;\Conid{A} }\rangle {}\<[E]%
\ColumnHook
\end{hscode}\resethooks
\end{minipage}

\noindent
% \paragraph{The Adjoints}
The left adjoint, functor \ensuremath{\mathit{Id}\rtimes\mathbin{-}}, maps objects and morphisms from \ensuremath{\mathbb{C}} to \ensuremath{\mathbb{P}\rtimes\mathbb{C}}.
Furthermore, \ensuremath{\mathit{Id}} itself is a pointed endofunctor with \ensuremath{\eta^{Id}\mathrel{=}id}.
The right adjoint, functor \ensuremath{\mathit{App}\kern+2pt}, does the opposite: it maps objects and morphisms
from \ensuremath{\mathbb{P}\rtimes\mathbb{C}} to \ensuremath{\mathbb{C}}, applying the pointed endofunctor on \ensuremath{\mathbb{C}} to
the object in \ensuremath{\mathbb{C}} and forgetting the natural transformation in the morphism.

\noindent
\begin{minipage}{0.5\textwidth}\begin{hscode}\SaveRestoreHook
\column{B}{@{}>{\hspre}l<{\hspost}@{}}%
\column{3}{@{}>{\hspre}l<{\hspost}@{}}%
\column{19}{@{}>{\hspre}l<{\hspost}@{}}%
\column{E}{@{}>{\hspre}l<{\hspost}@{}}%
\>[3]{}(\mathit{Id}\rtimes\mathbin{-})\;\Conid{A}{}\<[19]%
\>[19]{}\mathrel{=}\langle \mathit{Id},\Conid{A}\rangle {}\<[E]%
\\
\>[3]{}(\mathit{Id}\rtimes\mathbin{-})\;\Varid{f}{}\<[19]%
\>[19]{}\mathrel{=}\langle \eta^{Id},\mathit{Id}\;\Varid{f}\rangle \mathrel{=}\langle id_{  },\Varid{f}\rangle {}\<[E]%
\ColumnHook
\end{hscode}\resethooks
\end{minipage}%
\begin{minipage}{0.5\textwidth}\begin{hscode}\SaveRestoreHook
\column{B}{@{}>{\hspre}l<{\hspost}@{}}%
\column{3}{@{}>{\hspre}l<{\hspost}@{}}%
\column{23}{@{}>{\hspre}l<{\hspost}@{}}%
\column{E}{@{}>{\hspre}l<{\hspost}@{}}%
\>[3]{}\mathit{App}\kern+2pt\langle \Conid{F},\Conid{A}\rangle {}\<[23]%
\>[23]{}\mathrel{=}\Conid{F}\;\Conid{A}{}\<[E]%
\\
\>[3]{}\mathit{App}\kern+2pt\langle \alpha,\Varid{f}\rangle {}\<[23]%
\>[23]{}\mathrel{=}\Varid{f}{}\<[E]%
\ColumnHook
\end{hscode}\resethooks
\end{minipage}

% \paragraph{The Adjunction}
\noindent
The adjunction is witnessed by the following isomorphism, where the left adjunct
\ensuremath{\psi} and right adjunct \ensuremath{\psi^{-1}} are defined as follows:
\begin{hscode}\SaveRestoreHook
\column{B}{@{}>{\hspre}l<{\hspost}@{}}%
\column{4}{@{}>{\hspre}l<{\hspost}@{}}%
\column{64}{@{}>{\hspre}l<{\hspost}@{}}%
\column{E}{@{}>{\hspre}l<{\hspost}@{}}%
\>[4]{}\psi\mathbin{:}\mathbb{P}\rtimes\mathbb{C}\;\Bigl(\;(\mathit{Id}\rtimes\mathbin{-})\;\Conid{A},\langle \Conid{F},\Conid{B}\rangle \Bigr)\;{}\<[64]%
\>[64]{}\cong\;\mathbb{C}\;\Bigl(\;\Conid{A},\mathit{App}\kern+2pt\langle \Conid{F},\Conid{B}\rangle \Bigr)\mathbin{:}\psi^{-1}{}\<[E]%
\\
\>[4]{}\psi\langle \alpha,\Varid{f}\rangle \mathrel{=}\Varid{f}\;{}\<[64]%
\>[64]{}\psi^{-1}\;\Varid{g}\mathrel{=}\langle \eta^F,\Varid{g}\rangle {}\<[E]%
\ColumnHook
\end{hscode}\resethooks
% < psi :: bigl \< Id, A \> -> \< F, B\> bigr -> (A -> F B)
% < psi_inv :: (A -> F B) -> bigl \< Id, A \> -> \< F, B\> bigr

\noindent
These witnesses of the isomorphism satisfy the requisite
round-trip properties.

%-------------------------------------------------------------------------------
\subsection{The \ensuremath{\overline{\mathit{Free}}\dashv \overline{\mathit{U}}} Adjunction}

We now explain the free-forgetful adjunction on the left side of the diagram.
%
% \begin{center}
% \begin{tikzcd}
% {|catL|} & {|bot|} & {|pointed ></ catC|}
% \arrow["{|OFree|}"', from=1-3, to=1-1, shift right=3]
% \arrow["{|OU|}"', from=1-1, to=1-3, shift right=3]
% \end{tikzcd}
% \end{center}
%
In fact, category \ensuremath{\mathbb{P}\rtimes\mathbb{C}} is a special case of the product category,
for which the free-forgetful adjunction is an established construction.
We exploit this situation to define our free-forgetful adjunction as a special
instance of the product category.

\begin{center}
% https://q.uiver.app/?q=WzAsNCxbMCwxLCJ8Y2F0RHwiXSxbMiwxLCJ8cG9pbnRlZCA+PC8gY2F0Q3wiXSxbMiwwLCJ8Y2F0Q0MgPjwgY2F0Q3wiXSxbMCwwLCJ8SHwtQWxnIl0sWzEsMCwifE9GcmVlfCIsMix7Im9mZnNldCI6Mn1dLFswLDEsInxPVXwiLDIseyJvZmZzZXQiOjJ9XSxbMSwyLCJ8SXwiLDJdLFswLDMsInxKfCJdLFsyLDMsInxGcmVlfCIsMix7Im9mZnNldCI6Mn1dLFszLDIsInxVfCIsMSx7Im9mZnNldCI6Mn1dLFsxLDIsIlxcdGV4dGNvbG9ye2JsdWV9eygxKX0iLDAseyJzdHlsZSI6eyJib2R5Ijp7Im5hbWUiOiJub25lIn0sImhlYWQiOnsibmFtZSI6Im5vbmUifX19XSxbMiwzLCJ8Ym90fCBcXHRleHRjb2xvcntibHVlfXsoMil9IiwxLHsic3R5bGUiOnsiYm9keSI6eyJuYW1lIjoibm9uZSJ9LCJoZWFkIjp7Im5hbWUiOiJub25lIn19fV0sWzAsMSwifGJvdHwgXFx0ZXh0Y29sb3J7Ymx1ZX17KDQpfSIsMSx7InN0eWxlIjp7ImJvZHkiOnsibmFtZSI6Im5vbmUifSwiaGVhZCI6eyJuYW1lIjoibm9uZSJ9fX1dLFswLDMsIlxcdGV4dGNvbG9ye2JsdWV9eygzKX0iLDIseyJzdHlsZSI6eyJib2R5Ijp7Im5hbWUiOiJub25lIn0sImhlYWQiOnsibmFtZSI6Im5vbmUifX19XV0=
\begin{tikzcd}[row sep=small]
		{\ensuremath{H\mathit{-Alg}(\mathbb{C}^\mathbb{C} \times \mathbb{C})}} &&& {\ensuremath{\mathbb{C}^\mathbb{C}\times\mathbb{C}}} \\\\
		{\ensuremath{H\mathit{-Alg}(\mathbb{P}\rtimes\mathbb{C})}} &&& {\ensuremath{\mathbb{P}\rtimes\mathbb{C}}}
		\arrow["{\ensuremath{\overline{\mathit{Free}}}}"', shift right=3, from=3-4, to=3-1]
		\arrow["{\ensuremath{\overline{\mathit{U}}}}"', shift right=3, from=3-1, to=3-4]
		\arrow["{\ensuremath{\Conid{I}}}"', from=3-4, to=1-4]
		\arrow["{\ensuremath{\Conid{J}}}", from=3-1, to=1-1]
		\arrow["{\ensuremath{\Conid{Free}}}"', shift right=3, from=1-4, to=1-1]
		\arrow["{\ensuremath{\Conid{U}}}"', shift right=3, from=1-1, to=1-4]
		\arrow["{\textcolor{blue}{(1)}}", draw=none, from=3-4, to=1-4]
		\arrow["{\ensuremath{\bot} \textcolor{blue}{(2)}}"{description}, draw=none, from=1-4, to=1-1]
		\arrow["{\textcolor{blue}{(4)} \ensuremath{\bot}}"{description}, draw=none, from=3-1, to=3-4]
		\arrow["{\textcolor{blue}{(3)}}"', draw=none, from=3-1, to=1-1]
\end{tikzcd}
\end{center}

\noindent
In particular, we work in 4 steps:
\textcolor{blue}{(1)} we include category \ensuremath{\mathbb{P}\rtimes\mathbb{C}} in \ensuremath{\mathbb{C}^\mathbb{C}\times\mathbb{C}};
\textcolor{blue}{(2)} we discuss the free-forgetful adjunction of functor \ensuremath{\Conid{H}\mathbin{:}\mathbb{C}^\mathbb{C}\times\mathbb{C}\to \mathbb{C}^\mathbb{C}\times\mathbb{C}};
\textcolor{blue}{(3)} we present \ensuremath{H\mathit{-Alg}(\mathbb{P}\rtimes\mathbb{C})} as a restricted version of the category of \ensuremath{\Conid{H}}-algebras; and % |HAlg|.
\textcolor{blue}{(4)} we demonstrate how \ensuremath{\overline{\mathit{Free}}\dashv \overline{\mathit{U}}} is a composition of \ensuremath{\Conid{Free}\dashv \Conid{U}} and two inclusion functors
expressing the subcategory relations.

% In particular, we work in 4 steps:
% \begin{enumerate}
% \item We generalize category |pointed ></ catC| into |catCC >< catC|.
% \item We discuss the free-forgetful adjunction of functor |H : catCC >< catC -> catCC >< catC|.
% \item We present |catL| as a restricted version of the category of |H|-algebras. % |HAlg|.
% \item We demonstrate how |OFree -/ OU| is a composition of |Free -/ U| and two inclusion functors
% expressing the subcategory relations.
% \end{enumerate}

%% PART 1
%~  ~  ~  ~  ~  ~  ~  ~  ~  ~  ~  ~  ~  ~  ~  ~  ~  ~  ~  ~  ~  ~  ~  ~  ~  ~  ~
\paragraph{Generalizing the \ensuremath{\mathbb{P}\rtimes\mathbb{C}} Category}

\ensuremath{\mathbb{P}\rtimes\mathbb{C}} is isomorphic to a subcategory
of the product category \ensuremath{\mathbb{P}\times\mathbb{C}}.
Indeed, every object \ensuremath{\langle \Conid{F},\Conid{A}\rangle } in \ensuremath{\mathbb{P}\rtimes\mathbb{C}} corresponds to an object
\ensuremath{\langle \Conid{F},\Conid{F}\;\Conid{A}\rangle } in \ensuremath{\mathbb{P}\times\mathbb{C}}.
Moreover, \ensuremath{\mathbb{P}\times\mathbb{C}} is a subcategory\footnote{No isomorphism since not every natural
transformation preserves \ensuremath{\eta}.} of \ensuremath{\mathbb{C}^\mathbb{C}\times\mathbb{C}} such that
every object in \ensuremath{\mathbb{P}\times\mathbb{C}} is an object in \ensuremath{\mathbb{C}^\mathbb{C}\times\mathbb{C}},
forgetting \ensuremath{\Conid{F}}'s pointedness.
\begin{hscode}\SaveRestoreHook
\column{B}{@{}>{\hspre}l<{\hspost}@{}}%
\column{3}{@{}>{\hspre}l<{\hspost}@{}}%
\column{55}{@{}>{\hspre}l<{\hspost}@{}}%
\column{66}{@{}>{\hspre}l<{\hspost}@{}}%
\column{E}{@{}>{\hspre}l<{\hspost}@{}}%
\>[3]{}\mathbb{P}\rtimes\mathbb{C}\;\Bigl(\langle \Conid{F},\Conid{A}\rangle ,\langle \Conid{G},\Conid{B}\rangle \Bigr){}\<[55]%
\>[55]{}\mathrel{=}{}\<[66]%
\>[66]{}\mathbb{P}\times\mathbb{C}\;\Bigl(\langle \Conid{F},\Conid{F}\;\Conid{A}\rangle ,\langle \Conid{G},\Conid{G}\;\Conid{B}\rangle \Bigr){}\<[E]%
\\
\>[55]{}\subseteq\;{}\<[66]%
\>[66]{}\mathbb{C}^\mathbb{C}\times\mathbb{C}\;\Bigl(\langle \Conid{F},\Conid{F}\;\Conid{A}\rangle ,\langle \Conid{G},\Conid{G}\;\Conid{B}\rangle \Bigr){}\<[E]%
\ColumnHook
\end{hscode}\resethooks
By transitively combining these two relations, we
conclude that \ensuremath{\mathbb{P}\rtimes\mathbb{C}} is isomorphic to a subcategory of
\ensuremath{\mathbb{C}^\mathbb{C}\times\mathbb{C}}, captured by an inclusion functor
\ensuremath{\Conid{I}\mathbin{:}(\mathbb{P}\rtimes\mathbb{C})\to (\mathbb{C}^\mathbb{C}\times\mathbb{C})} such that
\ensuremath{\Conid{I}\langle \Conid{F},\Conid{A}\rangle \mathrel{=}\langle \Conid{F},\Conid{F}\;\Conid{A}\rangle } and
\ensuremath{\Conid{I}\langle \Varid{h}_{1},\Varid{h}_{2}\rangle \mathrel{=}\langle \Varid{h}_{1},\Varid{h}_{2}\rangle }.

% \noindent
% \begin{minipage}{0.5\textwidth}
% < I  \< F,   A \>    = \< F,   F A \>
% \end{minipage}%
% \begin{minipage}{0.5\textwidth}
% < I  \< h1,  h2  \>  = \< h1,  h2 \>
% \end{minipage}

% This fact is important as the latter category has an initial object, all finite
% coproducts and thus also least fixpoints. The former does not.

%% PART 2
%~  ~  ~  ~  ~  ~  ~  ~  ~  ~  ~  ~  ~  ~  ~  ~  ~  ~  ~  ~  ~  ~  ~  ~  ~  ~  ~
\paragraph{Generalized Adjunction}
We shift our focus to the \ensuremath{\mathbb{C}^\mathbb{C}\times\mathbb{C}} category.
Provided that \ensuremath{\mathbb{C}} has fixpoints of endofunctors, \ensuremath{\mathbb{C}^\mathbb{C}\times\mathbb{C}}
also does.
Hence, we can instantiate the free-forgetful adjunction over \ensuremath{\mathbb{C}^\mathbb{C}\times\mathbb{C}}, based
on an endofunctor \ensuremath{\Conid{H}\mathbin{:}(\mathbb{C}^\mathbb{C}\times\mathbb{C})\to (\mathbb{C}^\mathbb{C}\times\mathbb{C})}.
Here, \ensuremath{\Conid{H}^\star} is a free monad, defined as \ensuremath{\Conid{H}^\star\mathrel{=}\Conid{U}\;\Conid{Free}}.
Moreover, this fixpoint is equipped with a fold recursion scheme.

% \begin{center}
% \begin{tikzcd}
% 	{|HAlg|} & |bot| & {|catCC >< catC|}
% 	\arrow["{|Free|}"', shift right=3, from=1-3, to=1-1]
% 	\arrow["{|U|}"', shift right=3, from=1-1, to=1-3]
% \end{tikzcd}
% \end{center}

%% PART 3
%-------------------------------------------------------------------------------
\paragraph{Specializing the Adjunction}

In what follows we impose five restrictions that make the adjunction
act between categories \ensuremath{H\mathit{-Alg}(\mathbb{P}\rtimes\mathbb{C})} and \ensuremath{\mathbb{P}\rtimes\mathbb{C}}.

\begin{description}
\item [(1)]
We restrict ourselves to the subcategory of \ensuremath{\mathbb{C}^\mathbb{C}\times\mathbb{C}} where
objects are of the form \ensuremath{\langle \Conid{F},\Conid{F}\;\Conid{A}\rangle }.

\item [(2)]
We only consider endofunctors \ensuremath{\Conid{H}} created out of higher-order bifunctor
\ensuremath{\Conid{K}\mathbin{:}\mathbb{C}^\mathbb{C}\times\mathbb{C}\to \mathbb{C}} such that
\ensuremath{\Conid{H}\langle \Conid{F},\Conid{A}\rangle \mathrel{=}\langle \Conid{K}\;\Conid{F}\;\circ\;\Conid{F},\Conid{K}\;\Conid{F}\;\Conid{A}\rangle }.

Furthermore, algebra actions for carriers \ensuremath{\langle \Conid{F},\Conid{F}\;\Conid{A}\rangle } under
\ensuremath{\Conid{Free}\mathbin{:}\mathbb{C}^\mathbb{C}\times\mathbb{C}\to H\mathit{-Alg}(\mathbb{C}^\mathbb{C} \times \mathbb{C})}
have the form \ensuremath{\langle \alpha\mathbin{:}\Conid{K}\;\Conid{F}\;\circ\;\Conid{F}\to \Conid{F},\Varid{f}\mathbin{:}\Conid{K}\;\Conid{F}\;(\Conid{F}\;\Conid{A})\to \Conid{F}\;\Conid{A}\rangle }.

\item [(3)]
We restrict ourselves further to the subcategory of algebras where \ensuremath{\Varid{f}\mathrel{=}\alpha_A} and consider only
actions \ensuremath{\langle \alpha,\alpha_A\rangle } that are uniquely determined by their natural
transformation \ensuremath{\alpha}.

\item [(4)]
We only study \ensuremath{\mathbb{C}^\mathbb{C}} objects \ensuremath{\Conid{F}} that
are pointed functors, i.e., that are also objects in \ensuremath{\mathbb{P}}
with an associated \ensuremath{\eta^F\mathbin{:}\mathit{Id}\to \Conid{F}}.

\item [(5)]
Finally, we consider solely those morphisms \ensuremath{\langle \Varid{h}_{1},\Varid{h}_{2}\rangle } where
\ensuremath{\Varid{h}_{1}} is a pointed functor homomorphism.
\end{description}

\noindent
It turns out that under \textbf{(1)} and \textbf{(2)}, \ensuremath{\Conid{H}^\star\langle \Conid{F},\Conid{F}\;\Conid{A}\rangle } takes
the form \ensuremath{\langle Free_{\textsf{H}}\;\Conid{K}\;\Conid{F}}, \ensuremath{Free_{\textsf{H}}\;\Conid{K}\;\Conid{F}\;\Conid{A}\rangle }, where \ensuremath{\Conid{H}^\star} is the generalized free monad.
% We define |T| as follows:
\begin{hscode}\SaveRestoreHook
\column{B}{@{}>{\hspre}l<{\hspost}@{}}%
\column{3}{@{}>{\hspre}l<{\hspost}@{}}%
\column{E}{@{}>{\hspre}l<{\hspost}@{}}%
\>[3]{}Free_{\textsf{H}}\;\Conid{K}\;\Conid{F}\mathrel{=}\Lambda\;\Conid{X}\, .\,\Conid{F}\;\Conid{X}\mathbin{+}\Conid{K}\;Free_{\textsf{H}}\;(Free_{\textsf{H}}\;\Conid{X}){}\<[E]%
\ColumnHook
\end{hscode}\resethooks
Hence, \ensuremath{\Conid{H}^\star\langle \Conid{F},\Conid{F}\;\Conid{A}\rangle } is part of the subcategory of \textbf{(1)}.
Furthermore, the algebra action created by \ensuremath{\Conid{Free}} \textbf{(2)}
also resides in this subcategory \textbf{(3)}.
Notice that \ensuremath{Free_{\textsf{H}}} is not a monad in \ensuremath{\mathbb{P}\rtimes\mathbb{C}}, but that it is pointed \textbf{(4)}
because \ensuremath{\Conid{F}} is.
Also \textbf{(5)} holds, as functor \ensuremath{\Conid{Free}} respects it: the first component of
the fold with generator \ensuremath{\langle \Varid{h}_{1},\Varid{h}_{2}\rangle } and algebra \ensuremath{\langle \alpha,\alpha_A\rangle } is a
pointed functor morphism.
From these five restrictions on category \ensuremath{H\mathit{-Alg}(\mathbb{C}^\mathbb{C} \times \mathbb{C})},
we can define category \ensuremath{H\mathit{-Alg}(\mathbb{P}\rtimes\mathbb{C})}.
\emph{Objects} in \ensuremath{H\mathit{-Alg}(\mathbb{P}\rtimes\mathbb{C})} are tuples \ensuremath{\langle \langle \Conid{F},\Conid{A}\rangle ,\Varid{a}\mathbin{:}\Conid{K}\;\Conid{F}\;\circ\;\Conid{F}\to \Conid{F}\rangle },
where carrier \ensuremath{\langle \Conid{F},\Conid{A}\rangle } is an object in \ensuremath{\mathbb{P}\rtimes\mathbb{C}}
and action \ensuremath{\Varid{a}} is a morphism in \ensuremath{\mathbb{C}^\mathbb{C}}.
\emph{Morphisms} \ensuremath{\langle \Varid{h}_{1},\Varid{h}_{2}\rangle \mathbin{:}\langle \langle \Conid{F},\Conid{A}\rangle ,\Varid{a}\rangle \to \langle \langle \Conid{G},\Conid{B}\rangle ,\Varid{b}\rangle }
in \ensuremath{H\mathit{-Alg}(\mathbb{P}\rtimes\mathbb{C})} are homomorphisms \ensuremath{\langle \Varid{h}_{1},\Varid{h}_{2}\rangle \mathbin{:}\langle \Conid{F},\Conid{A}\rangle \to \langle \Conid{G},\Conid{B}\rangle }
in \ensuremath{\mathbb{P}\rtimes\mathbb{C}}:

\begin{center}
\begin{tikzcd}[row sep=small,column sep=scriptsize]
	\ensuremath{\Conid{K}\;\Conid{F}\;\circ\;\Conid{F}} && \ensuremath{\Conid{K}\;\Conid{G}\;\circ\;\Conid{G}} && \ensuremath{\Conid{K}\;\Conid{F}\;(\Conid{F}\;\Conid{A})} && \ensuremath{\Conid{K}\;\Conid{G}\;(\Conid{G}\;\Conid{B})} \\
	\\
	\ensuremath{\Conid{F}} && \ensuremath{\Conid{G}} && \ensuremath{\Conid{F}\;\Conid{A}} && \ensuremath{\Conid{G}\;\Conid{B}}
	\arrow["\ensuremath{\Varid{a}}"', from=1-1, to=3-1]
	\arrow["\ensuremath{\Varid{b}}"', from=1-3, to=3-3]
	\arrow["{\ensuremath{\Conid{K}\;\Varid{h}_{1}\;\Varid{h}_{1}}}", from=1-1, to=1-3]
	\arrow["{\ensuremath{\Varid{h}_{1}}}"', from=3-1, to=3-3]
	\arrow["{\ensuremath{\Varid{a}_{\Conid{A}}}}"', from=1-5, to=3-5]
	\arrow["{\ensuremath{\Conid{K}\;\Varid{h}_{1}\;\Varid{h}_{2}}}", from=1-5, to=1-7]
	\arrow["{\ensuremath{\Varid{b}_{\Conid{B}}}}", from=1-7, to=3-7]
	\arrow["{\ensuremath{\Varid{h}_{2}}}"', from=3-5, to=3-7]
\end{tikzcd}
\end{center}

\noindent
Observe that objects \ensuremath{\langle \langle \Conid{F},\Conid{A}\rangle ,\Varid{a}\rangle } in \ensuremath{H\mathit{-Alg}(\mathbb{P}\rtimes\mathbb{C})} are in
one-to-one correspondence with \ensuremath{\langle \langle \Conid{F},\Conid{F}\;\Conid{A}\rangle ,\langle \Varid{a},\Varid{a}_{\Conid{A}}\rangle \rangle }
of \ensuremath{H\mathit{-Alg}(\mathbb{C}^\mathbb{C} \times \mathbb{C})} that are subject to the above restrictions.
Similarly, morphisms \ensuremath{\langle \Varid{h}_{1},\Varid{h}_{2}\rangle } in \ensuremath{H\mathit{-Alg}(\mathbb{P}\rtimes\mathbb{C})} are also morphisms
in \ensuremath{H\mathit{-Alg}(\mathbb{C}^\mathbb{C} \times \mathbb{C})} that meet the restrictions.
The above two correspondences are witnessed by an inclusion functor \ensuremath{\Conid{J}\mathbin{:}H\mathit{-Alg}(\mathbb{P}\rtimes\mathbb{C})\to H\mathit{-Alg}(\mathbb{C}^\mathbb{C} \times \mathbb{C})}
such that
\ensuremath{\Conid{J}\langle \langle \Conid{F},\Conid{A}\rangle ,\Varid{a}\rangle \mathrel{=}\langle \langle \Conid{F},\Conid{F}\;\Conid{A}\rangle ,\langle \Varid{a},\Varid{a}_{\Conid{A}}\rangle \rangle } and
\ensuremath{\Conid{J}\langle \Varid{h}_{1},\Varid{h}_{2}\rangle \mathrel{=}\langle \Varid{h}_{1},\Varid{h}_{2}\rangle }.

% \noindent
% \begin{minipage}{0.5\textwidth}
% < J \< \< F, A \>, a \>   = \< \< F, F A \>, \< a, a_A \> \>
% \end{minipage}%
% \begin{minipage}{0.5\textwidth}
% < J \< h1, h2 \>          = \< h1, h2 \>
% \end{minipage}

%% PART 4
%- - - - - - - - - - - - - - - - - - - - - - - - - - - - - - - - - - - - - - - -
\paragraph{The Specialized Adjunction}

In categorical terms we define the \ensuremath{\overline{\mathit{Free}}\dashv \overline{\mathit{U}}}
adjunction as \ensuremath{J^{-1}\;\circ\;\Conid{Free}\;\circ\;\Conid{I}\dashv I^{-1}\;\circ\;\Conid{U}\;\circ\;\Conid{J}}.
%
% \noindent
% \begin{minipage}{0.5\textwidth}
% < OFree  = J1 circ Free circ I
% \end{minipage}%
% \begin{minipage}{0.5\textwidth}
% < OU     = I1 circ U circ J
% \end{minipage}
%
This implies that the \ensuremath{\overline{\mathit{Free}}\dashv \overline{\mathit{U}}} adjunction is the composition of the \ensuremath{\Conid{Free}\dashv \Conid{U}}
adjunction with the two adjunctions from isomorphisms \ensuremath{J^{-1}\dashv \Conid{J}} and \ensuremath{\Conid{I}\dashv I^{-1}}.
Moreover, the universal property of \ensuremath{\Conid{Free}} carries over to \ensuremath{\overline{\mathit{Free}}}.
Hence, \ensuremath{\overline{\mathit{Free}}\dashv \overline{\mathit{U}}} is also a free-forgetful adjunction
and \ensuremath{\overline{\mathit{U}}\;\circ\;\overline{\mathit{Free}}} is a free monad in \ensuremath{\mathbb{P}\rtimes\mathbb{C}}.

%-------------------------------------------------------------------------------
\subsection{Composing the Two Adjunctions}

When we compose the two adjunctions, the diagram is complete.

% https://q.uiver.app/?q=WzAsNSxbMCwxLCJ8Y2F0RHwiXSxbMiwxLCJ8cG9pbnRlZCA+PC8gY2F0Q3wiXSxbMiwwLCJ8Y2F0Q0MgPjwgY2F0Q3wiXSxbMCwwLCJ8SHwtQWxnIl0sWzQsMSwifGNhdEN8Il0sWzEsMCwifE9GcmVlfCIsMix7Im9mZnNldCI6Mn1dLFswLDEsInxPVXwiLDIseyJvZmZzZXQiOjJ9XSxbMSwyLCJ8SXwiLDJdLFswLDMsInxKfCJdLFsyLDMsInxGcmVlfCIsMix7Im9mZnNldCI6Mn1dLFszLDIsInxVfCIsMix7Im9mZnNldCI6Mn1dLFsyLDMsInxib3R8IiwxLHsic3R5bGUiOnsiYm9keSI6eyJuYW1lIjoibm9uZSJ9LCJoZWFkIjp7Im5hbWUiOiJub25lIn19fV0sWzAsMSwifGJvdHwiLDEseyJzdHlsZSI6eyJib2R5Ijp7Im5hbWUiOiJub25lIn0sImhlYWQiOnsibmFtZSI6Im5vbmUifX19XSxbNCwxLCJ8SWQgPjwvIC18IiwyLHsib2Zmc2V0IjozfV0sWzEsNCwifEFwcHwiLDIseyJvZmZzZXQiOjN9XSxbMSw0LCJ8Ym90fCIsMSx7InN0eWxlIjp7ImJvZHkiOnsibmFtZSI6Im5vbmUifSwiaGVhZCI6eyJuYW1lIjoibm9uZSJ9fX1dXQ==
\begin{center}
\begin{tikzcd}[column sep=large]
	{\ensuremath{H\mathit{-Alg}(\mathbb{C}^\mathbb{C} \times \mathbb{C})}} && {\ensuremath{\mathbb{C}^\mathbb{C}\times\mathbb{C}}} \\
	{\ensuremath{H\mathit{-Alg}(\mathbb{P}\rtimes\mathbb{C})}} && {\ensuremath{\mathbb{P}\rtimes\mathbb{C}}} && {\ensuremath{\mathbb{C}}}
	\arrow["{\ensuremath{\overline{\mathit{Free}}}}"', shift right=2, from=2-3, to=2-1]
	\arrow["{\ensuremath{\overline{\mathit{U}}}}"', shift right=2, from=2-1, to=2-3]
	\arrow["{\ensuremath{\Conid{I}}}"', from=2-3, to=1-3]
	\arrow["{\ensuremath{\Conid{J}}}", from=2-1, to=1-1]
	\arrow["{\ensuremath{\Conid{Free}}}"', shift right=2, from=1-3, to=1-1]
	\arrow["{\ensuremath{\Conid{U}}}"', shift right=2, from=1-1, to=1-3]
	\arrow["{\ensuremath{\bot}}"{description}, draw=none, from=1-3, to=1-1]
	\arrow["{\ensuremath{\bot}}"{description}, draw=none, from=2-1, to=2-3]
	\arrow["{\ensuremath{\mathit{Id}\rtimes\mathbin{-}}}"', shift right=2, from=2-5, to=2-3]
	\arrow["{\ensuremath{\mathit{App}\kern+2pt}}"', shift right=2, from=2-3, to=2-5]
	\arrow["{\ensuremath{\bot}}"{description}, draw=none, from=2-3, to=2-5]
\end{tikzcd}
\end{center}

\noindent
In order to get a free monad in \ensuremath{\mathbb{C}}, we apply the \ensuremath{\overline{\mathit{Free}}\;\circ\;(\mathit{Id}\rtimes\mathbin{-})\dashv \mathit{App}\kern+2pt\;\circ\;\overline{\mathit{U}}} adjunction.
We start with an object \ensuremath{\Conid{A}} in \ensuremath{\mathbb{C}}, and transform it to \ensuremath{\langle \mathit{Id},\Conid{A}\rangle }
in \ensuremath{\mathbb{P}\rtimes\mathbb{C}}, where the identity functor \ensuremath{\mathit{Id}} is pointed.
Using the free-forgetful adjunction and applying the above restrictions, we get
a free monad \ensuremath{\langle Free_{\textsf{H}}\;\Conid{K}\;\mathit{Id},Free_{\textsf{H}}\;\Conid{K}\;\mathit{Id}\;\Conid{A}\rangle } in \ensuremath{\mathbb{P}\rtimes\mathbb{C}}.
If we then apply the \ensuremath{\mathit{App}\kern+2pt} functor, we get a free monad \ensuremath{Free_{\textsf{H}}\;\Conid{K}\;\mathit{Id}\;\Conid{A}} in \ensuremath{\mathbb{C}}, which
is indeed the monad we have used for our instantiations.
Notice that \ensuremath{Free_{\textsf{H}}\;\Conid{K}\;\mathit{Id}} of this section is isomorphic to \ensuremath{Free_{\textsf{H}}\;\Conid{K}} of \Cref{sec:generic}.
% We have omitted the identity functor for convenience.
% Notice that, since we always have the identity functor in the monad, we omitted
% it for convenience.
%==================================================================
\section{Related Work}

% alternative approaches (mtl, non-alg ops as handlers)
\paragraph{Modularly Combining Side-Effects}
Alternative approaches to (modularly) combining different effects exist,
with the most prominent one being monad transformers \cite{mtl}.
Schrijvers et al. \cite{schrijvers19} have described in detail under which circumstances
or conditions the approach of algebraic effects \& handlers, on which our work is based,
is better or worse than the approach with monad transformers.
Usually, monad transformers appear more expressive, but algebraic effects \& handlers more modular.
% They also explain when and how algebraic effects can be encoded as monad
% transformers and vice versa.
Jaskelioff~\cite{ifl/Jaskelioff08,esop/Jaskelioff09} has investigated how to
make monad transformers more modular.
\paragraph{Non-Algebraic Effects}
We are not the first to realize that not all effectful operations fit in the approach
of algebraic effects \& handlers.
Plotkin and Power \cite{Plotkin03} first proposed to model so-called non-algebraic
operations as handlers. However, this approach comes at the cost of modularity:
it mixes syntax and semantics; the clear separation of both is one of the
coveted properties of algebraic effects \& handlers.
Wu et al. \cite{scope14} and Yang et al. \cite{esop22} have shown that the approach
of Plotkin and Power indeed loses modularity for scoped effects and handlers and
came up with a different, but very specific representation.
Xie et al. \cite{xie21} and van den Berg et al. \cite{vandenBerg21} have proposed
similar (ad hoc) representations for parallel and latent effects, respectively.
The latter two representations are, as far as we know, not yet put on formal footing.

% scoped effects have been discussed before:
% Of all kinds of effects in this paper, algebraic and scoped effects are the
% best-studied ones.
\paragraph{Adoption of Algebraic and Higher-Order Effects}
Algebraic effects \& handlers have been picked up in many
practical programming languages,
such as Links \cite{Hillerstrom16}, Koka \cite{Leijen14}, Effekt \cite{effekt}, and Frank \cite{frank},
and libraries (e.g., fused-effects \cite{fused-effects}, extensible-effects \cite{extensible-effects},
Eff in Ocaml \cite{eff-ocaml}).
Higher-order effects are following the same trend,
evidenced by their adoption at GitHub \cite{github} and in Haskell libraries such as
eff \cite{eff}, polysemy \cite{polysemy}, fused-effects \cite{fused-effects},
and in-other-words \cite{in-other-words}.

\paragraph{Scoped Effects \& Handlers as Higher-Order Effects}
Scoped effects \& handlers have been studied denotationally, with
Wu et al. \cite{scope14} also implementing the coproduct for scoped operations.
Furthermore, our instance of the framework for scoped effects corresponds to their monad \ensuremath{\Conid{E}}.
In their work, they suggest a higher-order syntax for effect signatures, with
explicit ``weaving'' (also sometimes called ``forwarding") of handlers:
threading of the handler through the higher-order syntax, distributing it for every
signature.
Their design is made for scoped effects specifically, and, compared to our framework,
requires extra verboseness (the weaving) in order to achieve the desired modularity.
% which gives their approach more freedom but also makes it verbose.
%
% the ESOP paper with a very similar monad but not all different instances.

Yang et al. \cite{esop22} have studied a free monad for scoped effects that approximates our
generic monad closely. However, their signature is tailored to algebraic and scoped effects
only, not generalizing to other higher-order effects.
Furthermore, they opt for two algebras to interpret scoped
effects: one for the computation in scope, another for the continuation. They have
only a single example in which these different algebras seem meaningful: alternating
search strategies (e.g., depth-first, breadth-first and depth-bounded search).

% about higher-order effects
\paragraph{A Calculus for Higher-Order Effects}
Poulsen and van der Rest \cite{higher-order} present a calculus that models
higher-order effects, of which the operational semantics contrasts
with the denotational and categorical semantics of our work.
Although Poulsen and van der Rest use a similar encoding of higher-order effects
with higher-order functors and so-called hefty algebras, their focus is different.
Their work is about the elaboration of scoped and latent effects into algebraic effects.
Whereas they only consider scoped and latent effects, we provide a broader range
of higher-order effects (parallel, writer, bracketing effects).
Furthermore, we back up our framework with a categorical model and show it theoretically correct.
%===============================================================================
\section{Conclusion}

In summary, this work has provided a generic way of encoding different
(algebraic and higher-order) effects,
retaining the coveted modularity of algebraic effects \& handlers.
%
% \birthe{Remove this until table?}
% The examples in this paper, show the practical usability of this work;
% the categorical foundations show its theoretical correctness.
% Not only have we shown that different existing effects (e.g., scoped, parallel,
% latent) fit the framework, we also came up with novel effects that were not
% characterized in terms of a free monad definition before (e.g., inner, \todo{bracketing}).
% An overview of the different effects can be found in the following table:
%
% \noindent
% \begin{tabular}{||l||lll||l||}
% \hline
% \textbf{Effect} & \multicolumn{3}{l}{\textbf{Bifunctor}} & \textbf{Examples} \\
% \hline
% Algebraic               &  |KAlg_Sig    | & |F A| & |=  Sigma A|    & State, nondeterminism   \\
% Scoped                  &  |KSc_Gam     | & |F A| & |=  Gamma (F A)| & Local variables, exceptions \\
% Parallel                &  |KPar_Rho    | & |F A| & |=  Rho (F B) >< (Rho B => A)| & For-loops \\
% Inner                   &  |KWrite_Phi  | & |F A| & |=  F (Phi A)| & Writer \\
% \multirow{2}{*}{Latent} & |KLat_Zet_l   | & |F A| & |=  intpc zet P C >< Lat 1 >< | & \multirow{2}{*}{Lazy evaluation, staging} \\
%           & \multicolumn{3}{l}{|(intx C X >< Lat 1 => F (Lat X)) >< (Lat P => A)|} & \\
%           \hline
% \end{tabular}

Some design choices impact the reusability and accessibility of this
work. For instance, our monad closely resembles the ``regular'' free monad,
as we know from algebraic effects, because we choose pointed endofunctors and
a custom definition of the product category.
This design choice implies that we have
a single, but consistent interpreter for the internal computation and the
continuation of our effects. Although this seems to limit the expressivity at first
sight, we found little meaningful examples in which a different interpretation is desired.

A possible strain of research to follow-up on this work is to investigate the
interactions and laws for the composition of different higher-order effects (e.g.,
what if we parallellize scoped effects?), in the same spirit as many works
\cite{justdoit,zhixuan21} have studied the interactions of state and nondeterminism.

%% References:
\newpage
\bibliographystyle{splncs04}
\bibliography{reference}

\newpage
\appendix

%-------------------------------------------------------------------------------
\section{Background on Category Theory}
\label{app:ct-background}

In this Appendix, we go over some basic definitions, concepts and terminology.
This subsection provides the necessary preliminaries for \Cref{sec:formalization},
which elaborates on the categorical foundations of our generic framework.

\begin{definition}[Category]
  A \emph{category} \ensuremath{\mathbb{C}} is defined in terms of four parts:
  \begin{enumerate}[topsep=0pt, partopsep=0pt]
     \item A collection of \emph{objects} \ensuremath{\Conid{A}}, \ensuremath{\Conid{B}}.
     \item For every two objects in \ensuremath{\mathbb{C}}, a collection of \emph{morphisms} between those two objects.
     This is often called the \emph{hom-set}, and we denote it by \ensuremath{{\mathbb{C}\kern-3pt}\;(\Conid{A},\Conid{B})}.
     \item For every object an \emph{identity morphism}:
     \ensuremath{\forall\kern-2pt\;\Conid{A}\;\kern-3pt\in\kern-3pt\;\mathbb{C}\mathbin{:}id_{ \Conid{A} }\;\kern-3pt\in\kern-3pt\;{\mathbb{C}\kern-3pt}\;(\Conid{A},\Conid{A})}
     \item For every two morphisms their \emph{composition}:
     \ensuremath{\forall\kern-2pt\;\Varid{f}\;\kern-3pt\in\kern-3pt\;{\mathbb{C}\kern-3pt}\;(\Conid{A},\Conid{A}),\Varid{g}\;\kern-3pt\in\kern-3pt\;{\mathbb{C}\kern-3pt}\;(\Conid{B},\Conid{C})\mathbin{:}\Varid{f}\;\circ\;\Varid{g}\;\kern-3pt\in\kern-3pt\;{\mathbb{C}\kern-3pt}\;(\Conid{A},\Conid{C})}
  \end{enumerate}
  Furthermore, two properties should be satisfied:
  % \emph{Associativity} of composition, and
  % \emph{left} and \emph{right unit} of composition.
  \begin{enumerate}
     \item \emph{Associativity} of composition.\begin{hscode}\SaveRestoreHook
\column{B}{@{}>{\hspre}l<{\hspost}@{}}%
\column{3}{@{}>{\hspre}l<{\hspost}@{}}%
\column{7}{@{}>{\hspre}l<{\hspost}@{}}%
\column{E}{@{}>{\hspre}l<{\hspost}@{}}%
\>[3]{}\forall\kern-2pt\;\Conid{A},\Conid{B},\Conid{C},\Conid{D}\;\kern-3pt\in\kern-3pt\;\mathbb{C}\mathbin{:}\forall\kern-2pt\;\Varid{f}\;\kern-3pt\in\kern-3pt\;{\mathbb{C}\kern-3pt}\;(\Conid{A},\Conid{B}),\Varid{g}\;\kern-3pt\in\kern-3pt\;{\mathbb{C}\kern-3pt}\;(\Conid{B},\Conid{C}),\Varid{h}\;\kern-3pt\in\kern-3pt\;{\mathbb{C}\kern-3pt}\;(\Conid{C},\Conid{D})\mathbin{:}{}\<[E]%
\\
\>[3]{}\hsindent{4}{}\<[7]%
\>[7]{}(\Varid{f}\;\circ\;\Varid{g})\;\circ\;\Varid{h}\;\equiv\;\Varid{f}\;\circ\;(\Varid{g}\;\circ\;\Varid{h}){}\<[E]%
\ColumnHook
\end{hscode}\resethooks
     \item \emph{Left unit}, \emph{right unit} of composition.\begin{hscode}\SaveRestoreHook
\column{B}{@{}>{\hspre}l<{\hspost}@{}}%
\column{3}{@{}>{\hspre}l<{\hspost}@{}}%
\column{E}{@{}>{\hspre}l<{\hspost}@{}}%
\>[3]{}\forall\kern-2pt\;\Conid{A},\Conid{B}\;\kern-3pt\in\kern-3pt\;\mathbb{C}\mathbin{:}\forall\kern-2pt\;\Varid{f}\;\kern-3pt\in\kern-3pt\;{\mathbb{C}\kern-3pt}\;(\Conid{A},\Conid{B})\mathbin{:}id_{ \Conid{B} }\;\circ\;\Varid{f}\;\equiv\;\Varid{f}\;\equiv\;\Varid{f}\;\circ\;id_{ \Conid{A} }{}\<[E]%
\ColumnHook
\end{hscode}\resethooks
  \end{enumerate}
  \qed
\end{definition}

\begin{definition}[Homomorphism]
  A \emph{homomorphism} is a morphism that preserves the algebraic structure of the objects.
\end{definition}

\begin{definition}[Initial Object]
  The \emph{initial object} of a category \ensuremath{\mathbb{C}} is an object in \ensuremath{\mathbb{C}} such that
  for all objects \ensuremath{\Conid{A}} in \ensuremath{\mathbb{C}} there exists a unique morphism from that initial object
  to \ensuremath{\Conid{A}}.
\end{definition}

\begin{definition}[Functor]
  A \emph{functor} \ensuremath{\Conid{F}\mathbin{:}\mathbb{C}\to \mathbb{D}} is a structure-preserving mapping between
  two categories \ensuremath{\mathbb{C}} and \ensuremath{\mathbb{D}}.
  It maps both the objects and the hom-set of category \ensuremath{\mathbb{C}} into category \ensuremath{\mathbb{D}},
  respecting identities and composition:
  \ensuremath{\Conid{F}\;id_{ \Conid{A} }\mathrel{=}id_{ \Conid{F}\;\Conid{A} }} and
  \ensuremath{\Conid{F}\;(\Varid{f}\;\circ\;\Varid{g})\mathrel{=}\Conid{F}\;\Varid{f}\;\circ\;\Conid{F}\;\Varid{g}}.
  \qed
\end{definition}

\begin{definition}[Endofunctor]
  An \emph{endofunctor} \ensuremath{\Conid{F}\mathbin{:}\mathbb{C}\to \mathbb{C}} is a functor from a category to itself.
  The category of endofunctors on \ensuremath{\mathbb{C}} is denoted by \ensuremath{\mathbb{C}^\mathbb{C}}.
  \qed
\end{definition}

\begin{definition}[Pointed Endofunctor]
  An endofunctor \ensuremath{\Conid{P}\mathbin{:}\mathbb{C}\to \mathbb{C}} is \emph{pointed} if it is equipped with
  a designated natural transformation from the identity functor \ensuremath{\eta^P\mathbin{:}\mathit{Id}\to \Conid{P}}.
  We denote the category of pointed endofunctors by \ensuremath{\mathbb{P}}.
  \qed
\end{definition}

% % - - - - - - - - - - - - - - - - - - - - - - - - - - - - - - - - - - - - - - - -
% \subsubsection{Adjunctions}
%
% Adjunctions give rise to monads and form an important building block for our
% theoretical framework.

\begin{definition}[Adjunction]
  An \emph{adjunction} \ensuremath{\Conid{L}\dashv \Conid{R}} is a relation between two functors
  \ensuremath{\Conid{L}\mathbin{:}\mathbb{C}\to \mathbb{D}} and \ensuremath{\Conid{R}\mathbin{:}\mathbb{D}\to \mathbb{C}}, with \ensuremath{\mathbb{C}} and \ensuremath{\mathbb{D}} categories, so
  that there exists a natural isomorphism between their hom-sets:\begin{hscode}\SaveRestoreHook
\column{B}{@{}>{\hspre}l<{\hspost}@{}}%
\column{3}{@{}>{\hspre}l<{\hspost}@{}}%
\column{E}{@{}>{\hspre}l<{\hspost}@{}}%
\>[3]{}\psi\mathbin{:}{\mathbb{C}\kern-3pt}\;\bigl(\;\Conid{L}\;\Conid{A},\Conid{B}\;\bigr)\;\cong\;{\mathbb{D}\kern-3pt}\;\bigl(\;\Conid{A},\Conid{R}\;\Conid{B}\;\bigr)\mathbin{:}\psi^{-1}{}\<[E]%
\ColumnHook
\end{hscode}\resethooks
  Here, \ensuremath{\psi} and \ensuremath{\psi^{-1}} are two natural transformations witnessing the isomorphism,
  which should satisfy the requisite roundtrip properties (\ensuremath{\psi\;\circ\;\psi^{-1}\mathrel{=}id\mathrel{=}\psi^{-1}\;\circ\;\psi}).
\qed
\end{definition}

\noindent
Often, an adjunction is denoted by a diagram and \ensuremath{\Conid{L}} and \ensuremath{\Conid{R}} are called the
\emph{left adjoint} and \emph{right adjoint}, respectively.
\begin{center}
\begin{tikzcd}
{\ensuremath{\mathbb{D}}} & {\ensuremath{\bot}} & {\ensuremath{\mathbb{C}}}
\arrow["{L}"', from=1-3, to=1-1, shift right=3]
\arrow["{R}"', from=1-1, to=1-3, shift right=3]
\end{tikzcd}
\end{center}

\noindent
An adjunction gives rise to a monad \ensuremath{\Conid{R}\;\circ\;\Conid{L}} in \ensuremath{\mathbb{C}}, that is equipped with
a \emph{unit} operator and a sequencing operator.\begin{hscode}\SaveRestoreHook
\column{B}{@{}>{\hspre}l<{\hspost}@{}}%
\column{3}{@{}>{\hspre}l<{\hspost}@{}}%
\column{8}{@{}>{\hspre}l<{\hspost}@{}}%
\column{E}{@{}>{\hspre}l<{\hspost}@{}}%
\>[3]{}\eta{}\<[8]%
\>[8]{}\mathbin{:}\Conid{A}\to \Conid{R}\;(\Conid{L}\;\Conid{A}){}\<[E]%
\\
\>[3]{}\textsf{join}{}\<[8]%
\>[8]{}\mathbin{:}(\Conid{R}\;\circ\;\Conid{L}\;\circ\;\Conid{R}\;\circ\;\Conid{L})\;\Conid{A}\to \Conid{R}\;(\Conid{L}\;\Conid{A}){}\<[E]%
\ColumnHook
\end{hscode}\resethooks
Adjunctions can be combined:
we can compose \ensuremath{\Conid{L}_{1}\dashv \Conid{R}_{1}} and \ensuremath{\Conid{L}_{2}\dashv \Conid{R}_{2}}, composing their functors
such that \ensuremath{\Conid{L}_{2}\;\circ\;\Conid{L}_{1}\dashv \Conid{R}_{1}\;\circ\;\Conid{R}_{2}}.

\noindent
\begin{tikzcd}[column sep=scriptsize]
	{\ensuremath{\mathbb{C}_1}} & {\ensuremath{\bot}} & {\ensuremath{\mathbb{C}_2}} & {\ensuremath{\bot}} & {\ensuremath{\mathbb{C}_3}} & {\ensuremath{\cong}} & {\ensuremath{\mathbb{C}_1}} & {\ensuremath{\bot}} & {\ensuremath{\mathbb{C}_3}}
	\arrow["{\ensuremath{\Conid{L}_{2}}}"', shift right=3, from=1-3, to=1-1]
	\arrow["{\ensuremath{\Conid{R}_{2}}}"', shift right=3, from=1-1, to=1-3]
	\arrow["{\ensuremath{\Conid{L}_{1}}}"', shift right=3, from=1-5, to=1-3]
	\arrow["{\ensuremath{\Conid{R}_{1}}}"', shift right=3, from=1-3, to=1-5]
	\arrow["{\ensuremath{\Conid{L}_{2}\;\circ\;\Conid{L}_{1}}}"', shift right=3, from=1-9, to=1-7]
	\arrow["{\ensuremath{\Conid{R}_{1}\;\circ\;\Conid{R}_{2}}}"', shift right=3, from=1-7, to=1-9]
\end{tikzcd}

Furthermore, we can define the product of adjunctions \ensuremath{\Conid{L}_{1}\dashv \Conid{R}_{1}} and \ensuremath{\Conid{L}_{2}\dashv \Conid{R}_{2}},
acting between their product categories and taking the product of the adjoints.

\begin{center}
\begin{minipage}{0.3\textwidth}
  \begin{tikzcd}[column sep=small]
  	{\ensuremath{\mathbb{C}_1}} & {\ensuremath{\bot}} & {\ensuremath{\mathbb{C}_2}}
  	\arrow["{\ensuremath{\Conid{L}_{1}}}"', shift right=3, from=1-3, to=1-1]
  	\arrow["{\ensuremath{\Conid{R}_{1}}}"', shift right=3, from=1-1, to=1-3]
  \end{tikzcd}
  and
\end{minipage}%
\begin{minipage}{0.3\textwidth}
  \begin{tikzcd}[column sep=small]
  	{\ensuremath{\mathbb{C}_3}} & {\ensuremath{\bot}} & {\ensuremath{\mathbb{C}_4}}
  	\arrow["{\ensuremath{\Conid{L}_{2}}}"', shift right=3, from=1-3, to=1-1]
  	\arrow["{\ensuremath{\Conid{R}_{2}}}"', shift right=3, from=1-1, to=1-3]
  \end{tikzcd}
  then
\end{minipage}%
\begin{minipage}{0.4\textwidth}
  \begin{tikzcd}[column sep=small]
  	{\ensuremath{\mathbb{C}_1\times\mathbb{C}_3}} & {\ensuremath{\bot}} & {\ensuremath{\mathbb{C}_2\times\mathbb{C}_4}}
  	\arrow["{\ensuremath{\Conid{L}_{1}\times\Conid{L}_{2}}}"', shift right=3, from=1-3, to=1-1]
  	\arrow["{\ensuremath{\Conid{R}_{1}\times\Conid{R}_{2}}}"', shift right=3, from=1-1, to=1-3]
  \end{tikzcd}
\end{minipage}%
\end{center}

\begin{definition}[Free-forgetful Adjunction]
  In general, a free-forgetful adjunction consists of a \emph{free functor} \ensuremath{\Conid{Free}},
  left adjoint of a \emph{forgetful functor} \ensuremath{\Conid{U}}, which forgets some structure.

  \begin{center}
  \begin{tikzcd}
  {\ensuremath{\Sigma\mathit{-Alg}(\mathbb{C})}} & {\ensuremath{\bot}} & {\ensuremath{\mathbb{C}}}
  \arrow["{\mathit{Free}}"', from=1-3, to=1-1, shift right=3]
  \arrow["{U}"', from=1-1, to=1-3, shift right=3]
  \end{tikzcd}
  \end{center}

  \noindent
  \begin{minipage}{0.65\textwidth}
  Here, we consider an endofunctor \ensuremath{\Sigma\mathbin{:}\mathbb{C}\to \mathbb{C}} in category \ensuremath{\mathbb{C}}.
  The category on the left is the category of \ensuremath{\Sigma}-algebras over \ensuremath{\mathbb{C}}, with
  as objects pairs \ensuremath{\langle \Conid{A},\Varid{alg}\rangle }, where \ensuremath{\Conid{A}} is called the \emph{carrier} of the
  algebra and \ensuremath{\Varid{alg}\mathbin{:}\Sigma\;\Conid{A}\to \Conid{A}} is the \emph{action}.
  Homomorphisms (Figure \ref{fig:comm}) \ensuremath{\Varid{f}\mathbin{:}\Conid{A}\to \Conid{B}} in \ensuremath{\Sigma\mathit{-Alg}(\mathbb{C})} map the carriers.
  \end{minipage}%
  \begin{minipage}{0.35\textwidth}
  \begin{center}
    \begin{tikzcd}[row sep=small]
    	{\ensuremath{\Sigma\;\Conid{A}}} & {\ensuremath{\Sigma\;\Conid{B}}} \\
    	{\ensuremath{\Conid{A}}} & {\ensuremath{\Conid{B}}}
    	\arrow["{\ensuremath{alg_B}}", from=1-2, to=2-2]
    	\arrow["{\ensuremath{alg_A}}"', from=1-1, to=2-1]
    	\arrow["{\ensuremath{\Varid{f}}}"', from=2-1, to=2-2]
    	\arrow["{\ensuremath{\Sigma\;\Varid{f}}}", from=1-1, to=1-2]
    \end{tikzcd}
    \label{fig:comm}
    \,\\
    \textbf{Fig. \ref{fig:comm}.} Actions are structure-preserving.
  \end{center}
  \end{minipage}
  \qed
\end{definition}

\noindent
Now, category \ensuremath{\Sigma\mathit{-Alg}(\mathbb{C})} also has an initial object.
Following Lambek's theorem, the action of the initial object is an isomorphism
(\ensuremath{\Sigma\;\Varid{i}\;\cong\;\Varid{i}} with \ensuremath{\Varid{i}} the initial object).
This means that the initial object is a fixed point of \ensuremath{\Sigma}. %, denoted by |SigStar|.
Therefore, we consider the algebra for which \ensuremath{\Sigma^\star\;\Conid{A}} is the carrier;
it has an action \ensuremath{\Conid{Op}\mathbin{:}\Sigma\;(\Sigma^\star\;\Conid{A})\to \Sigma^\star\;\Conid{A}}.
Furthermore, the definition of initial objects states that there must be a
unique homomorphism from the initial object to any other \ensuremath{\Sigma}-algebra.
We call this unique homomorphism a \emph{fold} and describe the uniqueness by
the following two diagrams, which must commute.

\begin{center}
\begin{tikzcd}[row sep=scriptsize]
	{\ensuremath{\Conid{A}}} & {\ensuremath{\Sigma^\star\;\Conid{A}}} && {\ensuremath{\Sigma\;(\Sigma^\star\;\Conid{A})}} & {\ensuremath{\Sigma^\star\;\Conid{A}}} \\
	{\ensuremath{\Conid{B}}} &&& {\ensuremath{\Sigma\;\Conid{B}}} & {\ensuremath{\Conid{B}}}
	\arrow["{\ensuremath{\Conid{Var}}}", from=1-1, to=1-2]
	\arrow["{\ensuremath{\Varid{gen}}}"', from=1-1, to=2-1]
	\arrow["{\ensuremath{\textsf{fold}}}", from=1-2, to=2-1]
	\arrow["{\ensuremath{\Conid{Op}}}", from=1-4, to=1-5]
	\arrow["{\ensuremath{\textsf{fold}}}", from=1-5, to=2-5]
	\arrow["{\ensuremath{\Varid{alg}}}"', from=2-4, to=2-5]
	\arrow["{\ensuremath{\Sigma\;\textsf{fold}}}"', from=1-4, to=2-4]
\end{tikzcd}
\end{center}

\noindent
This uniqueness is called the \emph{universal property} of fold.
This fold is determined in terms of the generator \ensuremath{\Varid{gen}\mathbin{:}\Conid{A}\to \Conid{B}}, which is a morphism
in the category of \ensuremath{\Sigma}-algebras, and the action of the target \ensuremath{\Sigma}-algebra.
% We often denote fold by |ll gen, alg rr|.
%
The free-forgetful adjunction gives rise to the free monad
in \ensuremath{\mathbb{C}}, generated by \ensuremath{\Sigma}; this free monad is often denoted by \ensuremath{\Sigma^\star}.
An alternative notation for a fold with generator \ensuremath{\Varid{gen}} and algebra \ensuremath{\Varid{alg}} is
\ensuremath{\llparenthesis\;\Varid{gen},\Varid{alg}\;\rrparenthesis}.
Notice the similarity between this adjunction and our free monad definition in \Cref{sec:aeh},
where we denote \ensuremath{\Sigma^\star} as \ensuremath{\Conid{Free}\;\sigma},
we use \ensuremath{\Conid{Var}} as the return of the monad, \ensuremath{\Conid{Op}} as the action of the initial object,
and \ensuremath{\textsf{fold}_{\textsf{Alg}}\;\Varid{gen}\;\Varid{alg}} as the structural recursion scheme to interpret \ensuremath{\Sigma^\star\;\Conid{A}}
into \ensuremath{\Conid{B}}, with the implementation matching the above commuting diagrams.

The model of \emph{algebraic effects \& handlers} revolves around the free monad
definition (effects) and its interpretation in terms of a fold-style recursion scheme (handlers).
Operations are captured by an endofunctor \ensuremath{\Sigma\mathbin{:}\mathbb{C}\to \mathbb{C}} on the base category \ensuremath{\mathbb{C}}.
This endofunctor is exactly the functor we have described in the above free-forgetful
adjunction.
Furthermore, it represents the effect signature and determines the shapes of
programs that call \ensuremath{\Sigma}-operations, and the shapes of handlers that interpret them.

%- - - - - - - - - - - - - - - - - - - - - - - - - - - - - - - - - - - - - - - -
\section{Two Ways of Modifying the Log with \ensuremath{\textsf{censor}}}
\label{app:censor}

We can write \ensuremath{\textsf{censor}} in terms of \ensuremath{\textsf{pass}}.
The difference between these two operations is that \ensuremath{\textsf{censor}} gets the modifying function
separately, whereas in \ensuremath{\textsf{pass}}, it is part of the computation result.
Thus, we can easily define \ensuremath{\textsf{censor}} in terms of \ensuremath{\textsf{pass}} as follows:
\begin{hscode}\SaveRestoreHook
\column{B}{@{}>{\hspre}l<{\hspost}@{}}%
\column{3}{@{}>{\hspre}l<{\hspost}@{}}%
\column{E}{@{}>{\hspre}l<{\hspost}@{}}%
\>[3]{}\textsf{censor}\mathbin{::}(\Varid{w}\to \Varid{w})\to \Varid{m}\;\Varid{a}\to \Varid{m}\;\Varid{a}{}\<[E]%
\\
\>[3]{}\textsf{censor}\;\Varid{f}\;\Varid{k}\mathrel{=}\textsf{pass}\mathbin{\$}\mathbf{do}\;\Varid{x}\leftarrow \Varid{k};\textsf{return}\;(\Varid{x},\Varid{f}){}\<[E]%
\ColumnHook
\end{hscode}\resethooks
From this definition, it follows that we can rewrite our \ensuremath{\Varid{reset}} example
in terms of \ensuremath{\textsf{censor}}.
\begin{hscode}\SaveRestoreHook
\column{B}{@{}>{\hspre}l<{\hspost}@{}}%
\column{3}{@{}>{\hspre}l<{\hspost}@{}}%
\column{E}{@{}>{\hspre}l<{\hspost}@{}}%
\>[3]{}\Varid{reset}\mathrel{=}\textsf{censor}\;(\Varid{const}\;\epsilon)\;(\textsf{return}\;()){}\<[E]%
\ColumnHook
\end{hscode}\resethooks
The result remains the same
\begin{hscode}\SaveRestoreHook
\column{B}{@{}>{\hspre}l<{\hspost}@{}}%
\column{3}{@{}>{\hspre}l<{\hspost}@{}}%
\column{E}{@{}>{\hspre}l<{\hspost}@{}}%
\>[3]{}\texttt{>>>}\;h_{\textsf{Write}}\;(\textsf{tell}\;\text{\ttfamily \char34 post\char34}\sequ \Varid{reset}\sequ \textsf{tell}\;\text{\ttfamily \char34 pre\char34}){}\<[E]%
\\
\>[3]{}((),\text{\ttfamily \char34 post\char34}){}\<[E]%
\ColumnHook
\end{hscode}\resethooks
Alternatively, we can think of this function as
the dual of \ensuremath{\textsf{local}} for reading from an environment,
where \ensuremath{\textsf{local}} first modifies the environment and then executes a computation.
Dually, \ensuremath{\textsf{censor}} executes a computation and then modifies the output.
Both \ensuremath{\textsf{local}} and \ensuremath{\textsf{censor}} can be modeled as a scoped computation.
For example, we can extend the above handler with scoped effect \ensuremath{\Conid{Censor}\;\Varid{w}}.
\begin{hscode}\SaveRestoreHook
\column{B}{@{}>{\hspre}l<{\hspost}@{}}%
\column{3}{@{}>{\hspre}l<{\hspost}@{}}%
\column{E}{@{}>{\hspre}l<{\hspost}@{}}%
\>[3]{}\mathbf{data}\;\Conid{Censor}\;\Varid{w}\;\Varid{a}\mathrel{=}\Conid{Censor}\;(\Varid{w}\to \Varid{w})\;\Varid{a}{}\<[E]%
\ColumnHook
\end{hscode}\resethooks

Furthermore, we extend our handler accordingly:
\begin{hscode}\SaveRestoreHook
\column{B}{@{}>{\hspre}l<{\hspost}@{}}%
\column{3}{@{}>{\hspre}l<{\hspost}@{}}%
\column{6}{@{}>{\hspre}l<{\hspost}@{}}%
\column{9}{@{}>{\hspre}l<{\hspost}@{}}%
\column{12}{@{}>{\hspre}l<{\hspost}@{}}%
\column{33}{@{}>{\hspre}l<{\hspost}@{}}%
\column{35}{@{}>{\hspre}l<{\hspost}@{}}%
\column{E}{@{}>{\hspre}l<{\hspost}@{}}%
\>[3]{}h_{\textsf{Censor}}{}\<[12]%
\>[12]{}\mathbin{::}(\Conid{Functor}\;\sigma,\Conid{Functor}\;\gamma,\Conid{Monoid}\;\Varid{w},\Conid{Functor}\;\varphi){}\<[E]%
\\
\>[3]{}\hsindent{3}{}\<[6]%
\>[6]{}\Rightarrow Free_{\textsf{H}}\;(K^{\textsf{Alg}}\;(\Conid{Tell}\;\Varid{w}\mathrel{{+}}\sigma){}\<[35]%
\>[35]{}\mathrel{\oplus}\colorbox{lightgray}{$K^{\textsf{Sc}}\;(\Conid{Censor}\;\Varid{w}\mathrel{{+}}\gamma)$}{}\<[E]%
\\
\>[35]{}\mathrel{\oplus}K^{\textsf{Write}}\;(\Conid{Listen}\;\Varid{w}\mathrel{{+}}\Conid{Pass}\;\Varid{w}\mathrel{{+}}\varphi))\;\Varid{a}{}\<[E]%
\\
\>[3]{}\hsindent{3}{}\<[6]%
\>[6]{}\to Free_{\textsf{H}}\;(K^{\textsf{Alg}}\;\sigma\mathrel{\oplus}K^{\textsf{Sc}}\;\gamma\mathrel{\oplus}K^{\textsf{Write}}\;\varphi)\;(\Varid{a},\Varid{w}){}\<[E]%
\\
\>[3]{}h_{\textsf{Censor}}{}\<[12]%
\>[12]{}\mathrel{=}\textsf{fold}\;\Varid{gen}\;(alg_{\textsf{Alg}}\kern+2pt\tikz[baseline=(char.base)]{ \node[circle,draw,inner sep=0pt,align=center,scale=0.3] (char) {\Huge{\#}};}\kern+2ptalg_{\textsf{Sc}}\kern+2pt\tikz[baseline=(char.base)]{ \node[circle,draw,inner sep=0pt,align=center,scale=0.3] (char) {\Huge{\#}};}\kern+2ptalg_{\textsf{Write}})\;\mathbf{where}{}\<[E]%
\\
\>[3]{}\hsindent{3}{}\<[6]%
\>[6]{}alg_{\textsf{Sc}}\;(\textsf{Enter}\;\Varid{sc})\mathrel{=}(alg_{\textsf{Censor}}\mathrel{{\#}}fwd_{\textsf{Censor}})\;\Varid{sc}\;\mathbf{where}{}\<[E]%
\\
\>[6]{}\hsindent{3}{}\<[9]%
\>[9]{}alg_{\textsf{Censor}}\;(\Conid{Censor}\;\Varid{f}\;\Varid{k}){}\<[33]%
\>[33]{}\mathrel{=}\mathbf{do}\;(\Varid{mx},\anonymous )\leftarrow \Varid{k};(\Varid{x},\Varid{w})\leftarrow \Varid{mx};\textsf{return}\;(\Varid{x},\Varid{f}\;\Varid{w}){}\<[E]%
\\
\>[6]{}\hsindent{3}{}\<[9]%
\>[9]{}fwd_{\textsf{Censor}}{}\<[33]%
\>[33]{}\mathrel{=}Op_{\textsf{H}}\, .\,\textsf{Enter}\, .\,\textsf{fmap}\;(\textsf{fmap}\;\Varid{fst}){}\<[E]%
\\
\>[3]{}\hsindent{3}{}\<[6]%
\>[6]{}\mathbin{...}{}\<[E]%
\ColumnHook
\end{hscode}\resethooks
The example computation gives the same result as before.
\begin{hscode}\SaveRestoreHook
\column{B}{@{}>{\hspre}l<{\hspost}@{}}%
\column{3}{@{}>{\hspre}l<{\hspost}@{}}%
\column{E}{@{}>{\hspre}l<{\hspost}@{}}%
\>[3]{}\texttt{>>>}\;\colorbox{lightgray}{$h_{\textsf{Censor}}$}\;(\textsf{tell}\;\text{\ttfamily \char34 post\char34}\sequ \Varid{reset}\sequ \textsf{tell}\;\text{\ttfamily \char34 pre\char34}){}\<[E]%
\\
\>[3]{}((),\text{\ttfamily \char34 post\char34}){}\<[E]%
\ColumnHook
\end{hscode}\resethooks

\section{Isomorphisms}
\label{app:iso}

\subsection{Algebraic Effects \& Handlers}

First, we define the isomorphisms:
\begin{hscode}\SaveRestoreHook
\column{B}{@{}>{\hspre}l<{\hspost}@{}}%
\column{3}{@{}>{\hspre}l<{\hspost}@{}}%
\column{17}{@{}>{\hspre}l<{\hspost}@{}}%
\column{E}{@{}>{\hspre}l<{\hspost}@{}}%
\>[3]{}iso_1\mathbin{::}\Conid{Functor}\;\sigma\Rightarrow \Conid{Free}\;\sigma\;\Varid{a}\to Free_{\textsf{H}}\;(K^{\textsf{Alg}}\;\sigma)\;\Varid{a}{}\<[E]%
\\
\>[3]{}iso_1\;(\Conid{Var}\;\Varid{x}){}\<[17]%
\>[17]{}\mathrel{=}Var_{\textsf{H}}\;\Varid{x}{}\<[E]%
\\
\>[3]{}iso_1\;(\Conid{Op}\;\Varid{op}){}\<[17]%
\>[17]{}\mathrel{=}Op_{\textsf{H}}\;(\textsf{Op}\;(\textsf{fmap}\;iso_1\;\Varid{op})){}\<[E]%
\ColumnHook
\end{hscode}\resethooks
\begin{hscode}\SaveRestoreHook
\column{B}{@{}>{\hspre}l<{\hspost}@{}}%
\column{3}{@{}>{\hspre}l<{\hspost}@{}}%
\column{24}{@{}>{\hspre}l<{\hspost}@{}}%
\column{E}{@{}>{\hspre}l<{\hspost}@{}}%
\>[3]{}iso_2\mathbin{::}\Conid{Functor}\;\sigma\Rightarrow Free_{\textsf{H}}\;(K^{\textsf{Alg}}\;\sigma)\;\Varid{a}\to \Conid{Free}\;\sigma\;\Varid{a}{}\<[E]%
\\
\>[3]{}iso_2\;(Var_{\textsf{H}}\;\Varid{x}){}\<[24]%
\>[24]{}\mathrel{=}\Conid{Var}\;\Varid{x}{}\<[E]%
\\
\>[3]{}iso_2\;(Op_{\textsf{H}}\;(\textsf{Op}\;\Varid{op})){}\<[24]%
\>[24]{}\mathrel{=}\Conid{Op}\;(\textsf{fmap}\;iso_2\;\Varid{op}){}\<[E]%
\ColumnHook
\end{hscode}\resethooks
And next, we show that the requisite roundtrip properties hold,
i.e., \ensuremath{iso_1\, .\,iso_2\mathrel{=}id\mathrel{=}iso_2\, .\,iso_2}.

\noindent
\begin{minipage}{0.5\textwidth}
\begin{hscode}\SaveRestoreHook
\column{B}{@{}>{\hspre}l<{\hspost}@{}}%
\column{3}{@{}>{\hspre}l<{\hspost}@{}}%
\column{4}{@{}>{\hspre}c<{\hspost}@{}}%
\column{4E}{@{}l@{}}%
\column{8}{@{}>{\hspre}l<{\hspost}@{}}%
\column{E}{@{}>{\hspre}l<{\hspost}@{}}%
\>[3]{}(iso_1\, .\,iso_2)\;(Var_{\textsf{H}}\;\Varid{x}){}\<[E]%
\\
\>[3]{}\hsindent{1}{}\<[4]%
\>[4]{}\equiv {}\<[4E]%
\>[8]{}iso_1\;(\Conid{Var}\;\Varid{x}){}\<[E]%
\\
\>[3]{}\hsindent{1}{}\<[4]%
\>[4]{}\equiv {}\<[4E]%
\>[8]{}Var_{\textsf{H}}\;\Varid{x}{}\<[E]%
\\[\blanklineskip]%
\>[3]{}(iso_1\, .\,iso_2)\;(Op_{\textsf{H}}\;(\textsf{Op}\;\Varid{op})){}\<[E]%
\\
\>[3]{}\hsindent{1}{}\<[4]%
\>[4]{}\equiv {}\<[4E]%
\>[8]{}iso_1\;(\Conid{Op}\;(\textsf{fmap}\;iso_2\;\Varid{op})){}\<[E]%
\\
\>[3]{}\hsindent{1}{}\<[4]%
\>[4]{}\equiv {}\<[4E]%
\>[8]{}Op_{\textsf{H}}\;(\textsf{Op}\;(\textsf{fmap}\;iso_1\;(\textsf{fmap}\;iso_2\;\Varid{op}))){}\<[E]%
\\
\>[3]{}\hsindent{1}{}\<[4]%
\>[4]{}\equiv {}\<[4E]%
\>[8]{}Op_{\textsf{H}}\;(\textsf{Op}\;(\textsf{fmap}\;(iso_1\, .\,iso_2)\;\Varid{op})){}\<[E]%
\\
\>[3]{}\hsindent{1}{}\<[4]%
\>[4]{}\equiv {}\<[4E]%
\>[8]{}Op_{\textsf{H}}\;(\textsf{Op}\;\Varid{op}){}\<[E]%
\ColumnHook
\end{hscode}\resethooks
\end{minipage}%
\begin{minipage}{0.5\textwidth}
\begin{hscode}\SaveRestoreHook
\column{B}{@{}>{\hspre}l<{\hspost}@{}}%
\column{3}{@{}>{\hspre}l<{\hspost}@{}}%
\column{4}{@{}>{\hspre}c<{\hspost}@{}}%
\column{4E}{@{}l@{}}%
\column{8}{@{}>{\hspre}l<{\hspost}@{}}%
\column{E}{@{}>{\hspre}l<{\hspost}@{}}%
\>[3]{}(iso_2\, .\,iso_1)\;(\Conid{Var}\;\Varid{x}){}\<[E]%
\\
\>[3]{}\hsindent{1}{}\<[4]%
\>[4]{}\equiv {}\<[4E]%
\>[8]{}iso_2\;(Var_{\textsf{H}}\;\Varid{x}){}\<[E]%
\\
\>[3]{}\hsindent{1}{}\<[4]%
\>[4]{}\equiv {}\<[4E]%
\>[8]{}\Conid{Var}\;\Varid{x}{}\<[E]%
\\[\blanklineskip]%
\>[3]{}(iso_2\, .\,iso_1)\;(\Conid{Op}\;\Varid{op}){}\<[E]%
\\
\>[3]{}\hsindent{1}{}\<[4]%
\>[4]{}\equiv {}\<[4E]%
\>[8]{}iso_2\;(Op_{\textsf{H}}\;(\textsf{Op}\;(\textsf{fmap}\;iso_1\;\Varid{op}))){}\<[E]%
\\
\>[3]{}\hsindent{1}{}\<[4]%
\>[4]{}\equiv {}\<[4E]%
\>[8]{}\Conid{Op}\;(\textsf{fmap}\;iso_2\;(\textsf{fmap}\;iso_1\;\Varid{op})){}\<[E]%
\\
\>[3]{}\hsindent{1}{}\<[4]%
\>[4]{}\equiv {}\<[4E]%
\>[8]{}\Conid{Op}\;(\textsf{fmap}\;(iso_2\, .\,iso_1)\;\Varid{op}){}\<[E]%
\\
\>[3]{}\hsindent{1}{}\<[4]%
\>[4]{}\equiv {}\<[4E]%
\>[8]{}\Conid{Op}\;\Varid{op}{}\<[E]%
\ColumnHook
\end{hscode}\resethooks
\end{minipage}%

\noindent
Furthermore, the handlers for these two recursive datatypes satisfy the following
equivalence: \ensuremath{\textsf{fold}_{\textsf{Alg}}\;\Varid{gen}\;\Varid{alg}\mathrel{=}h_{\textsf{Alg}}\;\Varid{gen}\;(\lambda (\textsf{Op}\;\Varid{op})\to \Varid{alg}\;\Varid{op})\, .\,iso_1}.

% < hAlg gen alg t = foldAlg gen (alg . Op_) (iso2 t)
%
% > fold' :: (Pointed g, Functor sig) => (a -> g b) -> (forall x . sig x -> x) -> Free sig a -> g b
% > fold' gen alg t = hAlg gen (\(Op_ op) -> alg op) (iso1 t)
\begin{hscode}\SaveRestoreHook
\column{B}{@{}>{\hspre}l<{\hspost}@{}}%
\column{3}{@{}>{\hspre}l<{\hspost}@{}}%
\column{4}{@{}>{\hspre}l<{\hspost}@{}}%
\column{E}{@{}>{\hspre}l<{\hspost}@{}}%
\>[3]{}h_{\textsf{Alg}}\;\Varid{gen}\;(\lambda (\textsf{Op}\;\Varid{op})\to \Varid{alg}\;\Varid{op})\;(iso_1\;(\Conid{Var}\;\Varid{x})){}\<[E]%
\\
\>[3]{}\hsindent{1}{}\<[4]%
\>[4]{}\equiv h_{\textsf{Alg}}\;\Varid{gen}\;(\lambda (\textsf{Op}\;\Varid{op})\to \Varid{alg}\;\Varid{op})\;(Var_{\textsf{H}}\;\Varid{x}){}\<[E]%
\\
\>[3]{}\hsindent{1}{}\<[4]%
\>[4]{}\equiv \Varid{gen}\;\Varid{x}{}\<[E]%
\\
\>[3]{}\hsindent{1}{}\<[4]%
\>[4]{}\equiv \textsf{fold}_{\textsf{Alg}}\;\Varid{gen}\;\Varid{alg}\;(\Conid{Var}\;\Varid{x}){}\<[E]%
\ColumnHook
\end{hscode}\resethooks
\begin{hscode}\SaveRestoreHook
\column{B}{@{}>{\hspre}l<{\hspost}@{}}%
\column{3}{@{}>{\hspre}l<{\hspost}@{}}%
\column{4}{@{}>{\hspre}l<{\hspost}@{}}%
\column{9}{@{}>{\hspre}l<{\hspost}@{}}%
\column{22}{@{}>{\hspre}l<{\hspost}@{}}%
\column{E}{@{}>{\hspre}l<{\hspost}@{}}%
\>[3]{}h_{\textsf{Alg}}\;\Varid{gen}\;(\lambda (\textsf{Op}\;\Varid{op})\to \Varid{alg}\;\Varid{op})\;(iso_1\;(\Conid{Op}\;\Varid{x})){}\<[E]%
\\
\>[3]{}\hsindent{1}{}\<[4]%
\>[4]{}\equiv h_{\textsf{Alg}}\;\Varid{gen}\;(\lambda (\textsf{Op}\;\Varid{op})\to \Varid{alg}\;\Varid{op})\;(Op_{\textsf{H}}\;(\textsf{Op}\;(\textsf{fmap}\;iso_1\;\Varid{x}))){}\<[E]%
\\
\>[3]{}\hsindent{1}{}\<[4]%
\>[4]{}\equiv (\lambda (\textsf{Op}\;\Varid{op})\to {}\<[22]%
\>[22]{}\Varid{alg}\;\Varid{op})\;{}\<[E]%
\\
\>[4]{}\hsindent{5}{}\<[9]%
\>[9]{}(\textsf{hmap}\;\textsf{fold}_{2}\;(\textsf{fmap}\;(h_{\textsf{Alg}}\;\Varid{gen}\;(\lambda (\textsf{Op}\;\Varid{op})\to \Varid{alg}\;\Varid{op}))\;(\textsf{Op}\;(\textsf{fmap}\;iso_1\;\Varid{x})))){}\<[E]%
\\
\>[3]{}\hsindent{1}{}\<[4]%
\>[4]{}\equiv (\lambda (\textsf{Op}\;\Varid{op})\to {}\<[22]%
\>[22]{}\Varid{alg}\;\Varid{op})\;{}\<[E]%
\\
\>[4]{}\hsindent{5}{}\<[9]%
\>[9]{}(\textsf{hmap}\;\textsf{fold}_{2}\;(\textsf{Op}\;(\textsf{fmap}\;(h_{\textsf{Alg}}\;\Varid{gen}\;(\lambda (\textsf{Op}\;\Varid{op})\to \Varid{alg}\;\Varid{op})\;(\textsf{fmap}\;iso_1\;\Varid{x}))))){}\<[E]%
\\
\>[3]{}\hsindent{1}{}\<[4]%
\>[4]{}\equiv (\lambda (\textsf{Op}\;\Varid{op})\to {}\<[22]%
\>[22]{}\Varid{alg}\;\Varid{op})\;{}\<[E]%
\\
\>[4]{}\hsindent{5}{}\<[9]%
\>[9]{}(\textsf{Op}\;(\textsf{fmap}\;(h_{\textsf{Alg}}\;\Varid{gen}\;(\lambda (\textsf{Op}\;\Varid{op})\to \Varid{alg}\;\Varid{op}))\;(\textsf{fmap}\;iso_1\;\Varid{x})))){}\<[E]%
\\
\>[3]{}\hsindent{1}{}\<[4]%
\>[4]{}\equiv \Varid{alg}\;(\textsf{fmap}\;(h_{\textsf{Alg}}\;\Varid{gen}\;(\lambda (\textsf{Op}\;\Varid{op})\to \Varid{alg}\;\Varid{op}))\;(\textsf{fmap}\;iso_1\;\Varid{x}))){}\<[E]%
\\
\>[3]{}\hsindent{1}{}\<[4]%
\>[4]{}\equiv \Varid{alg}\;(\textsf{fmap}\;(h_{\textsf{Alg}}\;\Varid{gen}\;(\lambda (\textsf{Op}\;\Varid{op})\to \Varid{alg}\;\Varid{op})\, .\,iso_1)\;\Varid{x})){}\<[E]%
\\
\>[3]{}\hsindent{1}{}\<[4]%
\>[4]{}\equiv \Varid{alg}\;(\textsf{fmap}\;(\textsf{fold}_{\textsf{Alg}}\;\Varid{gen}\;\Varid{alg})\;\Varid{x})){}\<[E]%
\\
\>[3]{}\hsindent{1}{}\<[4]%
\>[4]{}\equiv \textsf{fold}_{\textsf{Alg}}\;\Varid{gen}\;\Varid{alg}\;(\Conid{Op}\;\Varid{x}){}\<[E]%
\ColumnHook
\end{hscode}\resethooks
\subsection{Scoped Effects \& Handlers}

First, we define the isomorphisms:
\begin{hscode}\SaveRestoreHook
\column{B}{@{}>{\hspre}l<{\hspost}@{}}%
\column{3}{@{}>{\hspre}l<{\hspost}@{}}%
\column{20}{@{}>{\hspre}l<{\hspost}@{}}%
\column{E}{@{}>{\hspre}l<{\hspost}@{}}%
\>[3]{}iso_1\mathbin{::}\Conid{Functor}\;\gamma\Rightarrow Free_{\textsf{Sc}}\;\gamma\;\Varid{a}\to Free_{\textsf{H}}\;(K^{\textsf{Sc}}\;\gamma)\;\Varid{a}{}\<[E]%
\\
\>[3]{}iso_1\;(Var\;\Varid{x}){}\<[20]%
\>[20]{}\mathrel{=}Var_{\textsf{H}}\;\Varid{x}{}\<[E]%
\\
\>[3]{}iso_1\;(\Conid{Enter}\;\Varid{sc}){}\<[20]%
\>[20]{}\mathrel{=}Op_{\textsf{H}}\;(\textsf{Enter}\;(\textsf{fmap}\;(iso_1\, .\,\textsf{fmap}\;iso_1)\;\Varid{sc})){}\<[E]%
\ColumnHook
\end{hscode}\resethooks
\begin{hscode}\SaveRestoreHook
\column{B}{@{}>{\hspre}l<{\hspost}@{}}%
\column{3}{@{}>{\hspre}l<{\hspost}@{}}%
\column{27}{@{}>{\hspre}l<{\hspost}@{}}%
\column{E}{@{}>{\hspre}l<{\hspost}@{}}%
\>[3]{}iso_2\mathbin{::}\Conid{Functor}\;\gamma\Rightarrow Free_{\textsf{H}}\;(K^{\textsf{Sc}}\;\gamma)\;\Varid{a}\to Free_{\textsf{Sc}}\;\gamma\;\Varid{a}{}\<[E]%
\\
\>[3]{}iso_2\;(Var_{\textsf{H}}\;\Varid{x}){}\<[27]%
\>[27]{}\mathrel{=}Var\;\Varid{x}{}\<[E]%
\\
\>[3]{}iso_2\;(Op_{\textsf{H}}\;(\textsf{Enter}\;\Varid{sc})){}\<[27]%
\>[27]{}\mathrel{=}\Conid{Enter}\;(\textsf{fmap}\;(iso_2\, .\,\textsf{fmap}\;iso_2)\;\Varid{sc}){}\<[E]%
\ColumnHook
\end{hscode}\resethooks
And next, we show that the requisite roundtrip properties hold,
i.e., \ensuremath{iso_1\, .\,iso_2\mathrel{=}id\mathrel{=}iso_2\, .\,iso_2}.
\begin{hscode}\SaveRestoreHook
\column{B}{@{}>{\hspre}l<{\hspost}@{}}%
\column{3}{@{}>{\hspre}l<{\hspost}@{}}%
\column{4}{@{}>{\hspre}c<{\hspost}@{}}%
\column{4E}{@{}l@{}}%
\column{8}{@{}>{\hspre}l<{\hspost}@{}}%
\column{E}{@{}>{\hspre}l<{\hspost}@{}}%
\>[3]{}(iso_1\, .\,iso_2)\;(Var_{\textsf{H}}\;\Varid{x}){}\<[E]%
\\
\>[3]{}\hsindent{1}{}\<[4]%
\>[4]{}\equiv {}\<[4E]%
\>[8]{}iso_1\;(Var\;\Varid{x}){}\<[E]%
\\
\>[3]{}\hsindent{1}{}\<[4]%
\>[4]{}\equiv {}\<[4E]%
\>[8]{}Var_{\textsf{H}}\;\Varid{x}{}\<[E]%
\ColumnHook
\end{hscode}\resethooks
\begin{hscode}\SaveRestoreHook
\column{B}{@{}>{\hspre}l<{\hspost}@{}}%
\column{3}{@{}>{\hspre}l<{\hspost}@{}}%
\column{4}{@{}>{\hspre}c<{\hspost}@{}}%
\column{4E}{@{}l@{}}%
\column{8}{@{}>{\hspre}l<{\hspost}@{}}%
\column{E}{@{}>{\hspre}l<{\hspost}@{}}%
\>[3]{}(iso_1\, .\,iso_2)\;(Op_{\textsf{H}}\;(\textsf{Enter}\;\Varid{sc})){}\<[E]%
\\
\>[3]{}\hsindent{1}{}\<[4]%
\>[4]{}\equiv {}\<[4E]%
\>[8]{}iso_1\;(\Conid{Enter}\;(\textsf{fmap}\;(iso_2\, .\,\textsf{fmap}\;iso_2)\;\Varid{sc})){}\<[E]%
\\
\>[3]{}\hsindent{1}{}\<[4]%
\>[4]{}\equiv {}\<[4E]%
\>[8]{}Op_{\textsf{H}}\;(\textsf{Enter}\;(\textsf{fmap}\;(iso_1\, .\,\textsf{fmap}\;iso_1))\;(\textsf{fmap}\;(iso_2\, .\,\textsf{fmap}\;iso_2)\;\Varid{sc})){}\<[E]%
\\
\>[3]{}\hsindent{1}{}\<[4]%
\>[4]{}\equiv {}\<[4E]%
\>[8]{}Op_{\textsf{H}}\;(\textsf{Enter}\;(\textsf{fmap}\;(iso_1\, .\,\textsf{fmap}\;iso_1\, .\,iso_2\, .\,\textsf{fmap}\;iso_2)\;\Varid{sc})){}\<[E]%
\\
\>[3]{}\hsindent{1}{}\<[4]%
\>[4]{}\equiv {}\<[4E]%
\>[8]{}Op_{\textsf{H}}\;(\textsf{Enter}\;(\textsf{fmap}\;(iso_1\, .\,iso_2\, .\,\textsf{fmap}\;iso_1\, .\,\textsf{fmap}\;iso_2)\;\Varid{sc})){}\<[E]%
\\
\>[3]{}\hsindent{1}{}\<[4]%
\>[4]{}\equiv {}\<[4E]%
\>[8]{}Op_{\textsf{H}}\;(\textsf{Enter}\;(\textsf{fmap}\;(iso_1\, .\,iso_2\, .\,\textsf{fmap}\;(iso_1\, .\,iso_2))\;\Varid{sc})){}\<[E]%
\\
\>[3]{}\hsindent{1}{}\<[4]%
\>[4]{}\equiv {}\<[4E]%
\>[8]{}Op_{\textsf{H}}\;(\textsf{Enter}\;\Varid{sc}){}\<[E]%
\ColumnHook
\end{hscode}\resethooks
\begin{hscode}\SaveRestoreHook
\column{B}{@{}>{\hspre}l<{\hspost}@{}}%
\column{3}{@{}>{\hspre}l<{\hspost}@{}}%
\column{4}{@{}>{\hspre}c<{\hspost}@{}}%
\column{4E}{@{}l@{}}%
\column{8}{@{}>{\hspre}l<{\hspost}@{}}%
\column{E}{@{}>{\hspre}l<{\hspost}@{}}%
\>[3]{}(iso_2\, .\,iso_1)\;(Var\;\Varid{x}){}\<[E]%
\\
\>[3]{}\hsindent{1}{}\<[4]%
\>[4]{}\equiv {}\<[4E]%
\>[8]{}iso_2\;(Var_{\textsf{H}}\;\Varid{x}){}\<[E]%
\\
\>[3]{}\hsindent{1}{}\<[4]%
\>[4]{}\equiv {}\<[4E]%
\>[8]{}Var\;\Varid{x}{}\<[E]%
\ColumnHook
\end{hscode}\resethooks
\begin{hscode}\SaveRestoreHook
\column{B}{@{}>{\hspre}l<{\hspost}@{}}%
\column{3}{@{}>{\hspre}l<{\hspost}@{}}%
\column{4}{@{}>{\hspre}c<{\hspost}@{}}%
\column{4E}{@{}l@{}}%
\column{8}{@{}>{\hspre}l<{\hspost}@{}}%
\column{E}{@{}>{\hspre}l<{\hspost}@{}}%
\>[3]{}(iso_2\, .\,iso_1)\;(\Conid{Enter}\;\Varid{sc}){}\<[E]%
\\
\>[3]{}\hsindent{1}{}\<[4]%
\>[4]{}\equiv {}\<[4E]%
\>[8]{}iso_2\;(Op_{\textsf{H}}\;(\textsf{Enter}\;(\textsf{fmap}\;(iso_1\, .\,\textsf{fmap}\;iso_1)\;\Varid{sc}))){}\<[E]%
\\
\>[3]{}\hsindent{1}{}\<[4]%
\>[4]{}\equiv {}\<[4E]%
\>[8]{}\Conid{Enter}\;(\textsf{fmap}\;(iso_2\, .\,\textsf{fmap}\;iso_2)\;(\textsf{fmap}\;(iso_1\, .\,\textsf{fmap}\;iso_1)\;\Varid{sc})){}\<[E]%
\\
\>[3]{}\hsindent{1}{}\<[4]%
\>[4]{}\equiv {}\<[4E]%
\>[8]{}\Conid{Enter}\;(\textsf{fmap}\;(iso_2\, .\,\textsf{fmap}\;iso_2\, .\,iso_1\, .\,\textsf{fmap}\;iso_1)\;\Varid{sc}){}\<[E]%
\\
\>[3]{}\hsindent{1}{}\<[4]%
\>[4]{}\equiv {}\<[4E]%
\>[8]{}\Conid{Enter}\;(\textsf{fmap}\;(iso_2\, .\,iso_1\, .\,\textsf{fmap}\;iso_2\, .\,\textsf{fmap}\;iso_1)\;\Varid{sc}){}\<[E]%
\\
\>[3]{}\hsindent{1}{}\<[4]%
\>[4]{}\equiv {}\<[4E]%
\>[8]{}\Conid{Enter}\;(\textsf{fmap}\;(iso_2\, .\,iso_1\, .\,\textsf{fmap}\;(iso_2\, .\,iso_1))\;\Varid{sc}){}\<[E]%
\\
\>[3]{}\hsindent{1}{}\<[4]%
\>[4]{}\equiv {}\<[4E]%
\>[8]{}\Conid{Enter}\;\Varid{sc}{}\<[E]%
\ColumnHook
\end{hscode}\resethooks
\noindent
Furthermore, the handlers for these two recursive datatypes satisfy the following
equivalence: \ensuremath{\textsf{fold}_{\textsf{Sc}}\;\Varid{gen}\;alg_{E}\;alg_{E}\mathrel{=}h_{\textsf{Sc}}\;\Varid{gen}\;(\lambda (\textsf{Enter}\;\Varid{sc})\to enter_E\;alg_{E}\;\Varid{sc})\, .\,iso_1}
with \ensuremath{return_E\;alg_{E}\mathrel{=}\eta}.

% > hSc'  :: (Functor gam, Pointed g)
% >       => (a -> g b) -> (forall x . KSc gam g (g x) -> g x) -> T (KSc gam) a -> g b
% > hSc' gen alg = foldSc gen (EndoAlg eta (alg . Enter_)) (BaseAlg (alg . Enter_)) . iso2
%
% > foldSc' :: (Functor gam, Pointed g)
% >         => (a -> g b) -> EndoAlg gam g -> BaseAlg gam g (g b) -> FreeSc gam a -> g b
% > foldSc' gen ealg balg = hSc gen (\(Enter_ sc) -> enterE ealg sc) . iso1
\begin{hscode}\SaveRestoreHook
\column{B}{@{}>{\hspre}l<{\hspost}@{}}%
\column{3}{@{}>{\hspre}l<{\hspost}@{}}%
\column{4}{@{}>{\hspre}l<{\hspost}@{}}%
\column{E}{@{}>{\hspre}l<{\hspost}@{}}%
\>[3]{}h_{\textsf{Sc}}\;\Varid{gen}\;(\lambda (\textsf{Enter}\;\Varid{sc})\to enter_E\;alg_{E}\;\Varid{sc})\;(iso_1\;(Var\;\Varid{x})){}\<[E]%
\\
\>[3]{}\hsindent{1}{}\<[4]%
\>[4]{}\equiv h_{\textsf{Sc}}\;\Varid{gen}\;(\lambda (\textsf{Enter}\;\Varid{sc})\to enter_E\;alg_{E}\;\Varid{sc})\;(Var_{\textsf{H}}\;\Varid{x}){}\<[E]%
\\
\>[3]{}\hsindent{1}{}\<[4]%
\>[4]{}\equiv \Varid{gen}\;\Varid{x}{}\<[E]%
\\
\>[3]{}\hsindent{1}{}\<[4]%
\>[4]{}\equiv \textsf{fold}_{\textsf{Sc}}\;\Varid{gen}\;alg_{E}\;alg_{E}\;(Var\;\Varid{x}){}\<[E]%
\ColumnHook
\end{hscode}\resethooks
\begin{hscode}\SaveRestoreHook
\column{B}{@{}>{\hspre}l<{\hspost}@{}}%
\column{3}{@{}>{\hspre}l<{\hspost}@{}}%
\column{4}{@{}>{\hspre}l<{\hspost}@{}}%
\column{16}{@{}>{\hspre}l<{\hspost}@{}}%
\column{25}{@{}>{\hspre}l<{\hspost}@{}}%
\column{28}{@{}>{\hspre}l<{\hspost}@{}}%
\column{44}{@{}>{\hspre}l<{\hspost}@{}}%
\column{45}{@{}>{\hspre}l<{\hspost}@{}}%
\column{E}{@{}>{\hspre}l<{\hspost}@{}}%
\>[3]{}h_{\textsf{Sc}}\;\Varid{gen}\;(\lambda (\textsf{Enter}\;\Varid{sc})\to enter_E\;alg_{E}\;\Varid{sc})\;(iso_1\;(\Conid{Enter}\;\Varid{sc})){}\<[E]%
\\
\>[3]{}\hsindent{1}{}\<[4]%
\>[4]{}\equiv h_{\textsf{Sc}}\;\Varid{gen}\;{}\<[16]%
\>[16]{}(\lambda (\textsf{Enter}\;\Varid{sc})\to enter_E\;alg_{E}\;\Varid{sc})\;{}\<[E]%
\\
\>[16]{}(Op_{\textsf{H}}\;(\textsf{Enter}\;(\textsf{fmap}\;(iso_1\, .\,\textsf{fmap}\;iso_1)\;\Varid{sc}))){}\<[E]%
\\
\>[3]{}\hsindent{1}{}\<[4]%
\>[4]{}\equiv (\lambda (\textsf{Enter}\;\Varid{sc})\to {}\<[25]%
\>[25]{}enter_E\;alg_{E}\;\Varid{sc})\;{}\<[E]%
\\
\>[25]{}(\textsf{hmap}\;\textsf{fold}_{2}\;(\textsf{fmap}\;{}\<[44]%
\>[44]{}(h_{\textsf{Sc}}\;\Varid{gen}\;(\lambda (\textsf{Enter}\;\Varid{sc})\to enter_E\;alg_{E}\;\Varid{sc}))\;{}\<[E]%
\\
\>[44]{}(\textsf{Enter}\;(\textsf{fmap}\;(iso_1\, .\,\textsf{fmap}\;iso_1)\;\Varid{sc})))){}\<[E]%
\\
\>[3]{}\hsindent{1}{}\<[4]%
\>[4]{}\equiv (\lambda (\textsf{Enter}\;\Varid{sc})\to {}\<[25]%
\>[25]{}enter_E\;alg_{E}\;\Varid{sc})\;{}\<[E]%
\\
\>[25]{}(\textsf{hmap}\;\textsf{fold}_{2}\;(\textsf{Enter}\;(\textsf{fmap}\;(\textsf{fmap}\;(h_{\textsf{Sc}}\;\Varid{gen}{}\<[E]%
\\
\>[25]{}\hsindent{20}{}\<[45]%
\>[45]{}(\lambda (\textsf{Enter}\;\Varid{sc})\to enter_E\;alg_{E}\;\Varid{sc}))))\;{}\<[E]%
\\
\>[25]{}\hsindent{20}{}\<[45]%
\>[45]{}(\textsf{fmap}\;(iso_1\, .\,\textsf{fmap}\;iso_1)\;\Varid{sc}))){}\<[E]%
\\
\>[3]{}\hsindent{1}{}\<[4]%
\>[4]{}\equiv (\lambda (\textsf{Enter}\;\Varid{sc})\to {}\<[25]%
\>[25]{}enter_E\;alg_{E}\;\Varid{sc})\;{}\<[E]%
\\
\>[25]{}(\textsf{hmap}\;\textsf{fold}_{2}\;(\textsf{Enter}\;(\textsf{fmap}{}\<[E]%
\\
\>[25]{}\hsindent{3}{}\<[28]%
\>[28]{}(\textsf{fmap}\;(h_{\textsf{Sc}}\;\Varid{gen}\;(\lambda (\textsf{Enter}\;\Varid{sc})\to enter_E\;alg_{E}\;\Varid{sc}))\, .\,{}\<[E]%
\\
\>[28]{}\hsindent{17}{}\<[45]%
\>[45]{}iso_1\, .\,\textsf{fmap}\;iso_1)\;\Varid{sc}))){}\<[E]%
\\
\>[3]{}\hsindent{1}{}\<[4]%
\>[4]{}\equiv (\lambda (\textsf{Enter}\;\Varid{sc})\to {}\<[25]%
\>[25]{}enter_E\;alg_{E}\;\Varid{sc})\;{}\<[E]%
\\
\>[25]{}(\textsf{Enter}\;(\textsf{fmap}\;\textsf{fold}_{2}{}\<[E]%
\\
\>[25]{}\hsindent{3}{}\<[28]%
\>[28]{}(\textsf{fmap}\;(\textsf{fmap}\;(h_{\textsf{Sc}}\;\Varid{gen}\;(\lambda (\textsf{Enter}\;\Varid{sc})\to enter_E\;alg_{E}\;\Varid{sc}))\, .\,{}\<[E]%
\\
\>[28]{}\hsindent{17}{}\<[45]%
\>[45]{}iso_1\, .\,\textsf{fmap}\;iso_1)\;\Varid{sc}))){}\<[E]%
\\
\>[3]{}\hsindent{1}{}\<[4]%
\>[4]{}\equiv enter_E\;alg_{E}\;(\textsf{fmap}\;\textsf{fold}_{2}\;(\textsf{fmap}\;(\textsf{fmap}\;(h_{\textsf{Sc}}\;\Varid{gen}\;(\lambda (\textsf{Enter}\;\Varid{sc})\to enter_E\;alg_{E}\;\Varid{sc}))\, .\,{}\<[E]%
\\
\>[4]{}\hsindent{41}{}\<[45]%
\>[45]{}iso_1\, .\,\textsf{fmap}\;iso_1)\;\Varid{sc})){}\<[E]%
\\
\>[3]{}\hsindent{1}{}\<[4]%
\>[4]{}\equiv enter_E\;alg_{E}\;(\textsf{fmap}\;(\textsf{fold}_{2}\, .\,\textsf{fmap}\;(h_{\textsf{Sc}}\;\Varid{gen}\;(\lambda (\textsf{Enter}\;\Varid{sc})\to enter_E\;alg_{E}\;\Varid{sc}))\, .\,{}\<[E]%
\\
\>[4]{}\hsindent{41}{}\<[45]%
\>[45]{}iso_1\, .\,\textsf{fmap}\;iso_1)\;\Varid{sc})){}\<[E]%
\\
\>[3]{}\hsindent{1}{}\<[4]%
\>[4]{}\equiv enter_E\;alg_{E}\;(\textsf{fmap}\;(\textsf{fold}_{2}\, .\,\textsf{fmap}\;(h_{\textsf{Sc}}\;\Varid{gen}\;(\lambda (\textsf{Enter}\;\Varid{sc})\to enter_E\;alg_{E}\;\Varid{sc}))\, .\,{}\<[E]%
\\
\>[4]{}\hsindent{41}{}\<[45]%
\>[45]{}\textsf{fmap}\;iso_1\, .\,iso_1)\;\Varid{sc})){}\<[E]%
\\
\>[3]{}\hsindent{1}{}\<[4]%
\>[4]{}\equiv enter_E\;alg_{E}\;(\textsf{fmap}\;(\textsf{fold}_{2}\, .\,\textsf{fmap}\;(h_{\textsf{Sc}}\;\Varid{gen}\;(\lambda (\textsf{Enter}\;\Varid{sc})\to enter_E\;alg_{E}\;\Varid{sc})\, .\,{}\<[E]%
\\
\>[4]{}\hsindent{41}{}\<[45]%
\>[45]{}iso_1)\, .\,iso_1)\;\Varid{sc})){}\<[E]%
\\
\>[3]{}\hsindent{1}{}\<[4]%
\>[4]{}\equiv enter_E\;alg_{E}\;(\textsf{fmap}\;(\textsf{fold}_{2}\, .\,\textsf{fmap}\;(\textsf{fold}_{\textsf{Sc}}\;\Varid{gen}\;alg_{E}\;alg_{E})\, .\,iso_1)\;\Varid{sc})){}\<[E]%
\\
\>[3]{}\hsindent{1}{}\<[4]%
\>[4]{}\equiv enter_E\;alg_{E}\;(\textsf{fmap}\;(\textsf{fold}_{2}\, .\,iso_1\, .\,\textsf{fmap}\;(\textsf{fold}_{\textsf{Sc}}\;\Varid{gen}\;alg_{E}\;alg_{E}))\;\Varid{sc})){}\<[E]%
\\
\>[3]{}\hsindent{1}{}\<[4]%
\>[4]{}\equiv enter_E\;alg_{E}\;(\textsf{fmap}\;(h_{cata}\;alg_{E}\, .\,\textsf{fmap}\;(\textsf{fold}_{\textsf{Sc}}\;\Varid{gen}\;alg_{E}\;alg_{E}))\;\Varid{sc})){}\<[E]%
\\
\>[3]{}\hsindent{1}{}\<[4]%
\>[4]{}\equiv \textsf{fold}_{\textsf{Sc}}\;\Varid{gen}\;alg_{E}\;alg_{E}\;(\Conid{Enter}\;\Varid{sc}){}\<[E]%
\ColumnHook
\end{hscode}\resethooks
We need a helper lemma that states that \ensuremath{h_{cata}\;alg_{E}\mathrel{=}\textsf{fold}_{2}\, .\,iso_1} with \ensuremath{return_E\;alg_{E}\mathrel{=}\eta}.
\begin{hscode}\SaveRestoreHook
\column{B}{@{}>{\hspre}l<{\hspost}@{}}%
\column{3}{@{}>{\hspre}l<{\hspost}@{}}%
\column{4}{@{}>{\hspre}l<{\hspost}@{}}%
\column{E}{@{}>{\hspre}l<{\hspost}@{}}%
\>[3]{}h_{cata}\;alg_{E}\;(Var\;\Varid{x}){}\<[E]%
\\
\>[3]{}\hsindent{1}{}\<[4]%
\>[4]{}\equiv return_E\;alg_{E}\;\Varid{x}{}\<[E]%
\\
\>[3]{}\hsindent{1}{}\<[4]%
\>[4]{}\equiv \eta\;\Varid{x}{}\<[E]%
\\
\>[3]{}\hsindent{1}{}\<[4]%
\>[4]{}\equiv \textsf{fold}_{2}\;(Var_{\textsf{H}}\;\Varid{x}){}\<[E]%
\\
\>[3]{}\hsindent{1}{}\<[4]%
\>[4]{}\equiv \textsf{fold}_{2}\;(iso_1\;(Var\;\Varid{x})){}\<[E]%
\ColumnHook
\end{hscode}\resethooks
\begin{hscode}\SaveRestoreHook
\column{B}{@{}>{\hspre}l<{\hspost}@{}}%
\column{3}{@{}>{\hspre}l<{\hspost}@{}}%
\column{4}{@{}>{\hspre}l<{\hspost}@{}}%
\column{E}{@{}>{\hspre}l<{\hspost}@{}}%
\>[3]{}h_{cata}\;alg_{E}\;(\Conid{Enter}\;\Varid{sc}){}\<[E]%
\\
\>[3]{}\hsindent{1}{}\<[4]%
\>[4]{}\equiv (enter_E\;alg_{E}\, .\,\textsf{fmap}\;(h_{cata}\;alg_{E}\, .\,\textsf{fmap}\;(h_{cata}\;alg_{E})))\;\Varid{sc}{}\<[E]%
\\
\>[3]{}\hsindent{1}{}\<[4]%
\>[4]{}\equiv (enter_E\;alg_{E}\, .\,\textsf{fmap}\;(h_{cata}\;alg_{E}\, .\,\textsf{fmap}\;(h_{cata}\;alg_{E})))\;\Varid{sc}{}\<[E]%
\\
\>[3]{}\hsindent{1}{}\<[4]%
\>[4]{}\equiv (\Varid{alg}\, .\,\textsf{Enter}\, .\,\textsf{fmap}\;(h_{cata}\;alg_{E}\, .\,\textsf{fmap}\;(h_{cata}\;alg_{E})))\;\Varid{sc}{}\<[E]%
\\
\>[3]{}\hsindent{1}{}\<[4]%
\>[4]{}\equiv \Varid{alg}\;(\textsf{Enter}\;(\textsf{fmap}\;(h_{cata}\;alg_{E}\, .\,\textsf{fmap}\;(h_{cata}\;alg_{E}))\;\Varid{sc})){}\<[E]%
\\
\>[3]{}\hsindent{1}{}\<[4]%
\>[4]{}\equiv \Varid{alg}\;(\textsf{Enter}\;(\textsf{fmap}\;(\textsf{fold}_{2}\, .\,iso_1\, .\,\textsf{fmap}\;(h_{cata}\;alg_{E}))\;\Varid{sc})){}\<[E]%
\\
\>[3]{}\hsindent{1}{}\<[4]%
\>[4]{}\equiv \Varid{alg}\;(\textsf{Enter}\;(\textsf{fmap}\;(\textsf{fold}_{2}\, .\,\textsf{fmap}\;(h_{cata}\;alg_{E})\, .\,iso_1)\;\Varid{sc})){}\<[E]%
\\
\>[3]{}\hsindent{1}{}\<[4]%
\>[4]{}\equiv \Varid{alg}\;(\textsf{Enter}\;(\textsf{fmap}\;\textsf{fold}_{2}\;(\textsf{fmap}\;(\textsf{fmap}\;(h_{cata}\;alg_{E})\, .\,iso_1)\;\Varid{sc}))){}\<[E]%
\\
\>[3]{}\hsindent{1}{}\<[4]%
\>[4]{}\equiv \Varid{alg}\;(\textsf{hmap}\;\textsf{fold}_{2}\;(\textsf{Enter}\;(\textsf{fmap}\;(\textsf{fmap}\;(h_{cata}\;alg_{E})\, .\,iso_1)\;\Varid{sc}))){}\<[E]%
\\
\>[3]{}\hsindent{1}{}\<[4]%
\>[4]{}\equiv \Varid{alg}\;(\textsf{hmap}\;\textsf{fold}_{2}\;(\textsf{Enter}\;(\textsf{fmap}\;(\textsf{fmap}\;(\textsf{fold}_{2}\, .\,iso_1)\, .\,iso_1)\;\Varid{sc}))){}\<[E]%
\\
\>[3]{}\hsindent{1}{}\<[4]%
\>[4]{}\equiv \Varid{alg}\;(\textsf{hmap}\;\textsf{fold}_{2}\;(\textsf{Enter}\;(\textsf{fmap}\;(\textsf{fmap}\;\textsf{fold}_{2}\, .\,iso_1\, .\,\textsf{fmap}\;iso_1)\;\Varid{sc}))){}\<[E]%
\\
\>[3]{}\hsindent{1}{}\<[4]%
\>[4]{}\equiv \Varid{alg}\;(\textsf{hmap}\;\textsf{fold}_{2}\;(\textsf{Enter}\;(\textsf{fmap}\;(\textsf{fmap}\;\textsf{fold}_{2})\;(\textsf{fmap}\;(iso_1\, .\,\textsf{fmap}\;iso_1)\;\Varid{sc})))){}\<[E]%
\\
\>[3]{}\hsindent{1}{}\<[4]%
\>[4]{}\equiv \Varid{alg}\;(\textsf{hmap}\;\textsf{fold}_{2}\;(\textsf{fmap}\;\textsf{fold}_{2}\;(\textsf{Enter}\;(\textsf{fmap}\;(iso_1\, .\,\textsf{fmap}\;iso_1)\;\Varid{sc})))){}\<[E]%
\\
\>[3]{}\hsindent{1}{}\<[4]%
\>[4]{}\equiv \textsf{fold}_{2}\;(Op_{\textsf{H}}\;(\textsf{Enter}\;(\textsf{fmap}\;(iso_1\, .\,\textsf{fmap}\;iso_1)\;\Varid{sc}))){}\<[E]%
\\
\>[3]{}\hsindent{1}{}\<[4]%
\>[4]{}\equiv \textsf{fold}_{2}\;(iso_1\;(\Conid{Enter}\;\Varid{sc})){}\<[E]%
\ColumnHook
\end{hscode}\resethooks

\subsection{Parallel Effects \& Handlers}

First, we define the isomorphisms:
\begin{hscode}\SaveRestoreHook
\column{B}{@{}>{\hspre}l<{\hspost}@{}}%
\column{3}{@{}>{\hspre}l<{\hspost}@{}}%
\column{23}{@{}>{\hspre}l<{\hspost}@{}}%
\column{E}{@{}>{\hspre}l<{\hspost}@{}}%
\>[3]{}iso_1\mathbin{::}\Conid{Functor}\;\rho\Rightarrow Free_{\textsf{Par}}\;\rho\;\Varid{a}\to Free_{\textsf{H}}\;(K^{\textsf{Par}}\;\rho)\;\Varid{a}{}\<[E]%
\\
\>[3]{}iso_1\;(Var\;\Varid{x}){}\<[23]%
\>[23]{}\mathrel{=}Var_{\textsf{H}}\;\Varid{x}{}\<[E]%
\\
\>[3]{}iso_1\;(\Conid{For}\;\Varid{iters}\;\Varid{k}){}\<[23]%
\>[23]{}\mathrel{=}Op_{\textsf{H}}\;(\textsf{For}\;(\textsf{fmap}\;iso_1\;\Varid{iters})\;(\textsf{fmap}\;iso_1\;\Varid{k})){}\<[E]%
\ColumnHook
\end{hscode}\resethooks
\begin{hscode}\SaveRestoreHook
\column{B}{@{}>{\hspre}l<{\hspost}@{}}%
\column{3}{@{}>{\hspre}l<{\hspost}@{}}%
\column{30}{@{}>{\hspre}l<{\hspost}@{}}%
\column{E}{@{}>{\hspre}l<{\hspost}@{}}%
\>[3]{}iso_2\mathbin{::}\Conid{Functor}\;\rho\Rightarrow Free_{\textsf{H}}\;(K^{\textsf{Par}}\;\rho)\;\Varid{a}\to Free_{\textsf{Par}}\;\rho\;\Varid{a}{}\<[E]%
\\
\>[3]{}iso_2\;(Var_{\textsf{H}}\;\Varid{x}){}\<[30]%
\>[30]{}\mathrel{=}Var\;\Varid{x}{}\<[E]%
\\
\>[3]{}iso_2\;(Op_{\textsf{H}}\;(\textsf{For}\;\Varid{iters}\;\Varid{k})){}\<[30]%
\>[30]{}\mathrel{=}\Conid{For}\;(\textsf{fmap}\;iso_2\;\Varid{iters})\;(\textsf{fmap}\;iso_2\;\Varid{k}){}\<[E]%
\ColumnHook
\end{hscode}\resethooks
And next, we show that the requisite roundtrip properties hold,
i.e., \ensuremath{iso_1\, .\,iso_2\mathrel{=}id\mathrel{=}iso_2\, .\,iso_2}.
\begin{hscode}\SaveRestoreHook
\column{B}{@{}>{\hspre}l<{\hspost}@{}}%
\column{3}{@{}>{\hspre}l<{\hspost}@{}}%
\column{4}{@{}>{\hspre}c<{\hspost}@{}}%
\column{4E}{@{}l@{}}%
\column{8}{@{}>{\hspre}l<{\hspost}@{}}%
\column{E}{@{}>{\hspre}l<{\hspost}@{}}%
\>[3]{}(iso_1\, .\,iso_2)\;(Var_{\textsf{H}}\;\Varid{x}){}\<[E]%
\\
\>[3]{}\hsindent{1}{}\<[4]%
\>[4]{}\equiv {}\<[4E]%
\>[8]{}iso_1\;(Var\;\Varid{x}){}\<[E]%
\\
\>[3]{}\hsindent{1}{}\<[4]%
\>[4]{}\equiv {}\<[4E]%
\>[8]{}Var_{\textsf{H}}\;\Varid{x}{}\<[E]%
\ColumnHook
\end{hscode}\resethooks
\begin{hscode}\SaveRestoreHook
\column{B}{@{}>{\hspre}l<{\hspost}@{}}%
\column{3}{@{}>{\hspre}l<{\hspost}@{}}%
\column{4}{@{}>{\hspre}c<{\hspost}@{}}%
\column{4E}{@{}l@{}}%
\column{8}{@{}>{\hspre}l<{\hspost}@{}}%
\column{E}{@{}>{\hspre}l<{\hspost}@{}}%
\>[3]{}(iso_1\, .\,iso_2)\;(Op_{\textsf{H}}\;(\textsf{For}\;\Varid{iters}\;\Varid{k})){}\<[E]%
\\
\>[3]{}\hsindent{1}{}\<[4]%
\>[4]{}\equiv {}\<[4E]%
\>[8]{}iso_1\;(\Conid{For}\;(\textsf{fmap}\;iso_2\;\Varid{iters})\;(\textsf{fmap}\;iso_2\;\Varid{k})){}\<[E]%
\\
\>[3]{}\hsindent{1}{}\<[4]%
\>[4]{}\equiv {}\<[4E]%
\>[8]{}Op_{\textsf{H}}\;(\textsf{For}\;(\textsf{fmap}\;iso_1\;(\textsf{fmap}\;iso_2\;\Varid{iters}))\;(\textsf{fmap}\;iso_1\;(\textsf{fmap}\;iso_2\;\Varid{k}))){}\<[E]%
\\
\>[3]{}\hsindent{1}{}\<[4]%
\>[4]{}\equiv {}\<[4E]%
\>[8]{}Op_{\textsf{H}}\;(\textsf{For}\;(\textsf{fmap}\;(iso_1\, .\,iso_2)\;\Varid{iters})\;(\textsf{fmap}\;(iso_1\, .\,iso_2)\;\Varid{k})){}\<[E]%
\\
\>[3]{}\hsindent{1}{}\<[4]%
\>[4]{}\equiv {}\<[4E]%
\>[8]{}Op_{\textsf{H}}\;(\textsf{For}\;\Varid{iters}\;\Varid{k}){}\<[E]%
\ColumnHook
\end{hscode}\resethooks
\begin{hscode}\SaveRestoreHook
\column{B}{@{}>{\hspre}l<{\hspost}@{}}%
\column{3}{@{}>{\hspre}l<{\hspost}@{}}%
\column{4}{@{}>{\hspre}c<{\hspost}@{}}%
\column{4E}{@{}l@{}}%
\column{8}{@{}>{\hspre}l<{\hspost}@{}}%
\column{E}{@{}>{\hspre}l<{\hspost}@{}}%
\>[3]{}(iso_2\, .\,iso_1)\;(Var\;\Varid{x}){}\<[E]%
\\
\>[3]{}\hsindent{1}{}\<[4]%
\>[4]{}\equiv {}\<[4E]%
\>[8]{}iso_2\;(Var_{\textsf{H}}\;\Varid{x}){}\<[E]%
\\
\>[3]{}\hsindent{1}{}\<[4]%
\>[4]{}\equiv {}\<[4E]%
\>[8]{}Var\;\Varid{x}{}\<[E]%
\ColumnHook
\end{hscode}\resethooks
\begin{hscode}\SaveRestoreHook
\column{B}{@{}>{\hspre}l<{\hspost}@{}}%
\column{3}{@{}>{\hspre}l<{\hspost}@{}}%
\column{4}{@{}>{\hspre}c<{\hspost}@{}}%
\column{4E}{@{}l@{}}%
\column{8}{@{}>{\hspre}l<{\hspost}@{}}%
\column{E}{@{}>{\hspre}l<{\hspost}@{}}%
\>[3]{}(iso_2\, .\,iso_1)\;(\Conid{For}\;\Varid{iters}\;\Varid{k}){}\<[E]%
\\
\>[3]{}\hsindent{1}{}\<[4]%
\>[4]{}\equiv {}\<[4E]%
\>[8]{}iso_2\;(Op_{\textsf{H}}\;(\textsf{For}\;(\textsf{fmap}\;iso_1\;\Varid{iters})\;(\textsf{fmap}\;iso_1\;\Varid{k}))){}\<[E]%
\\
\>[3]{}\hsindent{1}{}\<[4]%
\>[4]{}\equiv {}\<[4E]%
\>[8]{}\Conid{For}\;(\textsf{fmap}\;iso_2\;(\textsf{fmap}\;iso_1\;\Varid{iters}))\;(\textsf{fmap}\;iso_2\;(\textsf{fmap}\;iso_1\;\Varid{k})){}\<[E]%
\\
\>[3]{}\hsindent{1}{}\<[4]%
\>[4]{}\equiv {}\<[4E]%
\>[8]{}\Conid{For}\;(\textsf{fmap}\;(iso_2\, .\,iso_1)\;\Varid{iters})\;(\textsf{fmap}\;(iso_2\, .\,iso_1)\;\Varid{k}){}\<[E]%
\\
\>[3]{}\hsindent{1}{}\<[4]%
\>[4]{}\equiv {}\<[4E]%
\>[8]{}\Conid{For}\;\Varid{iters}\;\Varid{k}{}\<[E]%
\ColumnHook
\end{hscode}\resethooks
We define a handler for \ensuremath{Free_{\textsf{Par}}\;\rho\;\Varid{a}} as follows:
\begin{hscode}\SaveRestoreHook
\column{B}{@{}>{\hspre}l<{\hspost}@{}}%
\column{3}{@{}>{\hspre}l<{\hspost}@{}}%
\column{33}{@{}>{\hspre}l<{\hspost}@{}}%
\column{34}{@{}>{\hspre}l<{\hspost}@{}}%
\column{42}{@{}>{\hspre}c<{\hspost}@{}}%
\column{42E}{@{}l@{}}%
\column{46}{@{}>{\hspre}l<{\hspost}@{}}%
\column{47}{@{}>{\hspre}l<{\hspost}@{}}%
\column{58}{@{}>{\hspre}l<{\hspost}@{}}%
\column{E}{@{}>{\hspre}l<{\hspost}@{}}%
\>[3]{}\mathbf{data}\;Alg_{\textsf{Par}}\;\rho\;\Varid{f}\mathrel{=}Alg_{\textsf{Par}}\;\{\mskip1.5mu {}\<[33]%
\>[33]{}h_{\textsf{Var}}{}\<[42]%
\>[42]{}\mathbin{::}{}\<[42E]%
\>[46]{}\forall\kern-2pt\;\Varid{a}{}\<[58]%
\>[58]{}\, .\,\Varid{a}\to \Varid{f}\;\Varid{a},{}\<[E]%
\\
\>[33]{}h_{\textsf{For}}{}\<[42]%
\>[42]{}\mathbin{::}{}\<[42E]%
\>[46]{}\forall\kern-2pt\;\Varid{a}\;\Varid{b}{}\<[58]%
\>[58]{}\, .\,\rho\;(\Varid{f}\;\Varid{b})\to (\rho\;\Varid{b}\to \Varid{f}\;\Varid{a})\to \Varid{f}\;\Varid{a}\mskip1.5mu\}{}\<[E]%
\\[\blanklineskip]%
\>[3]{}\textsf{fold}_{\textsf{Par}}\mathbin{::}(\Conid{Functor}\;\rho,\Conid{Pointed}\;\Varid{f})\Rightarrow (\Varid{a}\to \Varid{f}\;\Varid{b})\to Alg_{\textsf{Par}}\;\rho\;\Varid{f}\to Free_{\textsf{Par}}\;\rho\;\Varid{a}\to \Varid{f}\;\Varid{b}{}\<[E]%
\\
\>[3]{}\textsf{fold}_{\textsf{Par}}\;\Varid{gen}\;\Varid{alg}\;(Var\;\Varid{x}){}\<[34]%
\>[34]{}\mathrel{=}\Varid{gen}\;\Varid{x}{}\<[E]%
\\
\>[3]{}\textsf{fold}_{\textsf{Par}}\;\Varid{gen}\;\Varid{alg}\;(\Conid{For}\;\Varid{iters}\;\Varid{k}){}\<[34]%
\>[34]{}\mathrel{=}h_{\textsf{For}}\;\Varid{alg}\;{}\<[47]%
\>[47]{}(\textsf{fmap}\;(\textsf{fold}_{\textsf{Par}}\;(h_{\textsf{Var}}\;\Varid{alg})\;\Varid{alg})\;\Varid{iters})\;{}\<[E]%
\\
\>[47]{}(\textsf{fold}_{\textsf{Par}}\;\Varid{gen}\;\Varid{alg}\, .\,\Varid{k}){}\<[E]%
\ColumnHook
\end{hscode}\resethooks
Furthermore, we can prove that there is an isomorphism between the handlers \ensuremath{h_{\textsf{Par}}} and \ensuremath{\textsf{fold}_{\textsf{Par}}}:
\ensuremath{\textsf{fold}_{\textsf{Par}}\;\Varid{gen}\;\Varid{alg}\mathrel{=}h_{\textsf{Par}}\;\Varid{gen}\;(\lambda (\textsf{For}\;\Varid{iters}\;\Varid{k})\to h_{\textsf{For}}\;\Varid{alg}\;\Varid{iters}\;\Varid{k})\, .\,iso_1}.
\begin{hscode}\SaveRestoreHook
\column{B}{@{}>{\hspre}l<{\hspost}@{}}%
\column{3}{@{}>{\hspre}l<{\hspost}@{}}%
\column{4}{@{}>{\hspre}l<{\hspost}@{}}%
\column{E}{@{}>{\hspre}l<{\hspost}@{}}%
\>[3]{}h_{\textsf{Par}}\;\Varid{gen}\;(\lambda (\textsf{For}\;\Varid{iters}\;\Varid{k})\to h_{\textsf{For}}\;\Varid{alg}\;\Varid{iters}\;\Varid{k})\;(iso_1\;(Var\;\Varid{x})){}\<[E]%
\\
\>[3]{}\hsindent{1}{}\<[4]%
\>[4]{}\equiv h_{\textsf{Par}}\;\Varid{gen}\;(\lambda (\textsf{For}\;\Varid{iters}\;\Varid{k})\to h_{\textsf{For}}\;\Varid{alg}\;\Varid{iters}\;\Varid{k})\;(Var_{\textsf{H}}\;\Varid{x}){}\<[E]%
\\
\>[3]{}\hsindent{1}{}\<[4]%
\>[4]{}\equiv \Varid{gen}\;(Var_{\textsf{H}}\;\Varid{x}){}\<[E]%
\\
\>[3]{}\hsindent{1}{}\<[4]%
\>[4]{}\equiv \textsf{fold}_{\textsf{Par}}\;\Varid{gen}\;\Varid{alg}\;(Var\;\Varid{x}){}\<[E]%
\ColumnHook
\end{hscode}\resethooks
\begin{hscode}\SaveRestoreHook
\column{B}{@{}>{\hspre}l<{\hspost}@{}}%
\column{3}{@{}>{\hspre}l<{\hspost}@{}}%
\column{4}{@{}>{\hspre}l<{\hspost}@{}}%
\column{12}{@{}>{\hspre}l<{\hspost}@{}}%
\column{17}{@{}>{\hspre}l<{\hspost}@{}}%
\column{19}{@{}>{\hspre}l<{\hspost}@{}}%
\column{28}{@{}>{\hspre}l<{\hspost}@{}}%
\column{31}{@{}>{\hspre}l<{\hspost}@{}}%
\column{66}{@{}>{\hspre}l<{\hspost}@{}}%
\column{E}{@{}>{\hspre}l<{\hspost}@{}}%
\>[3]{}h_{\textsf{Par}}\;\Varid{gen}\;(\lambda (\textsf{For}\;\Varid{iters}\;\Varid{k})\to h_{\textsf{For}}\;\Varid{alg}\;\Varid{iters}\;\Varid{k})\;(iso_1\;(\Conid{For}\;\Varid{iters}\;\Varid{k})){}\<[E]%
\\
\>[3]{}\hsindent{1}{}\<[4]%
\>[4]{}\equiv h_{\textsf{Par}}\;\Varid{gen}\;{}\<[17]%
\>[17]{}(\lambda (\textsf{For}\;\Varid{iters}\;\Varid{k})\to h_{\textsf{For}}\;\Varid{alg}\;\Varid{iters}\;\Varid{k})\;{}\<[E]%
\\
\>[17]{}(Op_{\textsf{H}}\;(\textsf{For}\;(\textsf{fmap}\;iso_1\;\Varid{iters})\;(iso_1\, .\,\Varid{k}))){}\<[E]%
\\
\>[3]{}\hsindent{1}{}\<[4]%
\>[4]{}\equiv (\lambda (\textsf{For}\;\Varid{iters}\;\Varid{k})\to {}\<[28]%
\>[28]{}h_{\textsf{For}}\;\Varid{alg}\;\Varid{iters}\;\Varid{k})\;{}\<[E]%
\\
\>[4]{}\hsindent{8}{}\<[12]%
\>[12]{}(\textsf{hmap}\;\textsf{fold}_{2}\;(\textsf{fmap}\;{}\<[31]%
\>[31]{}(\textsf{fold}\;\Varid{gen}\;(\lambda (\textsf{For}\;\Varid{iters}\;\Varid{k})\to h_{\textsf{For}}\;\Varid{alg}\;\Varid{iters}\;\Varid{k}))\;{}\<[E]%
\\
\>[31]{}(\textsf{For}\;(\textsf{fmap}\;iso_1\;\Varid{iters})\;(iso_1\, .\,\Varid{k})))){}\<[E]%
\\
\>[3]{}\hsindent{1}{}\<[4]%
\>[4]{}\equiv (\lambda (\textsf{For}\;\Varid{iters}\;\Varid{k})\to {}\<[28]%
\>[28]{}h_{\textsf{For}}\;\Varid{alg}\;\Varid{iters}\;\Varid{k})\;{}\<[E]%
\\
\>[4]{}\hsindent{8}{}\<[12]%
\>[12]{}(\textsf{hmap}\;\textsf{fold}_{2}\;(\textsf{For}\;{}\<[31]%
\>[31]{}(\textsf{fmap}\;iso_1\;\Varid{iters})\;{}\<[E]%
\\
\>[31]{}(\textsf{fold}\;\Varid{gen}\;(\lambda (\textsf{For}\;\Varid{iters}\;\Varid{k})\to h_{\textsf{For}}\;\Varid{alg}\;\Varid{iters}\;\Varid{k})\, .\,iso_1\, .\,\Varid{k}))){}\<[E]%
\\
\>[3]{}\hsindent{1}{}\<[4]%
\>[4]{}\equiv (\lambda (\textsf{For}\;\Varid{iters}\;\Varid{k})\to {}\<[28]%
\>[28]{}h_{\textsf{For}}\;\Varid{alg}\;\Varid{iters}\;\Varid{k})\;{}\<[E]%
\\
\>[4]{}\hsindent{8}{}\<[12]%
\>[12]{}(\textsf{For}\;{}\<[19]%
\>[19]{}(\textsf{fmap}\;\textsf{fold}_{2}\;(\textsf{fmap}\;iso_1\;\Varid{iters}))\;{}\<[E]%
\\
\>[19]{}(\textsf{fold}\;\Varid{gen}\;(\lambda (\textsf{For}\;\Varid{iters}\;\Varid{k})\to h_{\textsf{For}}\;\Varid{alg}\;\Varid{iters}\;\Varid{k})\, .\,iso_1\, .\,\Varid{k})){}\<[E]%
\\
\>[3]{}\hsindent{1}{}\<[4]%
\>[4]{}\equiv h_{\textsf{For}}\;\Varid{alg}\;{}\<[17]%
\>[17]{}(\textsf{fmap}\;\textsf{fold}_{2}\;(\textsf{fmap}\;iso_1\;\Varid{iters}))\;{}\<[E]%
\\
\>[17]{}(\textsf{fold}\;\Varid{gen}\;(\lambda (\textsf{For}\;\Varid{iters}\;\Varid{k})\to h_{\textsf{For}}\;\Varid{alg}\;\Varid{iters}\;\Varid{k}){}\<[66]%
\>[66]{}\, .\,iso_1\, .\,\Varid{k}){}\<[E]%
\\
\>[3]{}\hsindent{1}{}\<[4]%
\>[4]{}\equiv h_{\textsf{For}}\;\Varid{alg}\;{}\<[17]%
\>[17]{}(\textsf{fmap}\;(\textsf{fold}_{2}\, .\,iso_1)\;\Varid{iters})\;{}\<[E]%
\\
\>[17]{}(\textsf{fold}\;\Varid{gen}\;(\lambda (\textsf{For}\;\Varid{iters}\;\Varid{k})\to h_{\textsf{For}}\;\Varid{alg}\;\Varid{iters}\;\Varid{k}){}\<[66]%
\>[66]{}\, .\,iso_1\, .\,\Varid{k}){}\<[E]%
\\
\>[3]{}\hsindent{1}{}\<[4]%
\>[4]{}\equiv h_{\textsf{For}}\;\Varid{alg}\;{}\<[17]%
\>[17]{}(\textsf{fmap}\;(\textsf{fold}_{\textsf{Par}}\;(h_{\textsf{Var}}\;\Varid{alg})\;\Varid{alg})\;\Varid{iters})\;{}\<[E]%
\\
\>[17]{}(\textsf{fold}\;\Varid{gen}\;(\lambda (\textsf{For}\;\Varid{iters}\;\Varid{k})\to h_{\textsf{For}}\;\Varid{alg}\;\Varid{iters}\;\Varid{k}){}\<[66]%
\>[66]{}\, .\,iso_1\, .\,\Varid{k}){}\<[E]%
\\
\>[3]{}\hsindent{1}{}\<[4]%
\>[4]{}\equiv h_{\textsf{For}}\;\Varid{alg}\;{}\<[17]%
\>[17]{}(\textsf{fmap}\;(\textsf{fold}_{\textsf{Par}}\;(h_{\textsf{Var}}\;\Varid{alg})\;\Varid{alg})\;\Varid{iters})\;{}\<[E]%
\\
\>[17]{}(\textsf{fold}_{\textsf{Par}}\;\Varid{gen}\;\Varid{alg}\, .\,\Varid{k}){}\<[E]%
\\
\>[3]{}\hsindent{1}{}\<[4]%
\>[4]{}\equiv \textsf{fold}_{\textsf{Par}}\;\Varid{gen}\;\Varid{alg}\;(\Conid{For}\;\Varid{iters}\;\Varid{k}){}\<[E]%
\ColumnHook
\end{hscode}\resethooks

We need a helper lemma that states that \ensuremath{\textsf{fold}_{\textsf{Par}}\;(h_{\textsf{Var}}\;\Varid{alg})\;\Varid{alg}\mathrel{=}\textsf{fold}_{2}\, .\,iso_1}.
\begin{hscode}\SaveRestoreHook
\column{B}{@{}>{\hspre}l<{\hspost}@{}}%
\column{3}{@{}>{\hspre}l<{\hspost}@{}}%
\column{4}{@{}>{\hspre}l<{\hspost}@{}}%
\column{E}{@{}>{\hspre}l<{\hspost}@{}}%
\>[3]{}\textsf{fold}_{2}\;(iso_1\;(Var\;\Varid{x})){}\<[E]%
\\
\>[3]{}\hsindent{1}{}\<[4]%
\>[4]{}\equiv \textsf{fold}_{2}\;(Var_{\textsf{H}}\;\Varid{x}){}\<[E]%
\\
\>[3]{}\hsindent{1}{}\<[4]%
\>[4]{}\equiv \eta\;\Varid{x}{}\<[E]%
\ColumnHook
\end{hscode}\resethooks
\begin{hscode}\SaveRestoreHook
\column{B}{@{}>{\hspre}l<{\hspost}@{}}%
\column{3}{@{}>{\hspre}l<{\hspost}@{}}%
\column{4}{@{}>{\hspre}l<{\hspost}@{}}%
\column{11}{@{}>{\hspre}l<{\hspost}@{}}%
\column{E}{@{}>{\hspre}l<{\hspost}@{}}%
\>[3]{}\textsf{fold}_{2}\;(iso_1\;(\Conid{For}\;\Varid{iters}\;\Varid{k})){}\<[E]%
\\
\>[3]{}\hsindent{1}{}\<[4]%
\>[4]{}\equiv \textsf{fold}_{2}\;(Op_{\textsf{H}}\;(\textsf{For}\;(\textsf{fmap}\;iso_1\;\Varid{iters})\;(\textsf{fmap}\;iso_1\;\Varid{k}))){}\<[E]%
\\
\>[3]{}\hsindent{1}{}\<[4]%
\>[4]{}\equiv (\lambda (\textsf{For}\;\Varid{iters}\;\Varid{k})\to h_{\textsf{For}}\;\Varid{alg}\;\Varid{iters}\;\Varid{k})\;{}\<[E]%
\\
\>[4]{}\hsindent{7}{}\<[11]%
\>[11]{}(\textsf{hmap}\;\textsf{fold}_{2}\;(\textsf{fmap}\;\textsf{fold}_{2}\;(\textsf{For}\;(\textsf{fmap}\;iso_1\;\Varid{iters})\;(iso_1\, .\,\Varid{k})))){}\<[E]%
\\
\>[3]{}\hsindent{1}{}\<[4]%
\>[4]{}\equiv (\lambda (\textsf{For}\;\Varid{iters}\;\Varid{k})\to h_{\textsf{For}}\;\Varid{alg}\;\Varid{iters}\;\Varid{k})\;{}\<[E]%
\\
\>[4]{}\hsindent{7}{}\<[11]%
\>[11]{}(\textsf{hmap}\;\textsf{fold}_{2}\;(\textsf{For}\;(\textsf{fmap}\;iso_1\;\Varid{iters})\;(\textsf{fold}_{2}\, .\,iso_1\, .\,\Varid{k}))){}\<[E]%
\\
\>[3]{}\hsindent{1}{}\<[4]%
\>[4]{}\equiv (\lambda (\textsf{For}\;\Varid{iters}\;\Varid{k})\to h_{\textsf{For}}\;\Varid{alg}\;\Varid{iters}\;\Varid{k})\;{}\<[E]%
\\
\>[4]{}\hsindent{7}{}\<[11]%
\>[11]{}(\textsf{For}\;(\textsf{fmap}\;\textsf{fold}_{2}\;(\textsf{fmap}\;iso_1\;\Varid{iters}))\;(\textsf{fold}_{2}\, .\,iso_1\, .\,\Varid{k})){}\<[E]%
\\
\>[3]{}\hsindent{1}{}\<[4]%
\>[4]{}\equiv h_{\textsf{For}}\;\Varid{alg}\;(\textsf{fmap}\;\textsf{fold}_{2}\;(\textsf{fmap}\;iso_1\;\Varid{iters}))\;(\textsf{fold}_{2}\, .\,iso_1\, .\,\Varid{k}){}\<[E]%
\\
\>[3]{}\hsindent{1}{}\<[4]%
\>[4]{}\equiv h_{\textsf{For}}\;\Varid{alg}\;(\textsf{fmap}\;(\textsf{fold}_{2}\, .\,iso_1)\;\Varid{iters})\;(\textsf{fold}_{2}\, .\,iso_1\, .\,\Varid{k}){}\<[E]%
\\
\>[3]{}\hsindent{1}{}\<[4]%
\>[4]{}\equiv h_{\textsf{For}}\;\Varid{alg}\;(\textsf{fmap}\;(\textsf{fold}_{\textsf{Par}}\;(h_{\textsf{Var}}\;\Varid{alg})\;\Varid{alg})\;\Varid{iters})\;(\textsf{fold}_{\textsf{Par}}\;(h_{\textsf{Var}}\;\Varid{alg})\;\Varid{alg}\, .\,\Varid{k}){}\<[E]%
\\
\>[3]{}\hsindent{1}{}\<[4]%
\>[4]{}\equiv \textsf{fold}_{\textsf{Par}}\;(h_{\textsf{Var}}\;\Varid{alg})\;\Varid{alg}\;(\Conid{For}\;\Varid{iters}\;\Varid{k}){}\<[E]%
\ColumnHook
\end{hscode}\resethooks
\subsection{Writer Effect \& Handler}

First, we define the isomorphisms:
\begin{hscode}\SaveRestoreHook
\column{B}{@{}>{\hspre}l<{\hspost}@{}}%
\column{3}{@{}>{\hspre}l<{\hspost}@{}}%
\column{19}{@{}>{\hspre}l<{\hspost}@{}}%
\column{E}{@{}>{\hspre}l<{\hspost}@{}}%
\>[3]{}iso_1\mathbin{::}\Conid{Functor}\;\varphi\Rightarrow Free_{\textsf{Write}}\;\varphi\;\Varid{a}\to Free_{\textsf{H}}\;(K^{\textsf{Write}}\;\varphi)\;\Varid{a}{}\<[E]%
\\
\>[3]{}iso_1\;(Var\;\Varid{x}){}\<[19]%
\>[19]{}\mathrel{=}Var_{\textsf{H}}\;\Varid{x}{}\<[E]%
\\
\>[3]{}iso_1\;(\Conid{Exec}\;\Varid{op}){}\<[19]%
\>[19]{}\mathrel{=}Op_{\textsf{H}}\;(\textsf{Exec}\;((iso_1\, .\,\textsf{fmap}\;(\textsf{fmap}\;iso_1))\;\Varid{op})){}\<[E]%
\ColumnHook
\end{hscode}\resethooks
\begin{hscode}\SaveRestoreHook
\column{B}{@{}>{\hspre}l<{\hspost}@{}}%
\column{3}{@{}>{\hspre}l<{\hspost}@{}}%
\column{26}{@{}>{\hspre}l<{\hspost}@{}}%
\column{E}{@{}>{\hspre}l<{\hspost}@{}}%
\>[3]{}iso_2\mathbin{::}\Conid{Functor}\;\varphi\Rightarrow Free_{\textsf{H}}\;(K^{\textsf{Write}}\;\varphi)\;\Varid{a}\to Free_{\textsf{Write}}\;\varphi\;\Varid{a}{}\<[E]%
\\
\>[3]{}iso_2\;(Var_{\textsf{H}}\;\Varid{x}){}\<[26]%
\>[26]{}\mathrel{=}Var\;\Varid{x}{}\<[E]%
\\
\>[3]{}iso_2\;(Op_{\textsf{H}}\;(\textsf{Exec}\;\Varid{op})){}\<[26]%
\>[26]{}\mathrel{=}\Conid{Exec}\;((iso_2\, .\,\textsf{fmap}\;(\textsf{fmap}\;iso_2))\;\Varid{op}){}\<[E]%
\ColumnHook
\end{hscode}\resethooks
And next, we show that the requisite roundtrip properties hold,
i.e., \ensuremath{iso_1\, .\,iso_2\mathrel{=}id\mathrel{=}iso_2\, .\,iso_2}.
\begin{hscode}\SaveRestoreHook
\column{B}{@{}>{\hspre}l<{\hspost}@{}}%
\column{3}{@{}>{\hspre}l<{\hspost}@{}}%
\column{4}{@{}>{\hspre}c<{\hspost}@{}}%
\column{4E}{@{}l@{}}%
\column{8}{@{}>{\hspre}l<{\hspost}@{}}%
\column{E}{@{}>{\hspre}l<{\hspost}@{}}%
\>[3]{}(iso_1\, .\,iso_2)\;(Var_{\textsf{H}}\;\Varid{x}){}\<[E]%
\\
\>[3]{}\hsindent{1}{}\<[4]%
\>[4]{}\equiv {}\<[4E]%
\>[8]{}iso_1\;(Var\;\Varid{x}){}\<[E]%
\\
\>[3]{}\hsindent{1}{}\<[4]%
\>[4]{}\equiv {}\<[4E]%
\>[8]{}Var_{\textsf{H}}\;\Varid{x}{}\<[E]%
\ColumnHook
\end{hscode}\resethooks
\begin{hscode}\SaveRestoreHook
\column{B}{@{}>{\hspre}l<{\hspost}@{}}%
\column{3}{@{}>{\hspre}l<{\hspost}@{}}%
\column{4}{@{}>{\hspre}c<{\hspost}@{}}%
\column{4E}{@{}l@{}}%
\column{8}{@{}>{\hspre}l<{\hspost}@{}}%
\column{E}{@{}>{\hspre}l<{\hspost}@{}}%
\>[3]{}(iso_1\, .\,iso_2)\;(Op_{\textsf{H}}\;(\textsf{Exec}\;\Varid{op})){}\<[E]%
\\
\>[3]{}\hsindent{1}{}\<[4]%
\>[4]{}\equiv {}\<[4E]%
\>[8]{}iso_1\;(\Conid{Exec}\;((iso_2\, .\,\textsf{fmap}\;(\textsf{fmap}\;iso_2))\;\Varid{op})){}\<[E]%
\\
\>[3]{}\hsindent{1}{}\<[4]%
\>[4]{}\equiv {}\<[4E]%
\>[8]{}Op_{\textsf{H}}\;(\textsf{Exec}\;((iso_1\, .\,\textsf{fmap}\;(\textsf{fmap}\;iso_1))\;((iso_2\, .\,\textsf{fmap}\;(\textsf{fmap}\;iso_2))\;\Varid{op}))){}\<[E]%
\\
\>[3]{}\hsindent{1}{}\<[4]%
\>[4]{}\equiv {}\<[4E]%
\>[8]{}Op_{\textsf{H}}\;(\textsf{Exec}\;((iso_1\, .\,\textsf{fmap}\;(\textsf{fmap}\;iso_1)\, .\,iso_2\, .\,\textsf{fmap}\;(\textsf{fmap}\;iso_2))\;\Varid{op})){}\<[E]%
\\
\>[3]{}\hsindent{1}{}\<[4]%
\>[4]{}\equiv {}\<[4E]%
\>[8]{}Op_{\textsf{H}}\;(\textsf{Exec}\;((iso_1\, .\,iso_2\, .\,\textsf{fmap}\;(\textsf{fmap}\;iso_1\, .\,\textsf{fmap}\;iso_2))\;\Varid{op})){}\<[E]%
\\
\>[3]{}\hsindent{1}{}\<[4]%
\>[4]{}\equiv {}\<[4E]%
\>[8]{}Op_{\textsf{H}}\;(\textsf{Exec}\;((iso_1\, .\,iso_2\, .\,\textsf{fmap}\;(\textsf{fmap}\;(iso_1\, .\,iso_2)))\;\Varid{op})){}\<[E]%
\\
\>[3]{}\hsindent{1}{}\<[4]%
\>[4]{}\equiv {}\<[4E]%
\>[8]{}Op_{\textsf{H}}\;(\textsf{Exec}\;\Varid{op}){}\<[E]%
\ColumnHook
\end{hscode}\resethooks
\begin{hscode}\SaveRestoreHook
\column{B}{@{}>{\hspre}l<{\hspost}@{}}%
\column{3}{@{}>{\hspre}l<{\hspost}@{}}%
\column{4}{@{}>{\hspre}c<{\hspost}@{}}%
\column{4E}{@{}l@{}}%
\column{8}{@{}>{\hspre}l<{\hspost}@{}}%
\column{E}{@{}>{\hspre}l<{\hspost}@{}}%
\>[3]{}(iso_2\, .\,iso_1)\;(Var\;\Varid{x}){}\<[E]%
\\
\>[3]{}\hsindent{1}{}\<[4]%
\>[4]{}\equiv {}\<[4E]%
\>[8]{}iso_2\;(Var_{\textsf{H}}\;\Varid{x}){}\<[E]%
\\
\>[3]{}\hsindent{1}{}\<[4]%
\>[4]{}\equiv {}\<[4E]%
\>[8]{}Var\;\Varid{x}{}\<[E]%
\ColumnHook
\end{hscode}\resethooks
\begin{hscode}\SaveRestoreHook
\column{B}{@{}>{\hspre}l<{\hspost}@{}}%
\column{3}{@{}>{\hspre}l<{\hspost}@{}}%
\column{4}{@{}>{\hspre}c<{\hspost}@{}}%
\column{4E}{@{}l@{}}%
\column{8}{@{}>{\hspre}l<{\hspost}@{}}%
\column{E}{@{}>{\hspre}l<{\hspost}@{}}%
\>[3]{}(iso_2\, .\,iso_1)\;(\Conid{Exec}\;\Varid{op}){}\<[E]%
\\
\>[3]{}\hsindent{1}{}\<[4]%
\>[4]{}\equiv {}\<[4E]%
\>[8]{}iso_2\;(Op_{\textsf{H}}\;(\textsf{Exec}\;((iso_1\, .\,\textsf{fmap}\;(\textsf{fmap}\;iso_1))\;\Varid{op}))){}\<[E]%
\\
\>[3]{}\hsindent{1}{}\<[4]%
\>[4]{}\equiv {}\<[4E]%
\>[8]{}\Conid{Exec}\;((iso_2\, .\,\textsf{fmap}\;(\textsf{fmap}\;iso_2))\;(iso_1\, .\,\textsf{fmap}\;(\textsf{fmap}\;iso_1))\;\Varid{op}){}\<[E]%
\\
\>[3]{}\hsindent{1}{}\<[4]%
\>[4]{}\equiv {}\<[4E]%
\>[8]{}\Conid{Exec}\;((iso_2\, .\,\textsf{fmap}\;(\textsf{fmap}\;iso_2)\, .\,iso_1\, .\,\textsf{fmap}\;(\textsf{fmap}\;iso_1))\;\Varid{op}){}\<[E]%
\\
\>[3]{}\hsindent{1}{}\<[4]%
\>[4]{}\equiv {}\<[4E]%
\>[8]{}\Conid{Exec}\;((iso_2\, .\,iso_1\, .\,\textsf{fmap}\;(\textsf{fmap}\;iso_2\, .\,\textsf{fmap}\;iso_1))\;\Varid{op}){}\<[E]%
\\
\>[3]{}\hsindent{1}{}\<[4]%
\>[4]{}\equiv {}\<[4E]%
\>[8]{}\Conid{Exec}\;((iso_2\, .\,iso_1\, .\,\textsf{fmap}\;(\textsf{fmap}\;(iso_2\, .\,iso_1)))\;\Varid{op}){}\<[E]%
\\
\>[3]{}\hsindent{1}{}\<[4]%
\>[4]{}\equiv {}\<[4E]%
\>[8]{}\Conid{Exec}\;\Varid{op}{}\<[E]%
\ColumnHook
\end{hscode}\resethooks
\subsection{Latent Effects \& Handlers}

First, we define the isomorphisms:
\begin{hscode}\SaveRestoreHook
\column{B}{@{}>{\hspre}l<{\hspost}@{}}%
\column{3}{@{}>{\hspre}l<{\hspost}@{}}%
\column{27}{@{}>{\hspre}l<{\hspost}@{}}%
\column{E}{@{}>{\hspre}l<{\hspost}@{}}%
\>[3]{}iso_1\mathbin{::}Free_{\textsf{Lat}}\;\zeta\;\ell\;\Varid{a}\to Free_{\textsf{H}}\;(K^{\textsf{Lat}}\;\zeta\;\ell)\;\Varid{a}{}\<[E]%
\\
\>[3]{}iso_1\;(\Conid{Leaf}\;\Varid{x}){}\<[27]%
\>[27]{}\mathrel{=}Var_{\textsf{H}}\;\Varid{x}{}\<[E]%
\\
\>[3]{}iso_1\;(\Conid{Node}\;\Varid{op}\;\Varid{l}\;\Varid{sub}\;\Varid{k}){}\<[27]%
\>[27]{}\mathrel{=}Op_{\textsf{H}}\;(\textsf{Node}\;\Varid{op}\;\Varid{l}\;(\textsf{fmap}\;(\textsf{fmap}\;iso_1)\;\Varid{sub})\;(\textsf{fmap}\;iso_1\;\Varid{k})){}\<[E]%
\ColumnHook
\end{hscode}\resethooks
\begin{hscode}\SaveRestoreHook
\column{B}{@{}>{\hspre}l<{\hspost}@{}}%
\column{3}{@{}>{\hspre}l<{\hspost}@{}}%
\column{34}{@{}>{\hspre}l<{\hspost}@{}}%
\column{E}{@{}>{\hspre}l<{\hspost}@{}}%
\>[3]{}iso_2\mathbin{::}Free_{\textsf{H}}\;(K^{\textsf{Lat}}\;\zeta\;\ell)\;\Varid{a}\to Free_{\textsf{Lat}}\;\zeta\;\ell\;\Varid{a}{}\<[E]%
\\
\>[3]{}iso_2\;(Var_{\textsf{H}}\;\Varid{x}){}\<[34]%
\>[34]{}\mathrel{=}\Conid{Leaf}\;\Varid{x}{}\<[E]%
\\
\>[3]{}iso_2\;(Op_{\textsf{H}}\;(\textsf{Node}\;\Varid{op}\;\Varid{l}\;\Varid{sub}\;\Varid{k})){}\<[34]%
\>[34]{}\mathrel{=}\Conid{Node}\;\Varid{op}\;\Varid{l}\;(\textsf{fmap}\;(\textsf{fmap}\;iso_2)\;\Varid{sub})\;(\textsf{fmap}\;iso_2\;\Varid{k}){}\<[E]%
\ColumnHook
\end{hscode}\resethooks
And next, we show that the requisite roundtrip properties hold,
i.e., \ensuremath{iso_1\, .\,iso_2\mathrel{=}id\mathrel{=}iso_2\, .\,iso_2}.
\begin{hscode}\SaveRestoreHook
\column{B}{@{}>{\hspre}l<{\hspost}@{}}%
\column{3}{@{}>{\hspre}l<{\hspost}@{}}%
\column{4}{@{}>{\hspre}c<{\hspost}@{}}%
\column{4E}{@{}l@{}}%
\column{8}{@{}>{\hspre}l<{\hspost}@{}}%
\column{E}{@{}>{\hspre}l<{\hspost}@{}}%
\>[3]{}(iso_1\, .\,iso_2)\;(Var_{\textsf{H}}\;\Varid{x}){}\<[E]%
\\
\>[3]{}\hsindent{1}{}\<[4]%
\>[4]{}\equiv {}\<[4E]%
\>[8]{}iso_1\;(\Conid{Leaf}\;\Varid{x}){}\<[E]%
\\
\>[3]{}\hsindent{1}{}\<[4]%
\>[4]{}\equiv {}\<[4E]%
\>[8]{}Var_{\textsf{H}}\;\Varid{x}{}\<[E]%
\ColumnHook
\end{hscode}\resethooks
\begin{hscode}\SaveRestoreHook
\column{B}{@{}>{\hspre}l<{\hspost}@{}}%
\column{3}{@{}>{\hspre}l<{\hspost}@{}}%
\column{4}{@{}>{\hspre}c<{\hspost}@{}}%
\column{4E}{@{}l@{}}%
\column{8}{@{}>{\hspre}l<{\hspost}@{}}%
\column{E}{@{}>{\hspre}l<{\hspost}@{}}%
\>[3]{}(iso_1\, .\,iso_2)\;(Op_{\textsf{H}}\;(\textsf{Node}\;\Varid{op}\;\Varid{l}\;\Varid{sub}\;\Varid{k})){}\<[E]%
\\
\>[3]{}\hsindent{1}{}\<[4]%
\>[4]{}\equiv {}\<[4E]%
\>[8]{}iso_1\;(\Conid{Node}\;\Varid{op}\;\Varid{l}\;(\textsf{fmap}\;(\textsf{fmap}\;iso_2)\;\Varid{sub})\;(\textsf{fmap}\;iso_2\;\Varid{k})){}\<[E]%
\\
\>[3]{}\hsindent{1}{}\<[4]%
\>[4]{}\equiv {}\<[4E]%
\>[8]{}Op_{\textsf{H}}\;(\textsf{Node}\;\Varid{op}\;\Varid{l}\;(\textsf{fmap}\;(\textsf{fmap}\;iso_1)\;(\textsf{fmap}\;(\textsf{fmap}\;iso_2)\;\Varid{sub}))\;(\textsf{fmap}\;iso_1\;(\textsf{fmap}\;iso_2\;\Varid{k}))){}\<[E]%
\\
\>[3]{}\hsindent{1}{}\<[4]%
\>[4]{}\equiv {}\<[4E]%
\>[8]{}Op_{\textsf{H}}\;(\textsf{Node}\;\Varid{op}\;\Varid{l}\;(\textsf{fmap}\;(\textsf{fmap}\;iso_1\, .\,iso_2)\;\Varid{sub})\;(\textsf{fmap}\;(iso_1\, .\,iso_2)\;\Varid{k})){}\<[E]%
\\
\>[3]{}\hsindent{1}{}\<[4]%
\>[4]{}\equiv {}\<[4E]%
\>[8]{}Op_{\textsf{H}}\;(\textsf{Node}\;\Varid{op}\;\Varid{l}\;\Varid{sub}\;\Varid{k}){}\<[E]%
\ColumnHook
\end{hscode}\resethooks
\begin{hscode}\SaveRestoreHook
\column{B}{@{}>{\hspre}l<{\hspost}@{}}%
\column{3}{@{}>{\hspre}l<{\hspost}@{}}%
\column{4}{@{}>{\hspre}c<{\hspost}@{}}%
\column{4E}{@{}l@{}}%
\column{8}{@{}>{\hspre}l<{\hspost}@{}}%
\column{E}{@{}>{\hspre}l<{\hspost}@{}}%
\>[3]{}(iso_2\, .\,iso_1)\;(\Conid{Leaf}\;\Varid{x}){}\<[E]%
\\
\>[3]{}\hsindent{1}{}\<[4]%
\>[4]{}\equiv {}\<[4E]%
\>[8]{}iso_2\;(Var_{\textsf{H}}\;\Varid{x}){}\<[E]%
\\
\>[3]{}\hsindent{1}{}\<[4]%
\>[4]{}\equiv {}\<[4E]%
\>[8]{}\Conid{Leaf}\;\Varid{x}{}\<[E]%
\ColumnHook
\end{hscode}\resethooks
\begin{hscode}\SaveRestoreHook
\column{B}{@{}>{\hspre}l<{\hspost}@{}}%
\column{3}{@{}>{\hspre}l<{\hspost}@{}}%
\column{4}{@{}>{\hspre}c<{\hspost}@{}}%
\column{4E}{@{}l@{}}%
\column{8}{@{}>{\hspre}l<{\hspost}@{}}%
\column{E}{@{}>{\hspre}l<{\hspost}@{}}%
\>[3]{}(iso_2\, .\,iso_1)\;(\Conid{Node}\;\Varid{op}\;\Varid{l}\;\Varid{sub}\;\Varid{k}){}\<[E]%
\\
\>[3]{}\hsindent{1}{}\<[4]%
\>[4]{}\equiv {}\<[4E]%
\>[8]{}iso_2\;(Op_{\textsf{H}}\;(\textsf{Node}\;\Varid{op}\;\Varid{l}\;(\textsf{fmap}\;(\textsf{fmap}\;iso_1)\;\Varid{sub})\;(\textsf{fmap}\;iso_1\;\Varid{k}))){}\<[E]%
\\
\>[3]{}\hsindent{1}{}\<[4]%
\>[4]{}\equiv {}\<[4E]%
\>[8]{}\Conid{Node}\;\Varid{op}\;\Varid{l}\;(\textsf{fmap}\;(\textsf{fmap}\;iso_2)\;(\textsf{fmap}\;(\textsf{fmap}\;iso_1)\;\Varid{sub}))\;(\textsf{fmap}\;iso_2\;(\textsf{fmap}\;iso_1\;\Varid{k})){}\<[E]%
\\
\>[3]{}\hsindent{1}{}\<[4]%
\>[4]{}\equiv {}\<[4E]%
\>[8]{}\Conid{Node}\;\Varid{op}\;\Varid{l}\;(\textsf{fmap}\;(\textsf{fmap}\;(iso_2\, .\,iso_1))\;\Varid{sub})\;(\textsf{fmap}\;(iso_2\, .\,iso_1)\;\Varid{k}){}\<[E]%
\\
\>[3]{}\hsindent{1}{}\<[4]%
\>[4]{}\equiv {}\<[4E]%
\>[8]{}\Conid{Node}\;\Varid{op}\;\Varid{l}\;\Varid{sub}\;\Varid{k}{}\<[E]%
\ColumnHook
\end{hscode}\resethooks
\subsection{Bracketing Effect \& Handler}

First, we define the isomorphisms:
\begin{hscode}\SaveRestoreHook
\column{B}{@{}>{\hspre}l<{\hspost}@{}}%
\column{3}{@{}>{\hspre}l<{\hspost}@{}}%
\column{23}{@{}>{\hspre}l<{\hspost}@{}}%
\column{E}{@{}>{\hspre}l<{\hspost}@{}}%
\>[3]{}iso_1\mathbin{::}Free_{\textsf{Res}}\;\Varid{a}\to Free_{\textsf{H}}\;K^{\textsf{Res}}\;\Varid{a}{}\<[E]%
\\
\>[3]{}iso_1\;(Var\;\Varid{x}){}\<[23]%
\>[23]{}\mathrel{=}Var_{\textsf{H}}\;\Varid{x}{}\<[E]%
\\
\>[3]{}iso_1\;(\Conid{Bracket}\;\Varid{res}){}\<[23]%
\>[23]{}\mathrel{=}Op_{\textsf{H}}\;(\textsf{Bracket}\;(iso_1\;(\textsf{fmap}\;(\lambda (\Varid{x},\Varid{y})\to (iso_1\;\Varid{x},\textsf{return}\;(iso_1\;\Varid{y})))\;\Varid{res}))){}\<[E]%
\ColumnHook
\end{hscode}\resethooks
\begin{hscode}\SaveRestoreHook
\column{B}{@{}>{\hspre}l<{\hspost}@{}}%
\column{3}{@{}>{\hspre}l<{\hspost}@{}}%
\column{30}{@{}>{\hspre}l<{\hspost}@{}}%
\column{E}{@{}>{\hspre}l<{\hspost}@{}}%
\>[3]{}iso_2\mathbin{::}Free_{\textsf{H}}\;K^{\textsf{Res}}\;\Varid{a}\to Free_{\textsf{Res}}\;\Varid{a}{}\<[E]%
\\
\>[3]{}iso_2\;(Var_{\textsf{H}}\;\Varid{x}){}\<[30]%
\>[30]{}\mathrel{=}Var\;\Varid{x}{}\<[E]%
\\
\>[3]{}iso_2\;(Op_{\textsf{H}}\;(\textsf{Bracket}\;\Varid{res})){}\<[30]%
\>[30]{}\mathrel{=}\Conid{Bracket}\;(iso_2\;(\textsf{fmap}\;(\lambda (\Varid{x},\Varid{y})\to (iso_2\;\Varid{x},iso_2\;(\textsf{join}\;\Varid{y})))\;\Varid{res})){}\<[E]%
\ColumnHook
\end{hscode}\resethooks
And next, we show that the requisite roundtrip properties hold,
i.e., \ensuremath{iso_1\, .\,iso_2\mathrel{=}id\mathrel{=}iso_2\, .\,iso_2}.
\begin{hscode}\SaveRestoreHook
\column{B}{@{}>{\hspre}l<{\hspost}@{}}%
\column{3}{@{}>{\hspre}l<{\hspost}@{}}%
\column{5}{@{}>{\hspre}c<{\hspost}@{}}%
\column{5E}{@{}l@{}}%
\column{9}{@{}>{\hspre}l<{\hspost}@{}}%
\column{E}{@{}>{\hspre}l<{\hspost}@{}}%
\>[3]{}(iso_1\, .\,iso_2)\;(Var_{\textsf{H}}\;\Varid{x}){}\<[E]%
\\
\>[3]{}\hsindent{2}{}\<[5]%
\>[5]{}\equiv {}\<[5E]%
\>[9]{}iso_1\;(Var\;\Varid{x}){}\<[E]%
\\
\>[3]{}\hsindent{2}{}\<[5]%
\>[5]{}\equiv {}\<[5E]%
\>[9]{}Var_{\textsf{H}}\;\Varid{x}{}\<[E]%
\ColumnHook
\end{hscode}\resethooks
\begin{hscode}\SaveRestoreHook
\column{B}{@{}>{\hspre}l<{\hspost}@{}}%
\column{3}{@{}>{\hspre}l<{\hspost}@{}}%
\column{5}{@{}>{\hspre}c<{\hspost}@{}}%
\column{5E}{@{}l@{}}%
\column{9}{@{}>{\hspre}l<{\hspost}@{}}%
\column{24}{@{}>{\hspre}l<{\hspost}@{}}%
\column{25}{@{}>{\hspre}l<{\hspost}@{}}%
\column{27}{@{}>{\hspre}l<{\hspost}@{}}%
\column{37}{@{}>{\hspre}l<{\hspost}@{}}%
\column{E}{@{}>{\hspre}l<{\hspost}@{}}%
\>[3]{}(iso_1\, .\,iso_2)\;(Op_{\textsf{H}}\;(\textsf{Bracket}\;\Varid{res})){}\<[E]%
\\
\>[3]{}\hsindent{2}{}\<[5]%
\>[5]{}\equiv {}\<[5E]%
\>[9]{}iso_1\;(\Conid{Bracket}\;{}\<[24]%
\>[24]{}(iso_2\;(\textsf{fmap}\;(\lambda (\Varid{x},\Varid{y})\to (iso_2\;\Varid{x},iso_2\;(\textsf{join}\;\Varid{y})))\;\Varid{res}))){}\<[E]%
\\
\>[3]{}\hsindent{2}{}\<[5]%
\>[5]{}\equiv {}\<[5E]%
\>[9]{}Op_{\textsf{H}}\;(\textsf{Bracket}\;{}\<[24]%
\>[24]{}(iso_1\;(\textsf{fmap}\;{}\<[37]%
\>[37]{}(\lambda (\Varid{x},\Varid{y})\to (iso_1\;\Varid{x},\textsf{return}\;(iso_1\;\Varid{y})))\;{}\<[E]%
\\
\>[37]{}(iso_2\;(\textsf{fmap}\;(\lambda (\Varid{x},\Varid{y})\to (iso_2\;\Varid{x},iso_2\;(\textsf{join}\;\Varid{y})))\;\Varid{res}))))){}\<[E]%
\\
\>[3]{}\hsindent{2}{}\<[5]%
\>[5]{}\equiv {}\<[5E]%
\>[9]{}Op_{\textsf{H}}\;(\textsf{Bracket}\;{}\<[24]%
\>[24]{}((iso_1\, .\,\textsf{fmap}\;(\lambda (\Varid{x},\Varid{y})\to (iso_1\;\Varid{x},\textsf{return}\;(iso_1\;\Varid{y})))\, .\,{}\<[E]%
\\
\>[24]{}\hsindent{3}{}\<[27]%
\>[27]{}iso_2\, .\,\textsf{fmap}\;(\lambda (\Varid{x},\Varid{y})\to (iso_2\;\Varid{x},iso_2\;(\textsf{join}\;\Varid{y}))))\;\Varid{res})){}\<[E]%
\\
\>[3]{}\hsindent{2}{}\<[5]%
\>[5]{}\equiv {}\<[5E]%
\>[9]{}Op_{\textsf{H}}\;(\textsf{Bracket}\;{}\<[24]%
\>[24]{}((iso_1\, .\,iso_2\, .\,{}\<[E]%
\\
\>[24]{}\hsindent{1}{}\<[25]%
\>[25]{}\textsf{fmap}\;(\lambda (\Varid{x},\Varid{y})\to ((iso_1\, .\,iso_2)\;\Varid{x},(\textsf{return}\, .\,iso_1\, .\,iso_2\, .\,\textsf{join})\;\Varid{y}))))\;{}\<[E]%
\\
\>[24]{}\hsindent{1}{}\<[25]%
\>[25]{}\Varid{res})){}\<[E]%
\\
\>[3]{}\hsindent{2}{}\<[5]%
\>[5]{}\equiv {}\<[5E]%
\>[9]{}Op_{\textsf{H}}\;(\textsf{Bracket}\;{}\<[24]%
\>[24]{}((\textsf{fmap}\;(\lambda (\Varid{x},\Varid{y})\to (\Varid{x},(\textsf{return}\, .\,\textsf{join})\;\Varid{y}))))\;\Varid{res})){}\<[E]%
\\
\>[3]{}\hsindent{2}{}\<[5]%
\>[5]{}\equiv {}\<[5E]%
\>[9]{}Op_{\textsf{H}}\;(\textsf{Bracket}\;{}\<[24]%
\>[24]{}(\textsf{fmap}\;(\lambda (\Varid{x},\Varid{y})\to (\Varid{x},\Varid{y})))\;\Varid{res})){}\<[E]%
\\
\>[3]{}\hsindent{2}{}\<[5]%
\>[5]{}\equiv {}\<[5E]%
\>[9]{}Op_{\textsf{H}}\;(\textsf{Bracket}\;{}\<[24]%
\>[24]{}\Varid{res}){}\<[E]%
\ColumnHook
\end{hscode}\resethooks
\begin{hscode}\SaveRestoreHook
\column{B}{@{}>{\hspre}l<{\hspost}@{}}%
\column{3}{@{}>{\hspre}l<{\hspost}@{}}%
\column{5}{@{}>{\hspre}c<{\hspost}@{}}%
\column{5E}{@{}l@{}}%
\column{9}{@{}>{\hspre}l<{\hspost}@{}}%
\column{E}{@{}>{\hspre}l<{\hspost}@{}}%
\>[3]{}(iso_2\, .\,iso_1)\;(Var\;\Varid{x}){}\<[E]%
\\
\>[3]{}\hsindent{2}{}\<[5]%
\>[5]{}\equiv {}\<[5E]%
\>[9]{}iso_2\;(Var_{\textsf{H}}\;\Varid{x}){}\<[E]%
\\
\>[3]{}\hsindent{2}{}\<[5]%
\>[5]{}\equiv {}\<[5E]%
\>[9]{}Var\;\Varid{x}{}\<[E]%
\ColumnHook
\end{hscode}\resethooks
\begin{hscode}\SaveRestoreHook
\column{B}{@{}>{\hspre}l<{\hspost}@{}}%
\column{3}{@{}>{\hspre}l<{\hspost}@{}}%
\column{5}{@{}>{\hspre}c<{\hspost}@{}}%
\column{5E}{@{}l@{}}%
\column{9}{@{}>{\hspre}l<{\hspost}@{}}%
\column{18}{@{}>{\hspre}l<{\hspost}@{}}%
\column{21}{@{}>{\hspre}l<{\hspost}@{}}%
\column{E}{@{}>{\hspre}l<{\hspost}@{}}%
\>[3]{}(iso_2\, .\,iso_1)\;(\Conid{Bracket}\;\Varid{res}){}\<[E]%
\\
\>[3]{}\hsindent{2}{}\<[5]%
\>[5]{}\equiv {}\<[5E]%
\>[9]{}iso_2\;(Op_{\textsf{H}}\;(\textsf{Bracket}\;(iso_1\;(\textsf{fmap}\;(\lambda (\Varid{x},\Varid{y})\to (iso_1\;\Varid{x},\textsf{return}\;(iso_1\;\Varid{y})))\;\Varid{res})))){}\<[E]%
\\
\>[3]{}\hsindent{2}{}\<[5]%
\>[5]{}\equiv {}\<[5E]%
\>[9]{}\Conid{Bracket}\;{}\<[18]%
\>[18]{}(iso_2\;(\textsf{fmap}\;(\lambda (\Varid{x},\Varid{y})\to (iso_2\;\Varid{x},iso_2\;(\textsf{join}\;\Varid{y}))){}\<[E]%
\\
\>[18]{}(iso_1\;(\textsf{fmap}\;(\lambda (\Varid{x},\Varid{y})\to (iso_1\;\Varid{x},\textsf{return}\;(iso_1\;\Varid{y})))\;\Varid{res})))){}\<[E]%
\\
\>[3]{}\hsindent{2}{}\<[5]%
\>[5]{}\equiv {}\<[5E]%
\>[9]{}\Conid{Bracket}\;{}\<[18]%
\>[18]{}((iso_2\, .\,\textsf{fmap}\;(\lambda (\Varid{x},\Varid{y})\to (iso_2\;\Varid{x},iso_2\;(\textsf{join}\;\Varid{y})))\, .\,{}\<[E]%
\\
\>[18]{}\hsindent{3}{}\<[21]%
\>[21]{}iso_1\, .\,\textsf{fmap}\;(\lambda (\Varid{x},\Varid{y})\to (iso_1\;\Varid{x},\textsf{return}\;(iso_1\;\Varid{y}))))\;\Varid{res}){}\<[E]%
\\
\>[3]{}\hsindent{2}{}\<[5]%
\>[5]{}\equiv {}\<[5E]%
\>[9]{}\Conid{Bracket}\;{}\<[18]%
\>[18]{}((iso_2\, .\,iso_1\, .\,{}\<[E]%
\\
\>[18]{}\hsindent{3}{}\<[21]%
\>[21]{}\textsf{fmap}\;(\lambda (\Varid{x},\Varid{y})\to ((iso_2\, .\,iso_1)\;\Varid{x},(iso_2\, .\,\textsf{join}\, .\,\textsf{return}\, .\,iso_1)\;\Varid{y})))\;{}\<[E]%
\\
\>[18]{}\hsindent{3}{}\<[21]%
\>[21]{}\Varid{res}){}\<[E]%
\\
\>[3]{}\hsindent{2}{}\<[5]%
\>[5]{}\equiv {}\<[5E]%
\>[9]{}\Conid{Bracket}\;{}\<[18]%
\>[18]{}((\textsf{fmap}\;(\lambda (\Varid{x},\Varid{y})\to (\Varid{x},(iso_2\, .\,iso_1)\;\Varid{y})))\;\Varid{res}){}\<[E]%
\\
\>[3]{}\hsindent{2}{}\<[5]%
\>[5]{}\equiv {}\<[5E]%
\>[9]{}\Conid{Bracket}\;{}\<[18]%
\>[18]{}(\textsf{fmap}\;(\lambda (\Varid{x},\Varid{y})\to (\Varid{x},\Varid{y}))\;\Varid{res}){}\<[E]%
\\
\>[3]{}\hsindent{2}{}\<[5]%
\>[5]{}\equiv {}\<[5E]%
\>[9]{}\Conid{Bracket}\;{}\<[18]%
\>[18]{}\Varid{res}{}\<[E]%
\ColumnHook
\end{hscode}\resethooks

\end{document}